# Sustainable Venture Capital

**Samuel James Johnston**

S P Jain School of Global Management

This dissertation is submitted for the degree of

***Master of Business Administration (Executive)***

*Specialisation in Finance*

2022

# Declaration

I hereby declare that except where specific reference is made to the work of others, the contents of this dissertation are original and have not been submitted in whole or in part for consideration for any other degree or qualification in this, or any other University. This dissertation is the result of my own work and includes nothing which is the outcome of work done in collaboration, except where specifically indicated in the text.

Samuel James Johnston

2022

# Abstract

Sustainability initiatives are set to benefit greatly from the growing involvement of venture capital, in the same way that other technological endeavours have been enabled and accelerated in the post-war period. With the spoils increasingly being shared between shareholders and other stakeholders, this requires a more nuanced view than the finance-first methodologies deployed to date. Indeed, it is possible for a venture-backed sustainability startup to deliver outstanding results to society in general without returning a cent to investors, though the most promising outcomes deliver profit with purpose, satisfying all stakeholders in ways that make existing "extractive" venture capital seem hollow.

To explore this nascent area, a review of related research was conducted and social entrepreneurs & investors interviewed to construct a questionnaire assessing the interests and intentions of current & future ecosystem participants. Analysis of 114 responses received via several sampling methods revealed statistically significant relationships between investing preferences and genders, generations, sophistication, and other variables, all the way down to the level of individual UN Sustainable Development Goals (SDGs).

Several surprising findings were made, including that older respondents were more likely to concern themselves with sustainability, while youth had more urgent issues relating to employment. Males were more interested in solving hunger, while females are driven by equality and justice. Climate action was consistently the top priority.

Tools are provided for others to continue this research or focus it on their own audiences, and the results will be applied in developing a pioneering sustainability venture studio and fund.

# Dedication

This work is wholeheartedly dedicated to Lucy, and to her upcoming generation who will be more directly affected by these sustainability challenges (especially climate change), despite having been caused by the unsustainable activities of generations prior.

It constitutes my modest contribution to understanding and addressing the changes ahead, many of which I will not be around to experience. It is an attempt to bring to bear one of the most potent tools in finance — venture capital — in solving humanity's greatest challenges.

This work is also dedicated in loving memory of my dear friend from my Swiss chapter, Bernino Lind, who left us prematurely on 28 Feburary 2022 during the final stages of its preparation. Some of his last writings to me were that "*Planet Earth is rich and there is enough for us all, if we just learn to compromise and to share*".

Bernino believed "*Planet Earth can house about 12 billion people with no problems, given we share*", and he had faith in the technology to make it possible; I just wish he were still here to see us build and deploy it.

> "*Put some load on your shoulders. Have a purpose. Be something for others. Create something powerful by being willing to fail in all your nakedness.*
> *My wish for you; Peace and Prosperity.*
> *And most of all – have the courage to live a life true to yourself, not the life others expect of you.*"
> — Bernino Lind

# Acknowledgements

I would like to acknowledge to my loving family, without whose tolerance — from 2020 through 2022 and during a global pandemic — the coursework, research, and writing would not have been possible to complete.

I acknowledge all those professors and classmates, whose participation in the program influenced my writings in large and small ways, but in particular to my academic mentor, SPR Vittal. I also owe a debt of gratitude to my former colleague and industry mentor, Stephan Seyboth, who provided critical input based on first-hand experience to focus my research.

Finally, in the closing months of the program I had to undergo a series of surgeries for suspected skin cancer and appreciate the support I received from everyone involved in the process, including the survey participants themselves who helped me collect the data required on a condensed schedule. While the outcome was positive, the procedures were disruptive and had the potential to derail the project.

# Contents

# List of Figures

# List of Tables

# Nomenclature

CV: Corporate Ventures

IRR: Internal Rate of Return

M&A: Mergers & Acquisitions

MOIC: Multiple of Invested Capital

NGO: Non-Governmental Organisation

PhVC: Philanthropic Venture Capital

SDG: United Nations (UN) Sustainable Development Goals (SDGs)

SME: Small-to-Medium Enterprise

SPV: Special Purpose Vehicle

VC: Venture Capital

# Chapter 1 Introduction

## 1.1 Objectives

This study aims to investigate sustainable venture capital, both in terms of that which is allocated based on a thesis of sustainability, and that which is itself sustainable, based on the performance of today's sustainable venture capital firms and portfolio companies.

Too much of today's venture capital is extractive in nature, farming our personal information, taking advantage human psychological weaknesses, or exploiting limited natural resources and often degrading our environment in the process.

It asks the question: What actions should venture capital (VC) firms take to ensure sustainability of their investments and themselves? This is itself a play on words, as investments that do not incorporate sustainability are likely to prove increasingly unsustainable in a world of rising interest rates and inflation, supply chain pressures and disruptions, and of all the UN Sustainable Development Goals (SDGs), the impacts of climate change, among other forces.

The study draws on the skills and knowledge of management acquired from the S P Jain Global Executive MBA program, with a particular focus on subjects from the finance specialisation. It explores an enduring strategic solution to the problem of applying an investment thesis to financing the right projects (e.g., startups and scaleups with a sustainability focus) in the right way (e.g., pre-seed and seed funding) under the right model (e.g., venture builder or studio).

It makes a modest contribution to applied research in management of venture capital and financing sustainability projects, and gives several recommendations for same. The results will be applied as the foundation of the business strategy for a new venture studio, and in the investment thesis for any future associated fund/s.

## 1.2 Project Background

Sustainable Venture Capital is a relatively new concept and in this context a play on words, referring to both meanings:

- Venture capital concerned with sustainability: "avoidance of the depletion of natural resources in order to maintain an ecological balance", and
- Venture capital that is itself sustainable: "able to be maintained at a certain rate or level".

It is proposed that to be sustainable in the long term, venture capital must concern itself with sustainability either directly (i.e., at the heart of the business model), or indirectly (i.e., while pursuing another business model).

The UN reports we are failing to meet 2030 sustainability targets, yet the $200bn venture capital (VC) industry growing at 16% CAGR has traditionally funded extractive companies and is just starting to think about sustainability. B Corp certified venture capital firms like Blisce (Paris, France), Foundry Group (Colorado, US), Fifth Wall (California, US), and Bethnal Green (London, UK), are starting to appear, but are they themselves sustainable?

Society's unlimited consumption of limited resources has become an existential threat, and yet its allocation of finite capital is skewed towards new endeavours that are creating an unhappy & unhealthy population despite living in an age of abundance. The confluence of a global pandemic and climate change has created an opportunity to revisit our societal resource allocation process such that we can achieve both progress and profit.

## 1.3 Research Application

The outcome of the study serves as the primary input to the business model for a new sustainable venture studio based in Singapore with portfolio companies in the Asia Pacific region (including Australia & New Zealand) targeting a global audience. It also forms the basis of an investment thesis for an associated venture capital fund with limited partners privately sourced primarily from the United States and Europe (given the professional network of the author), which is expected to be both the first in a series, and one of the first of its kind.

Given the confluence of readily available capital and an overdue but significant and durable demand for sustainability projects, it is anticipated that the resulting portfolio companies will enjoy a higher probability of success than their peers. Indeed, venture studios already enjoy an unfair advantage over standalone startups given "60% of all companies created out of studios make it to Series A" (Zasowski, 2020).

As investor sentiment is nonetheless likely to be a critical success factor, it follows that this area should be a primary focus of the research. However, just as venture capital is not readily accessible to the wider market, nor are limited partners readily available to researchers, despite often having started their journey as successful retail investors. As such, retail investors were surveyed, filtered on features like portfolio size and/or used as a proxy to research the sentiment of venture capital investors.

Lessons learned on the types of projects investors are willing to support, aligned to the UN Sustainable Development Goals (SDGs), provides valuable input into the selection and genesis of portfolio companies. For maximum impact, the demands of investors should be aligned with the requirements to build a sustainable future. For example, there is already ample investment in good health and well-being (goal 3), but little in climate action (goal 13).

## 1.4 Expected Outcomes and Implications

In the lead-up to writing the report it was anticipated that some of the first sustainable venture capital firms would have already established market-beating portfolios despite a long bull run in equities, while for others it would have been too soon to tell.

Although it was also expected that performance of sustainability investments would initially lag that of traditional ventures having no such concerns for the future, going forward we may expect to see stakeholder governance beating shareholder governance in the medium- to long-term.

The implications of the research are significant in the context of their application, as they will be used to justify the investment of financial and human capital in a sustainable venture studio and any future associated fund/s.

# Chapter 2 Literature Review

## 2.1 Web of Science

A recent search[1] of the Web of Science Core Collection database found 615 articles between 1945 and 2017, which were filtered down to 387 article and review papers in English, spanning 209 scientific journals, the top 5 being listed in Table 1 (Antarciuc et al., 2018).

Table 1 – Top 5 scientific journals for sustainability articles

| Journal | Articles |
| --- | --- |
| Journal of Business Venturing | 43 |
| Energy Policy | 11 |
| Strategic Management Journal | 11 |
| Entrepreneurship policy and Practice | 8 |
| Journal of Management Studies | 8 |

We can exclude Energy Policy as it focuses on only a few specific Sustainable Development Goals (SDGs) rather than the majority: affordable and clean energy (goal 7), and climate action (goal 13). We can further deprioritise the Strategic Management Journal as most references to sustainability were in fact referring to "sustainable competitive advantage".

However, there were several papers covering the impact of the listing of established firms on sustainability indexes like the Dow Jones Sustainability Index (DJSI), finding that such events have little impact on stock market reaction. Some solace can however be found in the fact that the percentage of shares held by long-term investors increases, suggesting analysts and professional investors will pay more attention to CSR firms over time (Hawn et al., 2018).

[1] TOPIC: (venture capital) and (sustainab*) or (green) or (environment*) or (social* responsib*) or (cleantech)

As a result, the initial focus of research was on articles in the Journal of Business Venturing.

## 2.2 Venture Capital

### 2.2.1 Business angel decision making

In "*Business angel early stage decision making*" (Maxwell et al., 2011) the author considers why investors are interested in high growth ventures despite many of them failing to achieve their growth potential, often due to a lack of early-stage funding. They compare and contrast the approach adopted by angels and venture capital firms, bearing in mind that many of the latter start out as the former.

The author also builds on several existing models in proposing their own model for the investment process in Table 2, including several stages as likely to adopted by an investor assessing an entrepreneur as a limited partner assessing a venture capital fund. They also draw on research in behavioural economics and decision making to infer that angels use heuristics to minimise decision making effort with criteria that are easiest to retrieve. Opportunities that are unlikely to achieve their aspirations or which have a higher risk of failure than their tolerance level are rejected. Contrary to normative assumptions they tend not to trade off risk for return.

For those opportunities not yet filtered out, investors analyse behaviours for signs of low or excess levels of capabilities, experiences, or traits, then trust behaviours that inform relationship risk. Those who build trust are more likely to succeed than those who damage or violate trust, however damaging trust is not fatal if appropriate behavioural controls can be introduced. Better prepared and behaved entrepreneurs are less likely to be rejected.

Table 2 – Stages of the BA investment process (Maxwell et al., 2011)

| Summary | Model |
|---|---|
| Identification | Deal origination |
| Screening | Selection |
| | Post selection – investment return |
| | Post selection – resident risk |
| Evaluation | Entrepreneur assessment – managerial risk |
| | Relationship assessment – relationship risk |
| Structuring | Deal structuring |
| | Due diligence |
| | Agreement |
| Managing | Managing |
| Exit | Harvesting |

### 2.2.2 Diversification, risk, and returns

In "*Diversification, risk, and returns in venture capital*" (Buchner et al., 2017) the authors challenge the assumption that fund specialisation delivers higher returns, using finance theory to explain the positive relationship between fund diversification and performance. They explore the interactions between risk, diversification, and performance, proposing a "Risk Hypothesis" positing that greater diversification reduces risk thus enabling risk-averse fund managers to participate in riskier investments. The higher risk on average of each venture should deliver higher average returns, which they supported through the analysis of 308 VC fund investments across 10,131 portfolio companies.

Multivariate analysis was used to show the two dimensions of diversification considered (i.e., industry and development stage) lead to higher returns (i.e., IRR). The exception to the rule was less experienced venture capital funds which demonstrated no such relationship. This may be because more experienced fund managers have greater access to deal flow and therefore more opportunities to diversify. Fund managers venturing into industries in which they lack experience was also found to have a negative overall impact despite or indeed due to increased diversification.

For example, a fund manager may be able to invest in more early-stage ventures, offsetting the increased risk compared to later-stage ventures through diversification; higher potential opportunities do tend to carry greater risk. It is however important for the fund manager to operate within their domains of expertise, as their capacity to assist investees, and those investees' perception that they have the right knowledge and skills to help, is dependent on industry experience far more so than development stage experience. This means that an early-stage investor will have more success in late-stage investing than a health expert picking e-commerce companies.

Ultimately, with access to a team of fund managers with diverse industry experience, better returns can be obtained through diversification. However, attempting to diversify without that breadth of industry insight can have a negative impact on fund performance, so new and emerging managers may achieve optimal results from fund specialisation.

Note that this analysis of diversification focuses on portfolio composition rather than number of ventures, so its effects may be experienced even for a smaller portfolio.

### 2.2.3 Power law distributions

In "*Power law distributions in entrepreneurship: Implications for theory and research*" (Crawford et al., 2015) the authors examine 12,000 nascent, young, and hyper-growth firms and the distribution of variables including input (e.g., human and financial resources) and outcome (e.g., revenue and growth) variables. Extant entrepreneurship research (and social science research in general) improperly relies upon the assumption that these variables aggregate around a stable and meaningful mean, and that valid observations can be made based on the mean and standard deviation alone. Indeed, the mean is meaningless in most cases for a power law distribution (where most observations are far left of the mean), which is represented in Figure 1.

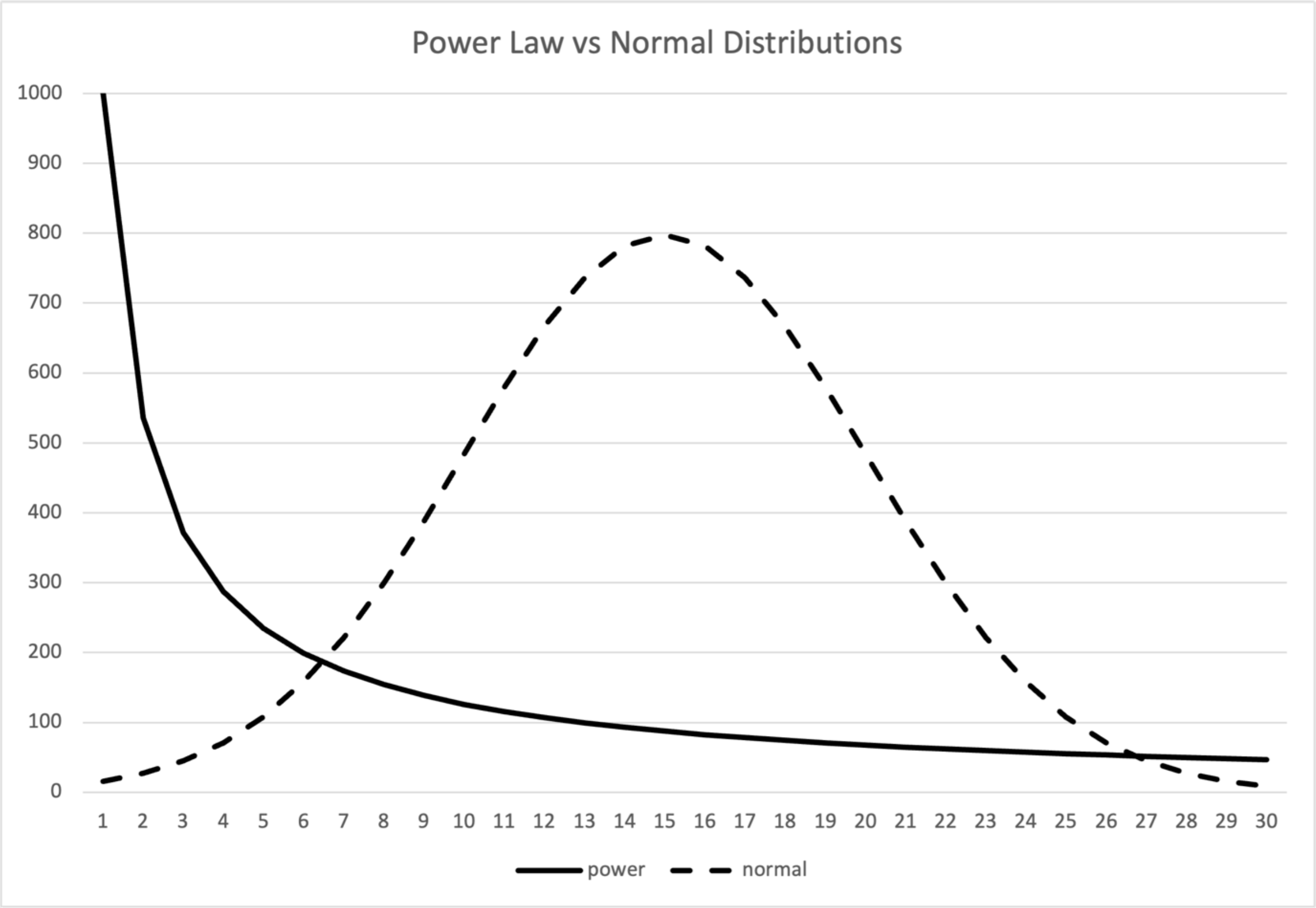

Figure 1 – Power law vs normal distributions

Examining firms across several industries, the researchers found 48 of 49 resource, cognition, action, and environment-based entrepreneurship variables playing central roles did in fact exhibit highly skewed power law distributions rather than the normal distributions that might have been expected. Should entrepreneurship research continue to rely on this invalid assumption, using data analytics tools including least squares regression and ANOVA rather than more appropriate techniques then the domain is unlikely to progress. Furthermore, practitioners relying on such research to identify very successful cases, which is the raison d'être of venture capital, will return sub-optimal results.

While venture capital portfolio management is built around the idea that very few of the firms (i.e., 1 in 20) will generate outsized returns (i.e., return the fund), this research suggests that input as well as outcome variables should be treated as power law distributions. The broad groups into which these 49 variables fall are given in Table 3, providing useful inspiration for development of both leading/input and lagging/outcome key performance indicators (KPIs).

Table 3 – Theoretical model inputs and outcome variables

| Input Variables | Outcome variables |
|---|---|
| Resources: human capital | Revenue ($000) |
| Resources: social capital | Employees |
| Resources: financial capital | Employee growth |
| Cognitions (i.e., future expectations) | |
| Actions (e.g. applying for a patent) | |
| Environment: industry sector ($000) | |

### 2.2.4 Corporate venture innovation

In "*Configurations for corporate venture innovation: Investigating the role of the dominant coalition*" (Waldkirch et al., 2021) the authors build on existing research into corporate ventures (CVs) relating to the "ambivalent role" of the dominant coalition (i.e., owners and managers) of the parent firms, furthering understanding of which configurations actually drive innovation; many do not, and many such initiatives ultimately fail to deliver innovation.

Firms invest resources in corporate ventures because they "can act freely outside the parent firm's boundaries and routines", enabling them to "exploit their creative potential". Successful endeavours typically not only involve the dominant coalition, but have them actively steering and reigning the venture as an important "pet project".

The resulting theoretical framework, represented in Table 4, has several parallels in the world of venture capital, and in particular venture studios which are effectively corporations solely dedicated to venturing. For example, the size and strategy of the firm as well as its intersection with the venture have a significant impact on the outcome, in addition to factors within the venture itself.

The research supports the top-down proactive driving of innovation typical of corporate ventures and venture studios, running counter to the assumption that innovation is typically reactive to opportunities from bottom-up. In cases with active dominant coalitions, it was external experiences with industry that led to innovation through the spotting of opportunities and simulating the venture to innovate. It also finds that high resource availability is a success factor, which supports structures where a firm or studio is paired with a dedicated fund.

One pertinent inhibitor identified was tight performance targets, without which corporate ventures were more able to experiment and innovate. In the context of venture capital, these typically take the form of premature revenue and profit targets, which incentivise the venture to prioritise growth often even before having identified a profitable business model. Balance also needs to be found between the influence of the dominant coalition and formalised venture planning processes, as the two can conflict.

Table 4 – Corporate venture innovation theoretical framework (Waldkirch et al., 2021)

| Causal Conditions | Outcome – CV Innovation |
|---|---|
| Parent firm level | New market focus |
| • Dominant coalition overall discretion | New product focus |
| ◦ CEO strength | New sales channel |
| ◦ Heir apparent | Degree of new knowledge |
| ◦ DC governance involvement | Relative effort required |
| ◦ Coalition visibility | |
| ◦ Market stability | |
| • Planned venturing strategy | |
| ◦ Venture planning mode | |
| • Resource availability | |
| ◦ Firm size (employees) | |
| Parent firm-venture intersection | |
| • Dominant coalition venture influence | |
| ◦ Supervisory board attention | |
| ◦ Board of management attention | |
| ◦ Venture leadership role | |
| ◦ Direct operational venture engagement | |
| Venture level | |
| • Planning autonomy | |
| • Operational autonomy | |
| • Operational independence | |
| • CV location | |
| • CV legal form | |

## 2.3 Sustainability

### 2.3.1 UN Sustainable Development Goals (SDGs)

In “*Do the United Nations' Sustainable Development Goals matter for social entrepreneurial ventures? A bottom-up perspective*” (Günzel-Jensen et al., 2020), the authors use a bottom-up approach to examine the contribution of social entrepreneurial ventures to the Sustainable Development Goal (SDG) framework in Figure 2. This brings said ventures great benefits, including facilitating resource mobilisation by bundling activities for addressing SDGs, and strengthening relationships with stakeholders by increasing legitimacy. The authors reveal the need for focusing governments and NGOs at the intersection of social entrepreneurs and the SDG framework, and for the social entrepreneurs themselves to know of, approve of, and materialise the SDGs so as to fully realise their potential in addressing them.

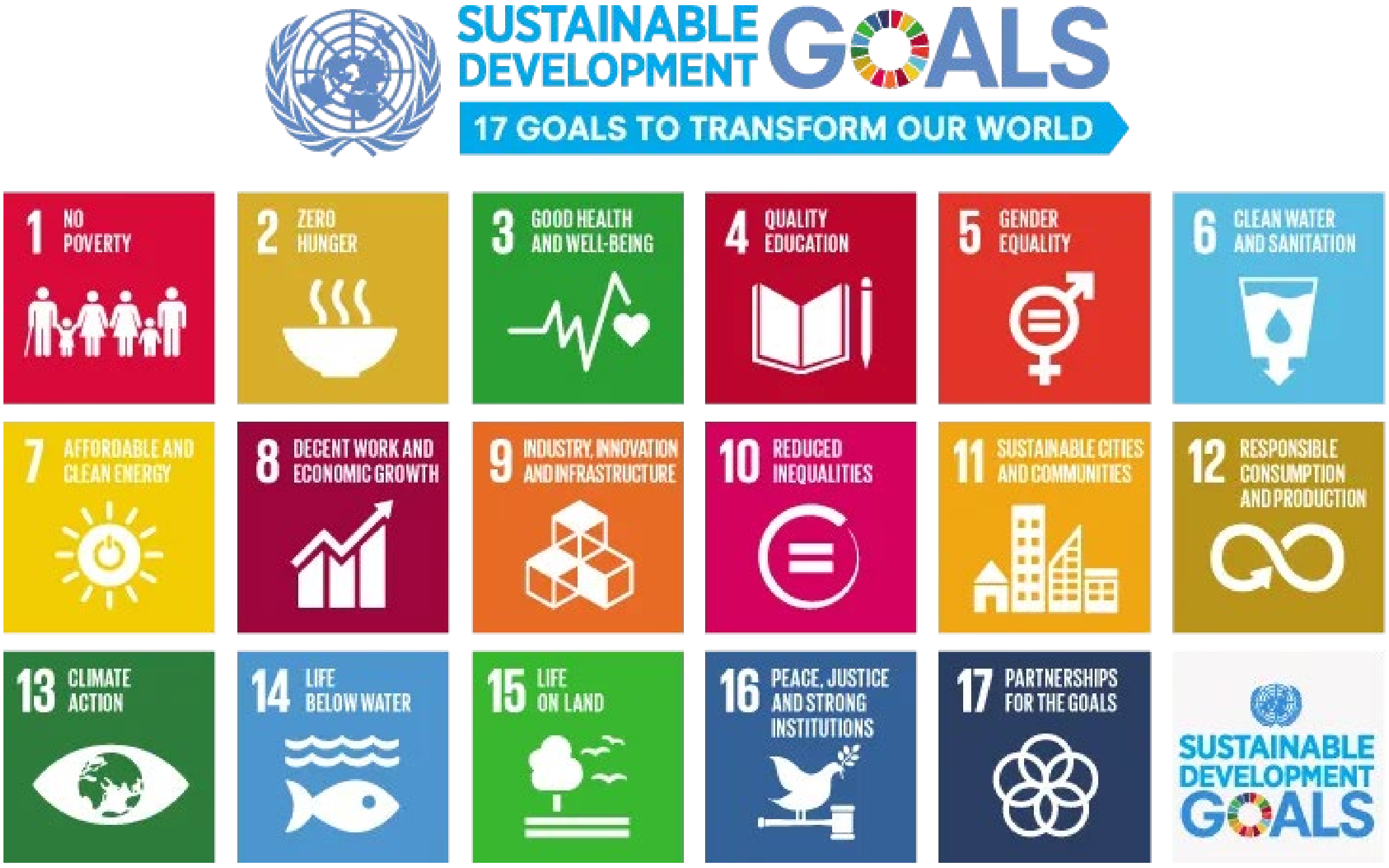

Figure 2 – UN Sustainable Development Goals (SDGs)

While research on societal grand challenges is still nascent, the authors dig deeper into social entrepreneurs' acceptance of the SDG framework and utilisation of same, identifying three types of utilisation summarised in Table 5.

Table 5 – SDG utilisation analytical procedure

| Type | Utilisation |
|---|---|
| SDG Evangelism | Organising frame (e.g. communication, organisation) |
| | Unifying frame (e.g. matching collaborators, signalling alignment) |
| SDG Opportunism | External requirements (e.g. resource mobilisation, performance) |
| | Non-binding manner (i.e., purely symbolic) |
| SDG Denial | Pragmatic reasons (e.g. operational irrelevance, impracticality) |
| | Idealistic reasons (e.g. grass-roots preference, distrust in NGOs) |

In terms of utilisation, of the 15 firms qualitatively evaluated:

- *SDG Evangelists* (*n*=5) considered working towards the goals an "affair of the heart", integrating them into work processes and developing tools around them to place them at the forefront of daily work. They became a common language to create mutual understanding, and to signal and communicate their performance to high-status stakeholders.
- *SDG Opportunists* (*n*=3) only used the goals as a means to an end, for example to mobilise resources from organisations that requested their use (e.g. government agencies). The SDGs were not applied as goals in and of themselves, nor were they integrated into operations due to their being symbolic in nature.
- *SDG Deniers* (*n*=7) formed the largest group for both pragmatic and idealistic reasons. Often they were too small to implement the framework, making it impractical at the micro- rather than macro-level. Others simply did not need a common language with the external world, being self-sustaining. Some even rejected the SDGs on idealistic reasons, whether due to their imperialistic nature or simply because of distrust in global organisations.

The UN SDGs serve an increasingly important purpose in the ecosystem, bringing significant benefits to those who embrace them fully (SDG Evangelists), working them into tools and processes and incorporating them into common language in daily work. That said, there remains benefit to be derived from using the SDGs tactically (SDG Opportunists), particularly for those to whom the framework is not relevant (for example, because they only tackle a single SDG), or who lack the resources to fully embrace them. It is less clear there is any benefit in rejecting the framework (SDG Deniers), particularly in light of its large and growing acceptance.

### 2.3.2 Sustainable venturing (B Corp certification)

In "*The double-edged sword of purpose-driven behaviour in sustainable venturing*" (Muñoz et al., 2018) the authors examine the entrepreneurial journey from idea to market, and particularly the impact of B Corp certification on organisations whose purpose extends beyond profit maximisation. They identify three paths for venture-driven entrepreneurs, shaped by distinct imprinting sequences. They also identify three critical sensitive windows: definition of scope of purpose, timing of purpose formalisation (i.e., B Corp certification), and shifts in feedback sources.

The authors ultimately challenge the linear relationship between purpose and purposeful organising, as well as the assumption that seeking B Corp certification at foundation of the firm is productive for the firm or society, regardless of when it is actually achieved. While it has been previously asserted that from a societal perspective, requiring (including via regulation) all new enterprises imprint purpose at foundation, premature imprinting of purpose can be detrimental if done prior to validating the business model with the market.

The three paths identified are as follows:

1. **Path 1** firms imprint purpose and receive B Corp certification early, with a long idea stage prior to having a business model and clear value proposition. They receive recognition via awards and media, which can be misconstrued as validation. This drives an orientation towards growth and expansion while the focus should be on business model validation.
2. **Path 2** firms exhibit an absence of purpose and unintended impact in the early stages, having a short idea stage and applying for B Corp certification comparatively later. Venturing is not necessarily triggered by a business idea that responds to social or environmental problems, with narrow purpose being built in as sustainable practices implemented start paying off. This purpose is expanded to cover additional sustainability challenges, with the market providing feedback in addition to non-market actors.
3. **Path 3** firms exhibit a change initiative and unintended emergence of a social movement, seeking to challenge the way other social enterprises, including B Corp certified firms, operate, considering their project a "vehicle for change" rather than a business. While this circular purpose is grandiose, neither market nor non-market actors understand the motives and mechanisms, so they do not engage beyond anecdotal information. The purpose is formalised, but deemed operationally redundant, and the business model and portfolio remain relevant only in so far as they further the objectives of that change initiative. This bounds value creation and delivery, informing a business model change within a general planning framework and transitioning to SME phase.

While an early imprinting of broad purpose (Path 1) may seem optimal, it can be "just a world-changing narrative that drives attention". Similarly, full commitment to a change initiative that happens to be instantiated in the form of a business with purpose formalised at a later date (Path 3) risks being detached from the market. Incremental integration of a narrower and more concrete purpose (Path 2) is more likely to be successful. That said, care should be taken to only transition from purpose to purposeful organisation and legally commit to a purpose that has already been market tested, even if it means missing the short 1-year window for startups to attain "B Corp Pending" status.

## 2.4 Related Topics

### 2.4.1 Philanthropic Venture Capital (PhVC)

In "*Deal Structuring in Philanthropic Venture Capital Investments: Financing Instrument, Valuation and Covenants*" (Scarlata & Alemany, 2011) the authors examine an alternative financing option for social entrepreneurs that, like traditional venture capital, provides both capital and value-added services to portfolio companies. Rather than seeking to maximise the financial returns however, PhVC adds an ethical dimension with a view to maximising social returns in addition to, or at the exclusion of, financial returns.

Where companies in a PhVC portfolio are non-profit entities they are subject to non-distribution constraints that prevent distribution of proceeds to owners, thereby better aligning the interests of investors and investees and avoiding moral hazard risks. In the event of a potential exit, non-profit proceeds would effectively be distributed to society, whether via ongoing services and activity or by way of transfer to a similar non-profit organisation.

Where portfolio companies are for-profit, two levels of analysis need to be considered. In the event the PhVC is itself for-profit, the proceeds can be distributed to the investors, but in the event it is structured as a non-profit, the non-distribution constraint applies and typically demands proceeds be redistributed within the fund or to similar organisations outside of it.

### 2.4.2 Innovation

Henry Chesbrough's "Open Business Models: how to thrive in the new innovation landscape" was also reviewed, which also comes recommended by Clayton M Christensen, the inspiration for the author's work on disruptive innovations. In "*The Open Business Model: Understanding an Emerging Concept*" (Weiblen, 2016), conceptual clarity was further refined with the introduction of a framework, a set of differentiation criteria, and a proposed definition: "An open business model describes the design or architecture of the value creation and value capturing of a focal firm, in which collaborative relationships with the ecosystem are central to explaining the overall logic."

This book itself focuses on the history of innovative businesses with an "inside-out" mindset, in which innovation is created internally and shared with the world for a price, typically as products, via licensing, or as is more and more the case, as a service (e.g. an electricity generator vs the utility grid). This is typical of a company like IBM, who was disrupted by client-server computing in the early 90s and again by cloud computing from around 2010.

The book addresses the shift towards open innovation which is less about building an innovation first and maintaining the trade secret, and more about appreciating that the overwhelming majority of innovation is outside the firm and needs to be incorporated via an "outside-in" process. This incorporates models where you are the provider of said innovations, which are licensed to others to sell potentially in competition with your own products.

Companies like P&G have adopted this approach, which is recommended by the book. Indeed, while startups "stand on the shoulders of giants" by building on innovations like cloud computing, the common strategy throughout the portfolio in the context of this research is to apply emerging technologies in disruptive business models. Licensing intellectual property from and/or to others (e.g., universities and integrators respectively) is also a valid model, albeit less relevant to this strategy which delivers services directly to businesses and consumers.

# Chapter 3 Research

This project incorporates both qualitative and quantitative data analysis, with a view to understanding reasons, motivations, and trends, and to quantify data from a sample and make generalisations about the population of interest.

The primary critical success factor for any venture capital initiative is the availability of capital for venturing. Thus, the question ("What actions should venture capital (VC) firms take to ensure sustainability of their investments and themselves?") is both about how to build and operate a startup, which has been extensively treated by academia and industry, and about how to obtain and ensure continued access to capital (i.e., by delivering outsized returns).

Initial access to venture capital is typically obtained by way of privately pitching an investment thesis to scores of limited partners, where the more attractive and relevant the pitch the more likely they are to invest. Continued access to capital, whether in a current or future fund, is then increasingly based on the track record of performance, which is not available for new managers.

Thus there are two stages for potential consideration: features that are of most interest in an initial pitch, and then metrics that are most indicative of success beyond traditional metrics which have already been extensively covered.

## 3.1 Methodology

This research called for the collection of qualitative and quantitative data from a representative sample of actual or potential sustainability investors to ascertain their preferences in relation to sustainable investments, per Table 6.

Qualitative interviews were used to determine the scope of the survey, which was deliberately restrained to 2–3 minutes (10–15 questions) in length. This decision was made to optimise for sample size over scope, as opting for a longer or more involved survey may well have made it difficult to obtain enough responses for the target minimum margin of error (i.e., 10%).

A draft survey was piloted with several interviewees and refined based on their feedback. For example, "other" options were incorporated, and several questions were made optional. The third question covering risk appetite nonetheless had a higher than normal abandonment rate.

Table 6 – Summary of research methodology

| Aspect | Approach |
|---|---|
| Data | Data on venture capital portfolio performance down to the individual company level, organized by firm, thesis, etc. |
| Sampling Design | Convenience sampling from network VCs (e.g., Antler) & cluster sampling from databases (e.g., by first letter of name). |
| Data collection methods | Structured data exports, qualitative interviews, quantitative surveys, and search/export from company databases. |
| Data analysis methods | Descriptive (e.g., multiple of invested capital) & inferential statistics (correlation, crosstabs, hypothesis testing, etc.). |

## 3.2 Analytics

To minimise double-handling, data was collected directly in a usable format via links to online survey tools (i.e., Typeform) in in public and private invitations sent via email and social media. It was analysed using common data analytics tools (i.e., IBM SPSS, Microsoft Excel).

Descriptive statistics were used to summarise the information and inferential statistics to draw conclusions from it, such as the propensity to invest depending on sustainability topics addressed. Data was obtained that was representative within a given margin of error (i.e., 10%) for an infinitely large population (i.e., retail investors), which was used as a more accessible proxy for the target population (i.e., venture capital investors, being a specialisedd subset of the larger population).

## 3.3 Sampling

Convenience, purposive, and snowball sampling methods were used to ensure that the constitution of the sample surveyed was such that the results were representative within the targeted margin of error (i.e., 10%), per Table 7.

Table 7 – Sampling methods

| Sampling method | Application |
|---|---|
| Convenience sampling | Approximately 35,000 individuals connected to the author on social media (e.g., LinkedIn, Twitter, Facebook) were invited to self-select for survey completion, with an additional filter function in the first questions. |
| Purposive sampling | Individuals were hand-picked from a professional network of over 10,000 contacts (i.e., LinkedIn), told about the context of the research, offered access to the results, and given a link to participate in the research. |
| Snowball sampling | Those included in the purposive sampling were also invited to refer colleagues who they believe may be interested in participating. |

### 3.3.1 Sample size

The target margin of error for an infinitely large population was less than 10%, with a confidence interval of 95%. This called for a minimum sample size of 97.

The sample analysed ($n$=114) resulted in a margin of error of 9.2% for an infinitely large population at a 95% confidence interval, which was within the 10% target.

## 3.4 Preparation

The data was prepared prior to analysis using several processes, which were automated using IBM SPSS Syntax in Appendix A.2.

### 3.4.1 Data cleansing

- "Other" options were provided for several questions (e.g., investor type, gender, generation) to catch any responses that had not been offered.

- Three respondents who indicated advanced investing strategies (i.e., startups, crypto, and managed funds) were classified as “Accredited (e.g., $1m assets/$200k income)”.
- One respondent who indicated that their year of birth (i.e., 1980) fell between two generations was classified as “Gen Y/Millennials (1981–1996)”.
- One respondent who indicated that their country (Germany) was not listed by the software (Botswana) was manually corrected.

### 3.4.2 Dichotomisation

Several variables were able to be reduced to two classifications to increase the sample size of each category, per Table 8.

Recoding was done using SPSS’s RECODE command, resulting variables being prefixed with “d_” to group dichotomised variables.

Table 8 – Dichotomised variables

| Dichotomised Variable | False (0) | True (1) |
|---|---|---|
| d_investor_type_pro | I don’t invest | Accredited |
| | Inactive | Qualified |
| | Retail | Institutional |
| d_risk_appetite_high | Neutral | Very high |
| | Low | High |
| | Very low | |
| d_holding_period_long | 1–5 years | More than 10 years |
| | Less than 1 year | 6–10 years |
| d_sustainability_preferred | Neutral | Very likely |
| | Unlikely | Likely |
| | Very unlikely | |
| d_generation_older | Gen Y/Millennials | Silent Generation |
| | Gen Z | Baby Boomers |
| | Gen A | Gen X |

## 3.5 Results

Of the approximately 35,000 participants invited to complete the survey via social media, only a small sample of which will have received the request due to selective news feed algorithms, 246 (0.7%) clicked through to the survey start page, 176 (71.5%) of them commenced the survey, and 114 (64.8%) of those completed it (*N*=114).

The average time to complete the survey was 4 minutes and 34 seconds, having been estimated by the software to take 3 minutes.

### 3.5.1 Demographics

#### 3.5.1.1 Gender

Asked for their gender ("*To which gender do you most identify?*"), 83 (74.1%) identified as male and 29 (25.4%) identified as female, *n* = 112. None identified as "Other".

Compared to the wider population (*M*=1.5), significantly more males than females self-selected to participate in the survey, *M*=1.74, $t(111) = 5.80$, $p < .001$.

Females were significantly more likely than males to:

- Recommend sustainable investments to others, $t(72) = -2.50$, $p = .007$.
- Hold an investment for longer, $t(40) = 2.34$, $p = .012$.

#### 3.5.1.2 Generation

Asked for their generation ("*Which generation do you primarily identify with?*"), respondents were primarily "Gen Y/Millennials (1981-1996" and secondarily "Gen X (1965-1979)" per Table 9, *n*=113.

Table 9 – Generation

| Generation | *f* | % | Cumulative % |
|---|---|---|---|
| Baby Boomers (1946-1964) | 6 | 5.3% | 5.3% |
| Gen X (1965-1979) | 36 | 31.9% | 37.2% |
| Gen Y/Millennials (1981-1996) | 70 | 61.9% | 99.1% |
| Gen Z (1997-2012) | 1 | 0.9% | 100.0% |

Older generations (d_generation_older) were significantly more likely to:

- Recommend sustainable investments to others (*M*=7.57, *SD*=1.96) than younger generations (*M*=6.42, *SD*=2.68), with medium effect, $t(105.8) = -2.62$, $p = .005$, $d = -.471$.
- Prefer sustainable investments over others (*M*=4.07, *SD*=.74) than younger generations (*M*=3.75, *SD*=.91), with medium effect, $t(111) = -1.96$, $p = .026$, $d = -.382$.

- Be a professional investor, with large effect, $t(55.6) = -3.97$, $p < .001$, $d = -.890$.

### 3.5.2 Country

Asked where they were from ("*Where you you live (country)?*"), participants were primarily from Singapore and secondarily from Australia, followed by India, United States, and Switzerland, per Figure 3.

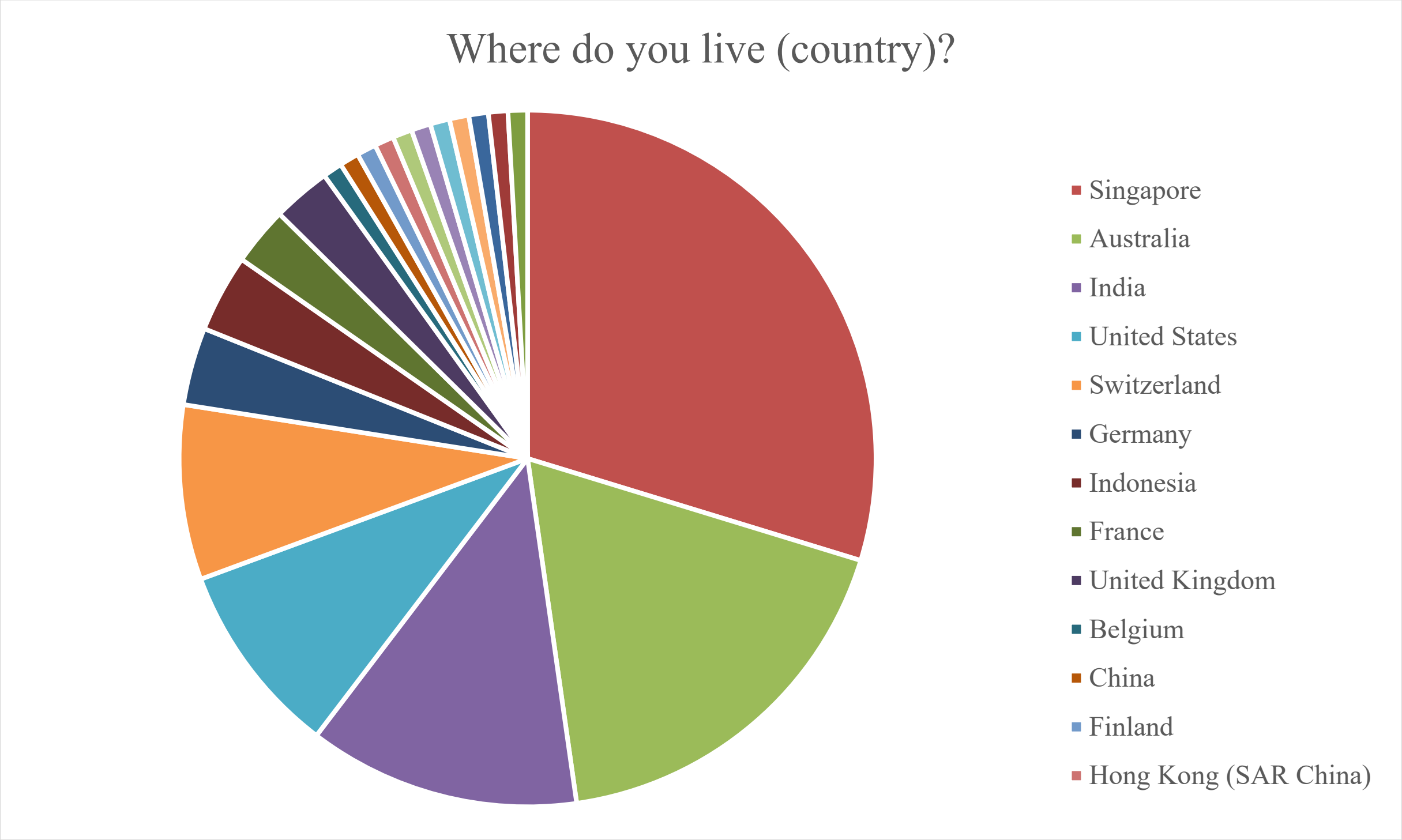

Figure 3 – Participants by country

No relationship was found between the country and other dependent variables studied using a one-way ANOVA test.

### 3.5.3 Investing

#### 3.5.3.1 Investor Type

Asked to classify their investing activities ("*What type of investor are you?*"), respondents were primarily "Retail (e.g., shares)" per Table 9, $n$=114.

Table 10 – Investor Type

| Type | *f* | % | Cumulative % |
|---|---|---|---|
| I don't invest | 6 | 5.3% | 5.3% |
| Inactive (e.g., pension) | 14 | 12.3% | 17.5% |
| Retail (e.g., shares) | 69 | 60.5% | 78.1% |
| Accredited (e.g., $1m assets/$200k income) | 16 | 14.0% | 92.1% |
| Qualified (e.g., $5m investments) | 3 | 2.6% | 94.7% |
| Institutional (e.g., corporate, pension) | 6 | 5.3% | 100.0% |

Professional investors (d_investor_type_pro) were significantly more likely to:

- Have a higher risk appetite ($M$=3.84, $SD$=.75) than amateurs ($M$=3.44, $SD$=.95), with medium effect, $t(48) = -2.23$, $p = .015$, $d = -.440$.

### 3.5.4 Risk Appetite

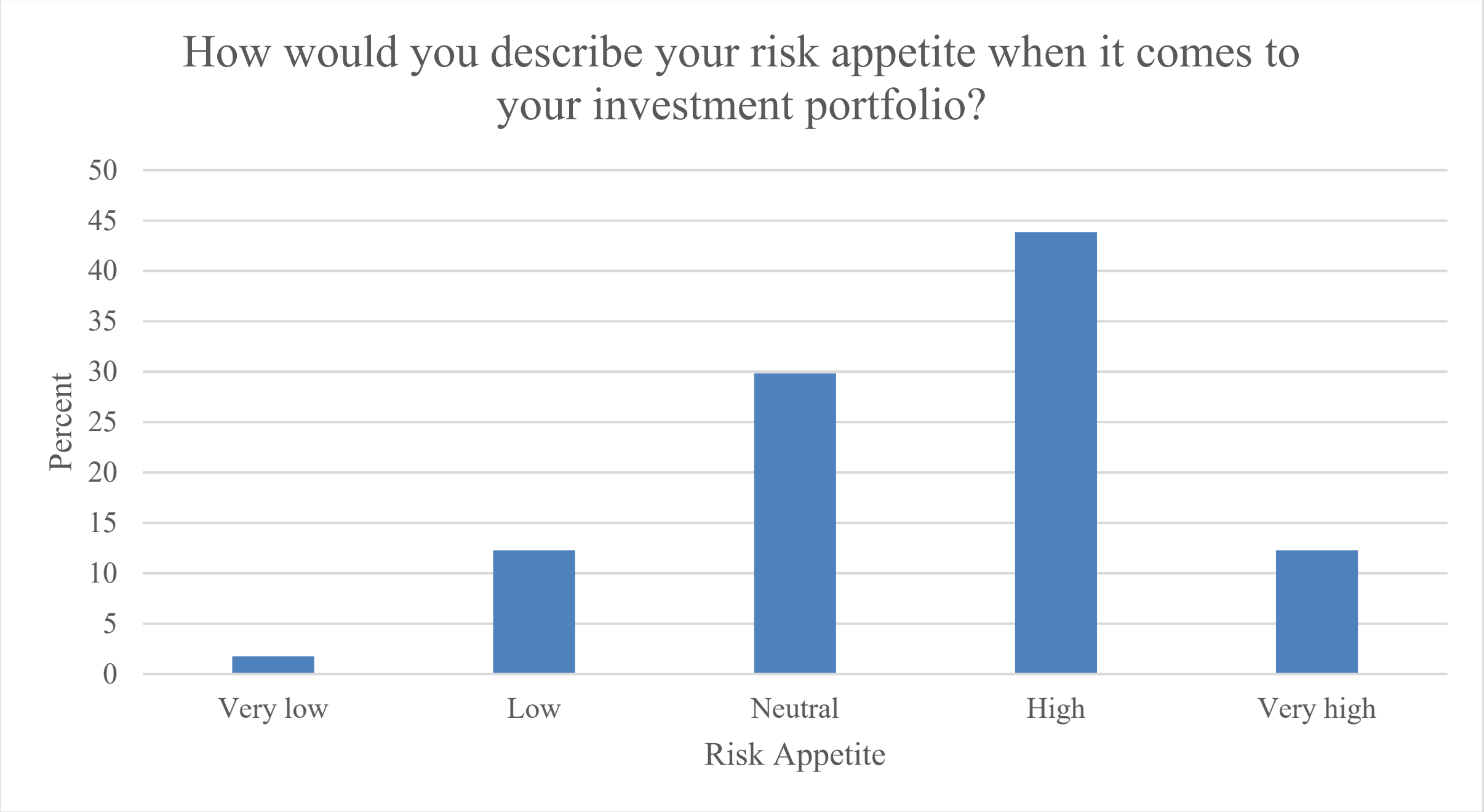

Figure 4 – Risk Appetite

Investors with a higher risk appetite (d_risk_appetite_high) were significantly more likely to:

- Be a professional investor (*M*=2.34, *SD*=1.03) than those with a low risk appetite (*M*=1.84, *SD*=.955), with medium effect, *t*(112) = -2.68, *p* = .004, *d* = -.506.
- Hold investments for shorter periods (*M*=2.34, *SD*=.877) than those with a low risk appetite (*M*=2.64, *SD*=.802), with medium effect, *t*(112) = 1.86, *p* = .033, *d* = -.351.
- Have a higher hurdle rate (*M*=13.33, *SD*=9.54) than those with a low risk appetite (*M*=8.85, *SD*=7.20), with medium effect, *t*(89) = -2.45, *p* = .008, *d* = -.520.

### 3.5.5 Holding Period

Asked for their holding period (“*How long would you typically hold an investment?*”) half of the respondents answered 1–5 years, per Figure 5.

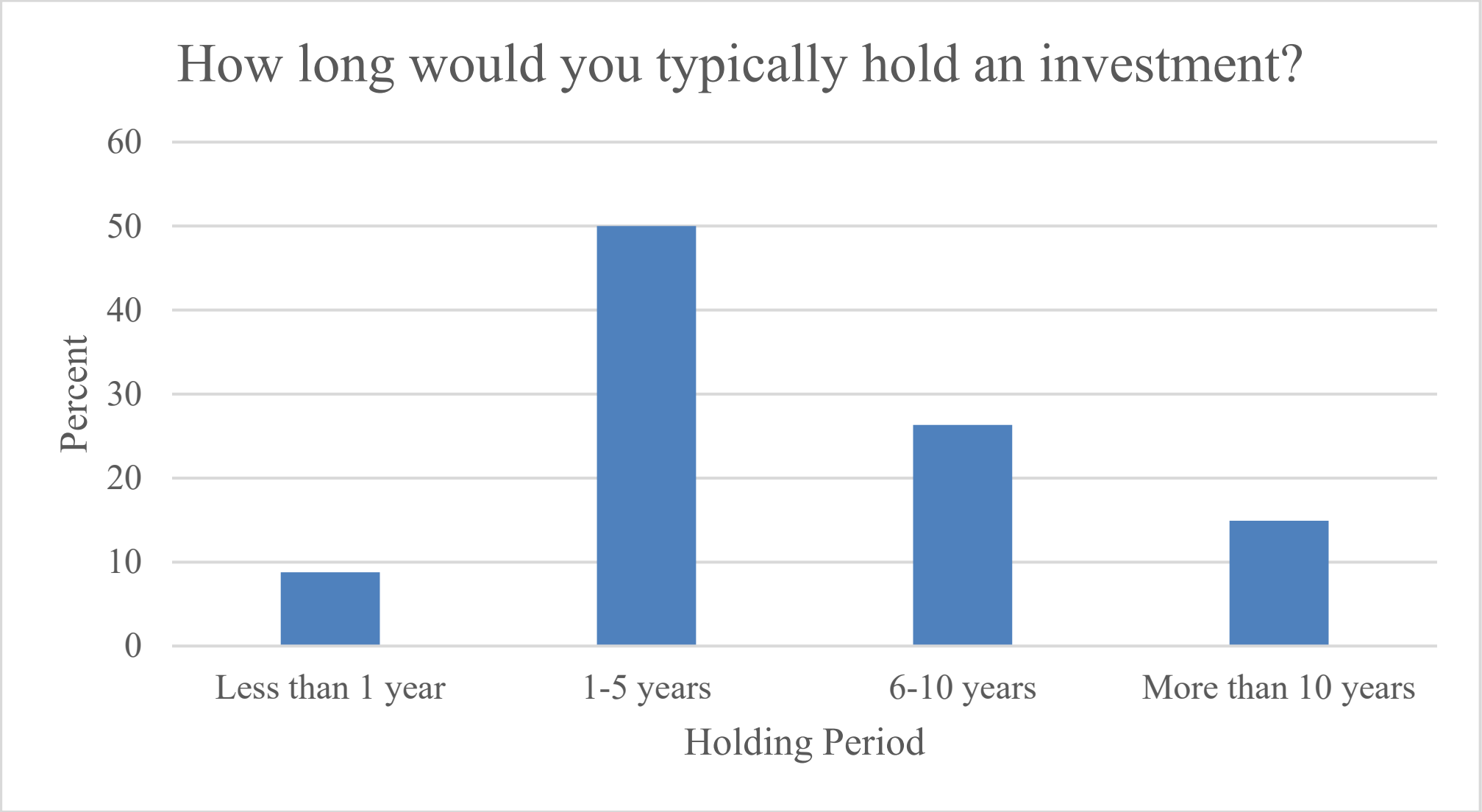

Figure 5 – Holding Period

### 3.5.6 Hurdle Rate

Asked for their hurdle rate ("*If you have one, what is your required rate of return or hurdle rate (%) for investments?*") on average respondents answered 11.41%, *Mdn*=10%, *N*=91, 95% CI [9.56%, 13.25%].

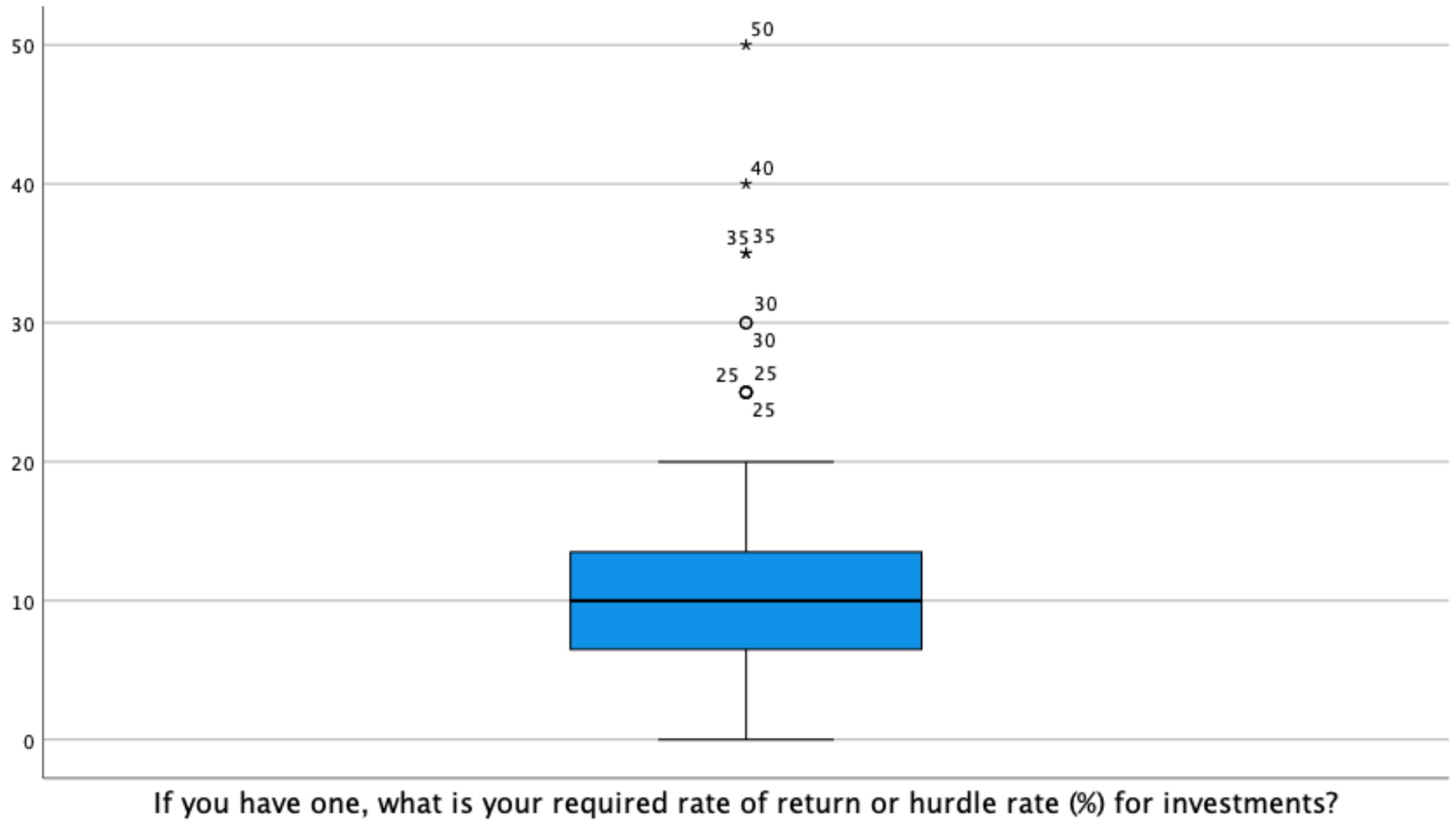

Figure 6 – Hurdle rate box plot

### 3.5.7 Sustainability Preference

Asked about their sustainability preference ("*How likely are you to prefer a sustainable investment over a traditional investment?*"), two thirds of participants were likely (40.4%, *f*=46) or very likely (26.3%, *f*=30) to prefer sustainable investments.

Almost 3 in 10 were neutral (28.1%, *f*=32) and barely 1 in 20 were unlikely to prefer sustainable investments (5.3%, *f*=6), per Figure 7.

None were very unlikely to prefer sustainable investments.

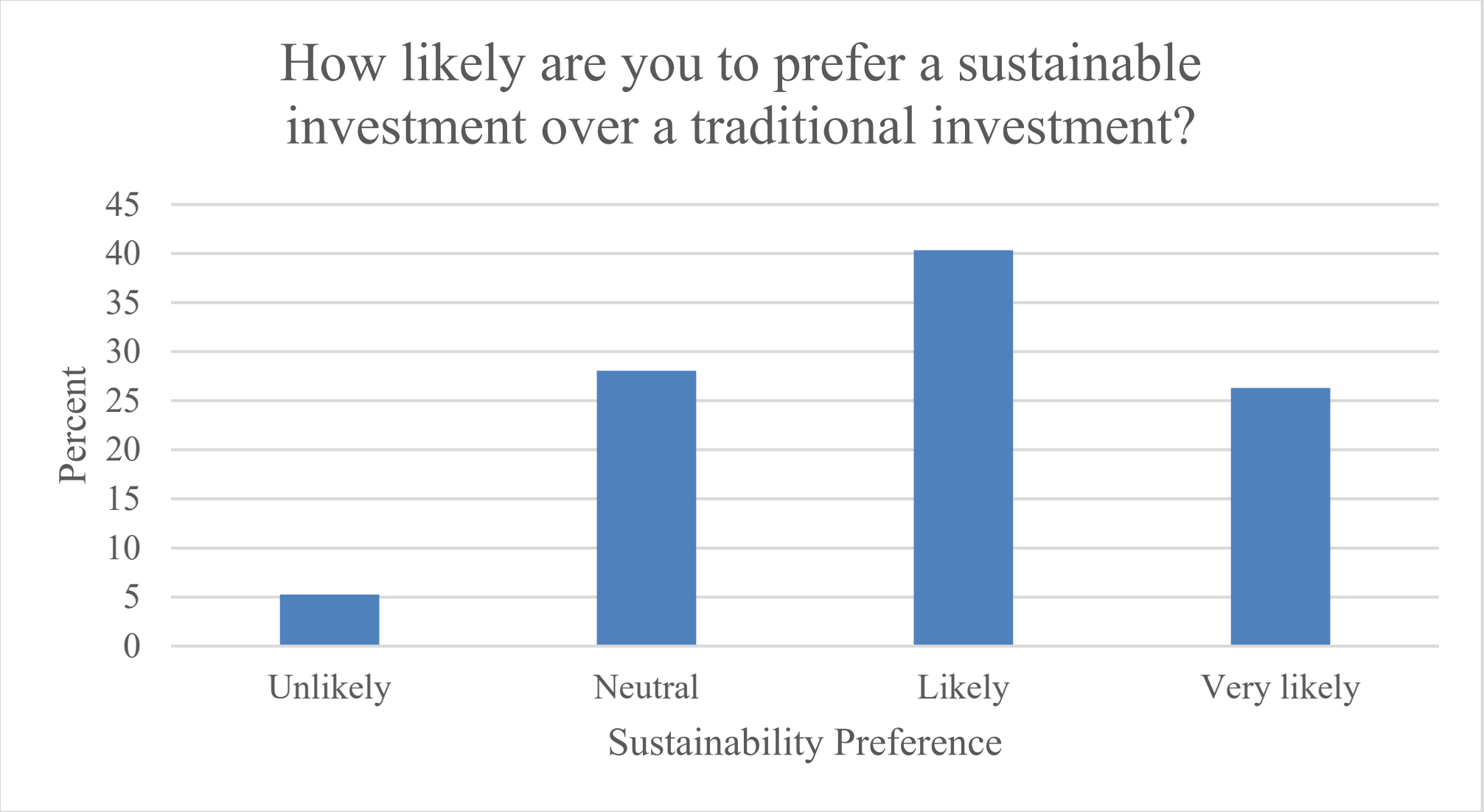

Figure 7 – Sustainability preference

### 3.5.8 Sustainability Returns

Asked about their expected returns for sustainable investments ("*What returns do you expect from a sustainable investment compared to the wider market?*"), a slight majority (57.9%, *f*=66) expected equivalent returns to the wider market, *N*=114.

The remainder was split between somewhat higher (22.8%, *f*=26) and somewhat lower (17.5%, *f*=20), with only one participant (0.9%, *f*=1) being extremely optimistic or pessimistic in selecting much lower or much higher, per Figure 8.

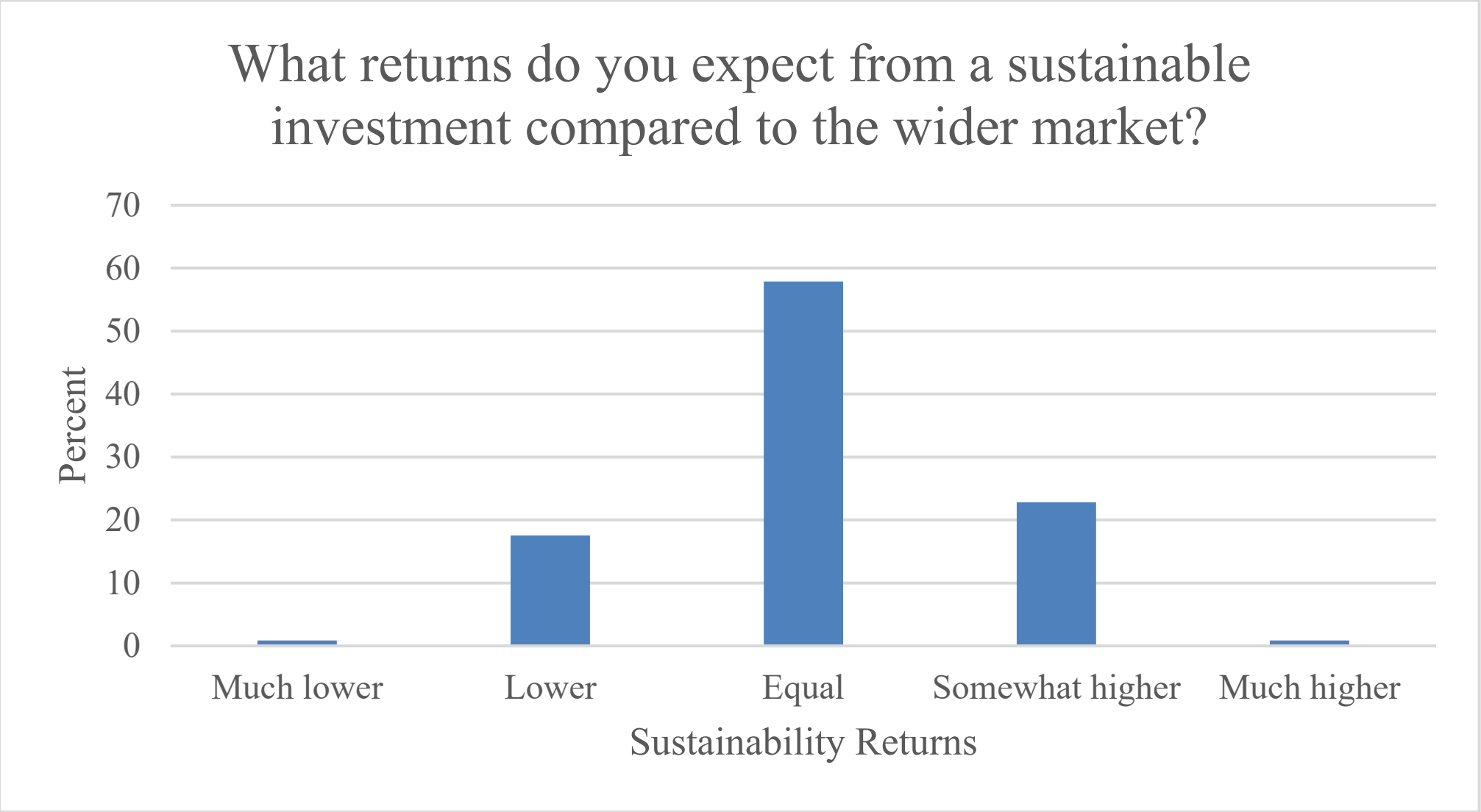

Figure 8 – Sustainability returns

### 3.5.9 Net Promoter Score (NPS)

In this survey, sustainable investments have a Net Promoter Score (NPS) of negative seven (-7), *N*=114.

The Net Promoter Score (NPS) is calculated by subtracting the number of detractors (score 0 – 7) from the number of promoters (9 – 10), per Table 11.

Table 11 – Net Promoter Score (NPS) categories

| NPS Type | *x* | *f* | % |
|---|---|---|---|
| Promoters | 9 – 10 | 36 | 37.7% |
| Passives | 8 – 9 | 35 | 30.7% |
| Detractors | 0 – 7 | 43 | 31.6% |

Table 12 – Net Promoter Score (NPS) histogram

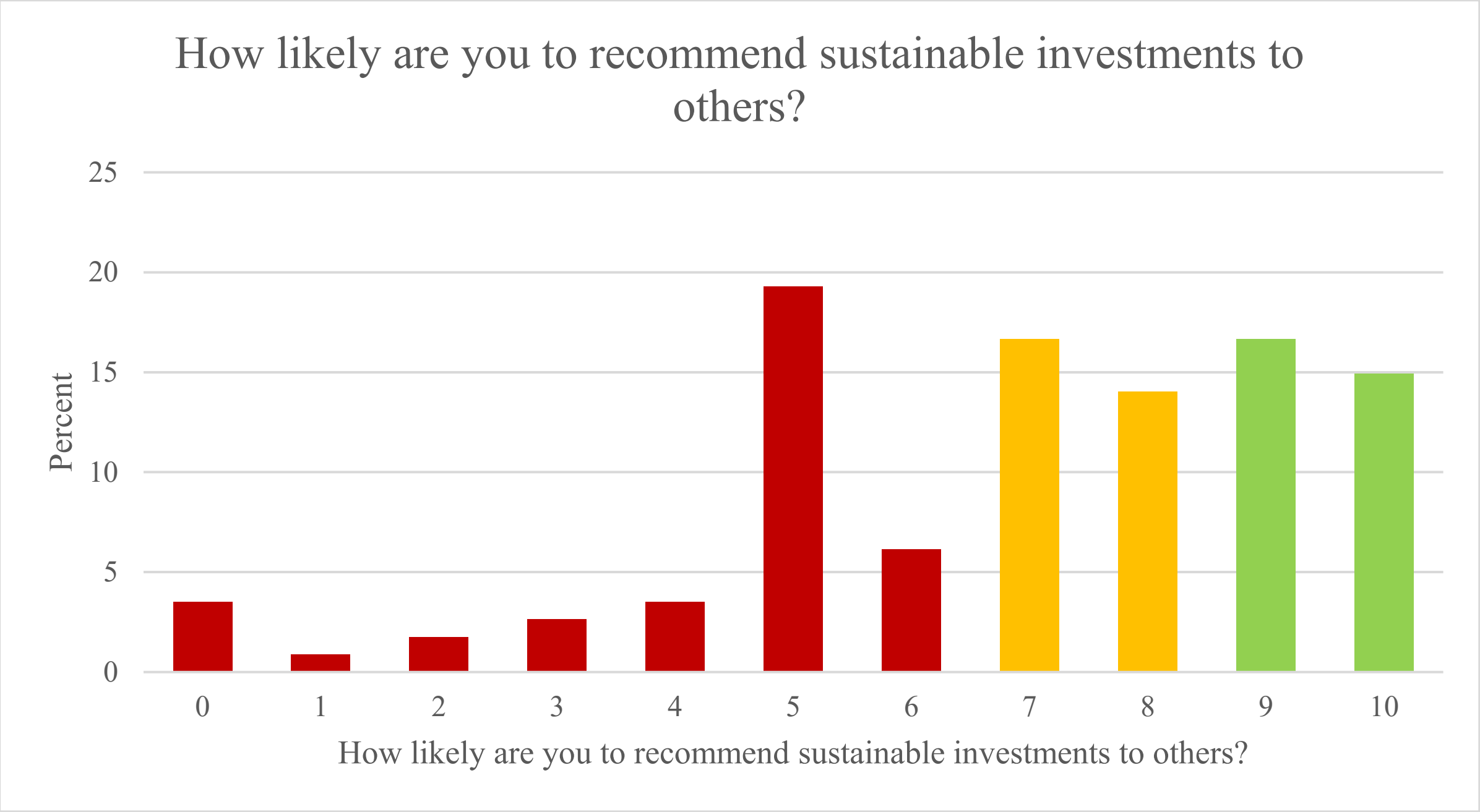

### 3.5.10 Motivation Rankings

Most participants were primarily motivated by environment, per Table 13, $f = 49$, $n = 110$.

As a ranking question, no "Other" option was given. Future research may want to attempt to identify "other" motivators, though for this study the advice of a sustainability professional was sought.

Table 13 – Motivation Rankings

| Motivation | *f (top ranking)* | *M* | *SD* |
|---|---|---|---|
| Environment | 49 | 3.95 | 1.15 |
| Resilience | 16 | 3.22 | 1.13 |
| Performance | 25 | 2.97 | 1.46 |
| Legacy | 16 | 2.91 | 1.35 |
| Empowerment | 4 | 1.95 | 1.22 |

### 3.5.11 Sustainable Development Goals (SDGs)

Participants were given the opportunity to "vote" for any three SDGs, and several statistically significant preferences based on gender and dichotomised generation were identified, per Table 14.

Table 14 – Sustainable Development Goal (SDG) preferences

| SDG | *f* | *Preferred* | $\chi 2$ | *p* |
|---|---|---|---|---|
| 1: No Poverty | 23 | | | |
| 2: Zero Hunger | 33 | Males | 6.88 | .009 |
| 3: Good Health & Well-Being | 24 | | | |
| 4: Quality Education | 38 | | | |
| 5: Gender Equality | 12 | Females | 30.30 | <.001 |
| 6: Clean Water & Sanitation | 33 | | | |
| 7: Affordable & Clean Energy | 32 | | | |
| 8: Decent Work & Economic Growth | 10 | Younger | 4.11 | .038 |
| 9: Industry, Innovation & Infrastructure | 11 | | | |
| 10: Reduced Inequalities | 17 | | | |
| 11: Sustainable Cities & Communities | 16 | | | |
| 12: Responsible Consumption & Prod. | 22 | | | |
| 13: Climate Action | 38 | Older | 6.17 | .010 |
| 14: Life Below Water | 11 | | | |
| 15: Life on Land | 4 | | | |
| 16: Peace, Justice & Strong Institutions | 11 | Females | 5.22 | .033 |
| 17: Partnership for the Goals | 1 | | | |

The top Sustainable Development Goals (SDGs) were, in order of preference:

1. SDG13: Climate Action
1. SDG4: Quality Education
2. SDG2: Zero Hunger
3. SDG7: Affordable and Clean Energy
3. SDG6: Clean Water and Sanitation
4. SDG8: Good Health and Well-Being
5. SDG1: No Poverty
5. SDG12: Responsible Consumption and Production

The frequencies of “voting” for each Sustainable Development Goal (SDG) are shown in Table 15.

Table 15 – Sustainable Development Goal (SDG) histogram

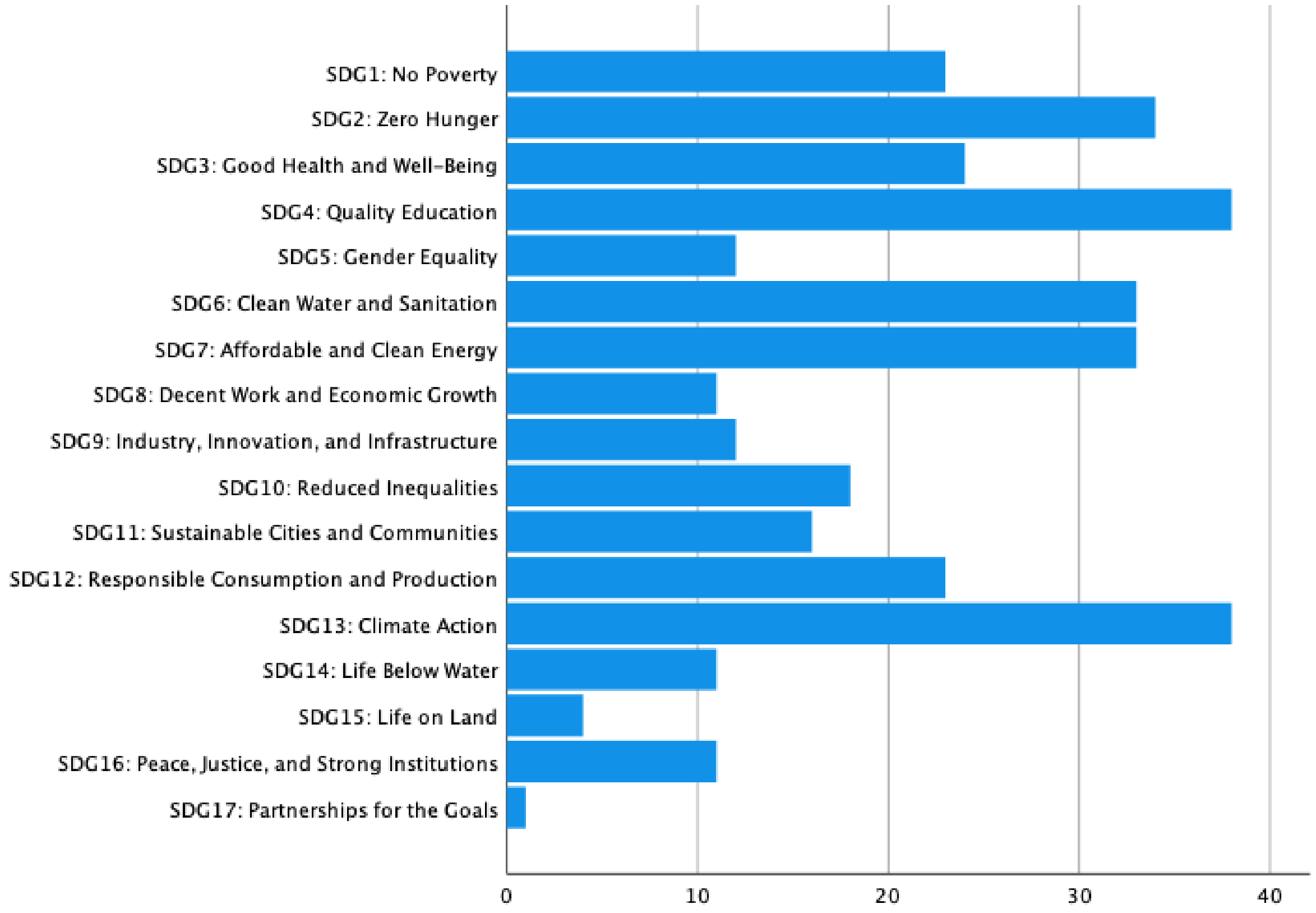

# Chapter 4 Company Databases

Several sources of live company data are maintained by private companies and made accessible via the Internet, including Crunchbase and PitchBook. Neither company readily offers research access, though while Crunchbase's Academic Research Access Program has a waitlist, they do offer commercial access.

Caution must be exercised in assuming normal (i.e., Gaussian) distributions apply to entrepreneurship (Crawford et al., 2015), as most relevant variables follow a power law distribution that skews heavily left of the mean. This finding significantly impacted this phase of the research, rendering many of the planned statistics (e.g., averages of any input or outcome variables) largely meaningless. While it is well understood that company performance and valuations in venture capital follow power laws, the same applies to almost all other entrepreneurial measurements.

## 4.1 Company rankings

Crunchbase includes a proprietary ranking system that enables pairwise comparison and ordering of companies in the database:

> "Crunchbase Rank (CB Rank) uses Crunchbase's intelligent algorithms to score and rank entities (e.g. Company, People, Investors, etc.) so you can quickly see what matters most in real time."

According to their algorithm, the top 10 sustainability companies are those in Figure 9, which includes links to the relevant database entries.

Sustainability companies were significantly more likely to have a higher CB Rank ranking ($M$=351,617) than others ($M$=456,105), with medium effect, $t(568.6) = 3.612$, $p < .001$, $d = .216$. The 90–day trend score was unrelated.

| Company | Country | Founded | Founders | Funding ($m) |
|---|---|---|---|---|
| Ola Electric | India | 2017 | 3 | 861 |
| BlocPower | US | 2014 | 2 | 111 |
| Climeworks | Switzerland | 2009 | 2 | 784 |
| The EVERY Company | US | 2015 | 2 | 240 |
| Northvolt | Sweden | 2016 | 2 | 6,040 |
| Blue Origin | US | 2000 | 1 | 167 |
| Natural Fiber Welding | US | 2015 | 1 | 156 |
| TemperPack | France | 2015 | 3 | 209 |
| Fluence | US | 2018 | 1 | 125 |
| Gogoro | US | 2011 | 2 | 480 |

Figure 9 – Top 10 sustainability companies by CB Rank

## 4.2 Sustainability taxonomy

Of over a million companies in the database, some 39,590 were classified in a "Sustainability" industry group including Biofuel, Biomass Energy, Clean Energy,

CleanTech, Energy Efficiency, Environmental Engineering, Green Building, Green Consumer Goods, GreenTech, Natural Resources, Organic, Pollution Control, Recycling, Renewable Energy, Solar, Sustainability, Waste Management, Water Purification, and Wind Energy, per Figure 10.

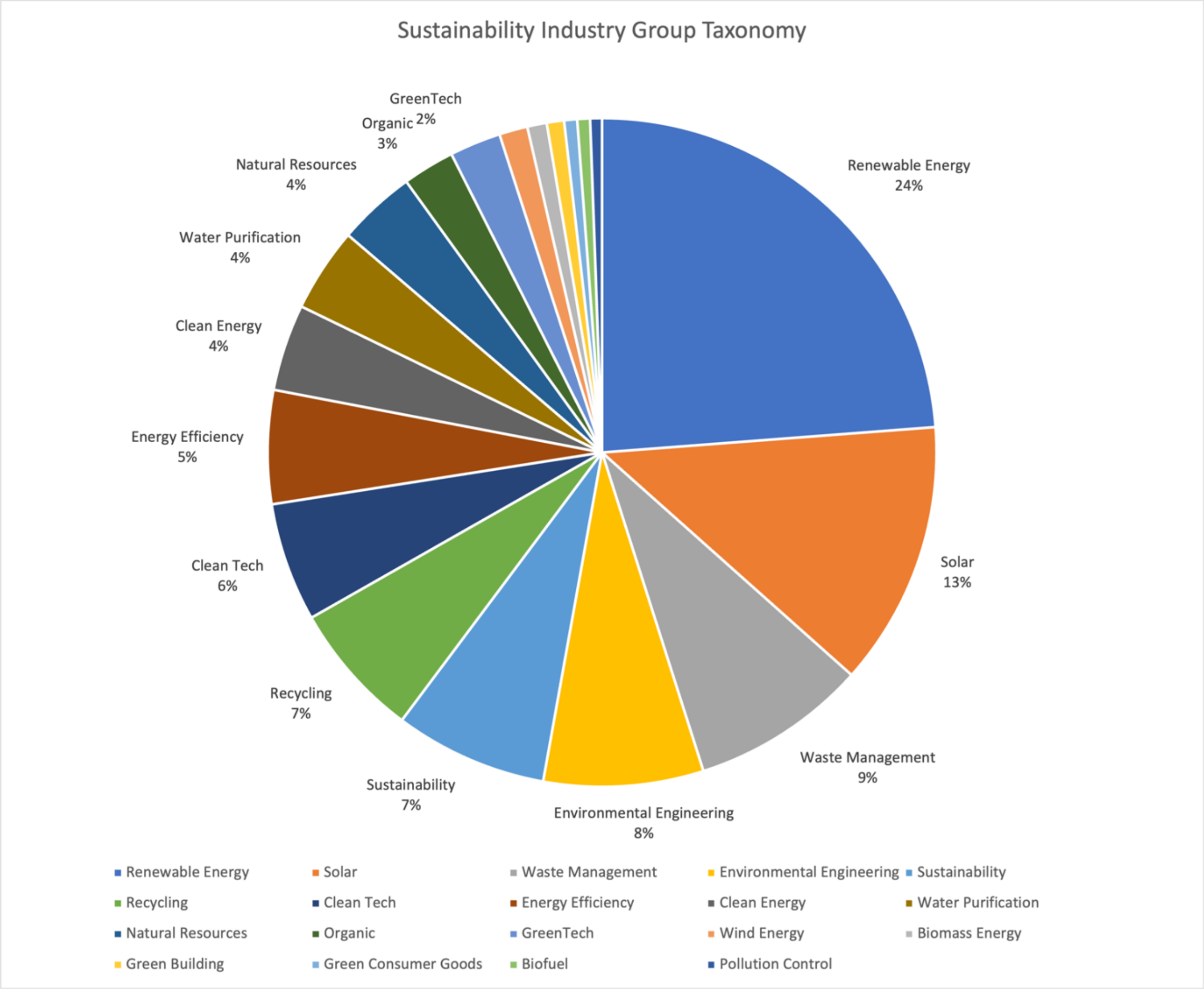

Figure 10 – Sustainability industry group taxonomy

## 4.3 Regional startup activity

Over the past 5 years (i.e., founding date from 2017 inclusive) the majority of the 38,112 sustainability startups identified by region were in North America (*f*=17,438), followed by Europe (*f*=13,645), Asia (*f*=4,370), Oceania (*f*=1,085), and Africa (*f*=577), per Figure 11.

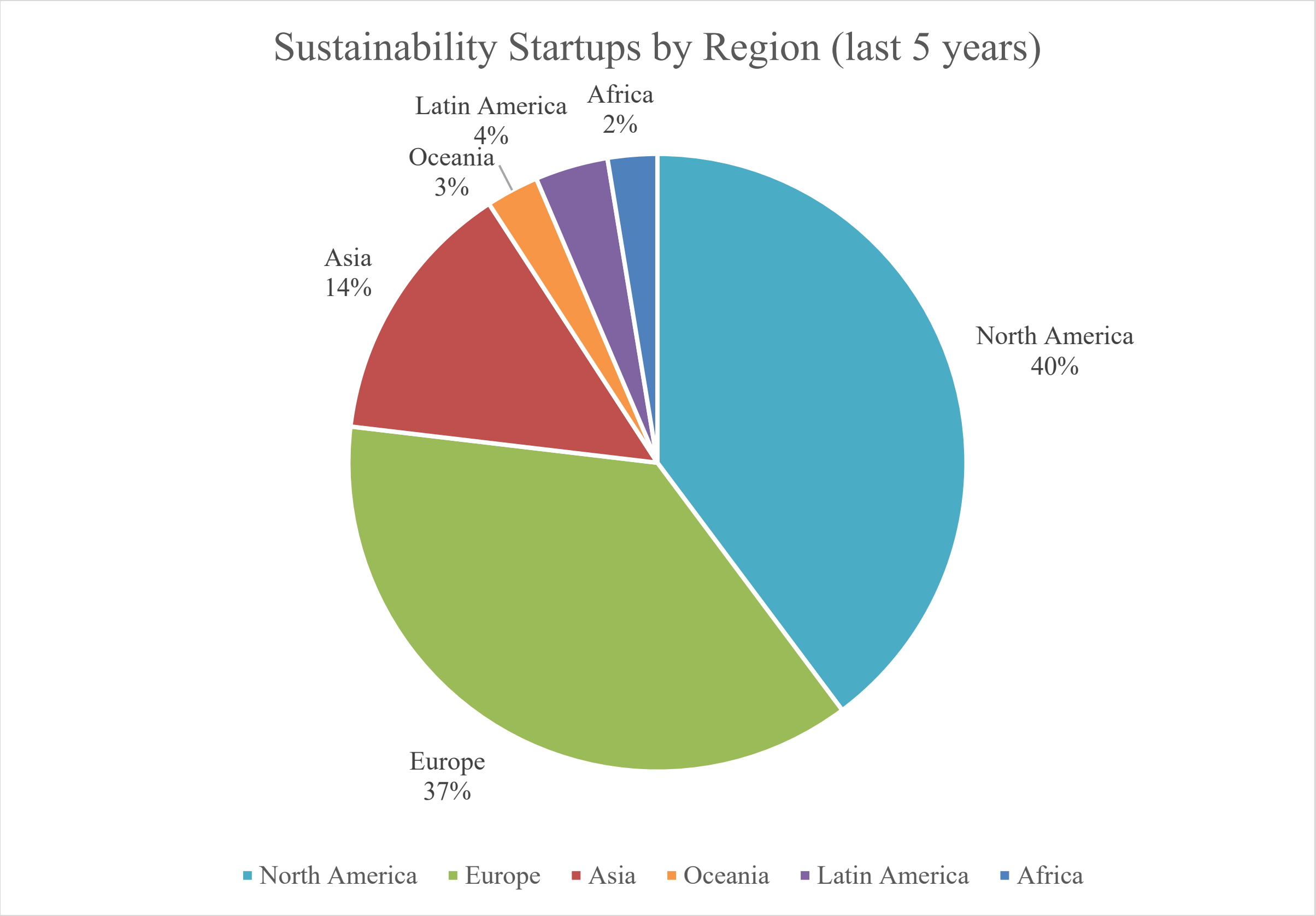

Figure 11 – Sustainability startups by region

## 4.4 Share of startups

Over the decade between 2011–2020 inclusive, the share of sustainability startups in the industry has averaged around 1–in–42 (*M*=2.37%), per Figure 12.

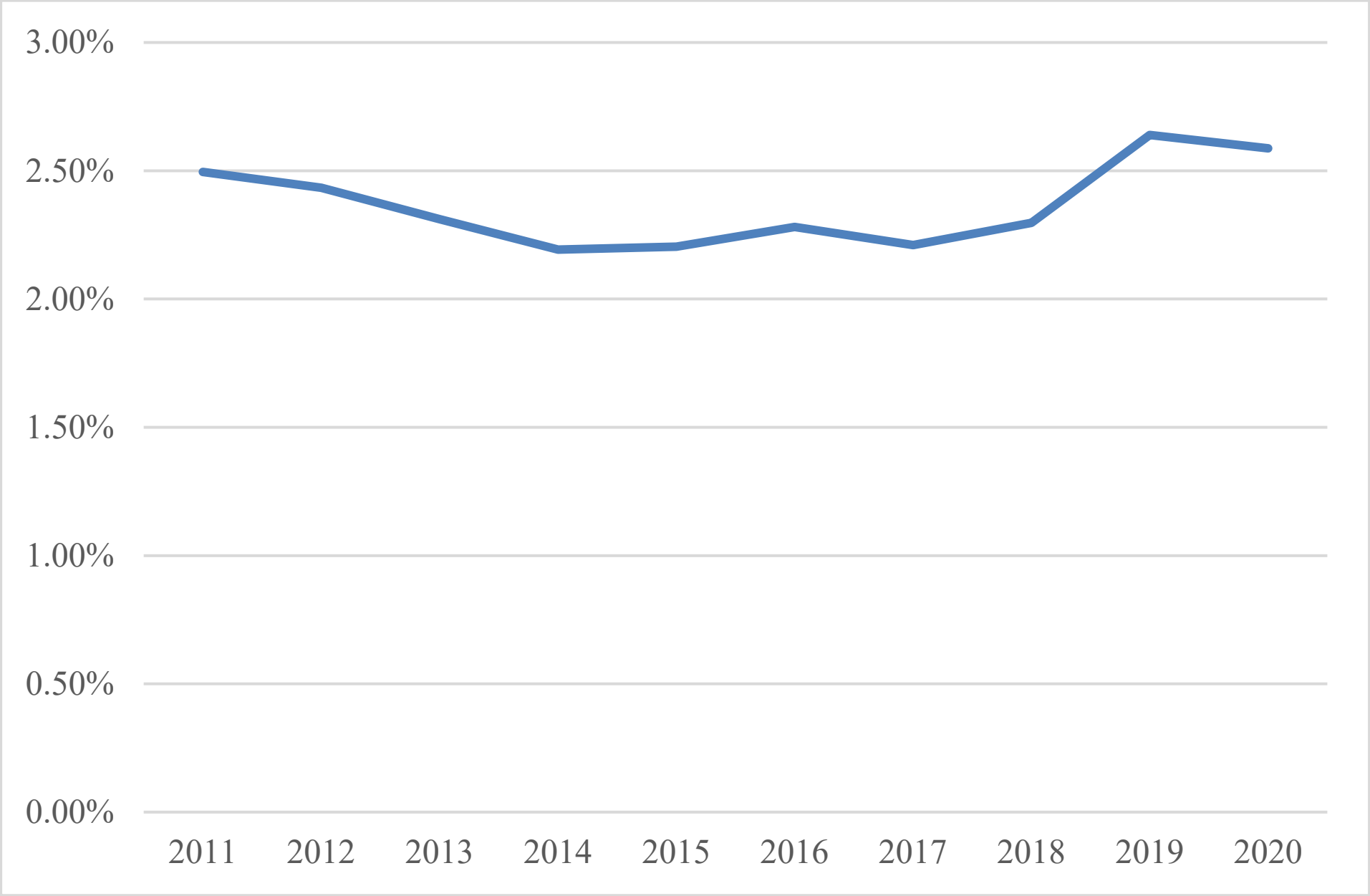

Figure 12 – Share of sustainability startups

## 4.5 Share of funded organisations

Where detailed funding information was recorded, sustainability startups are representing an outsized and growing share of funded organisations, reaching 4.94% in 2020, *f*=72, *n*=1,458, with a share of 4.34%, *f*=20,675, *n*=476,360.

The year predicted the number of funded organisations, $R^2$=.594, $F(1,10)$=14.6, $p$=.003.

The resulting linear regression predicts 187, 263, and 338 startups in 2030, 2040, and 2050.

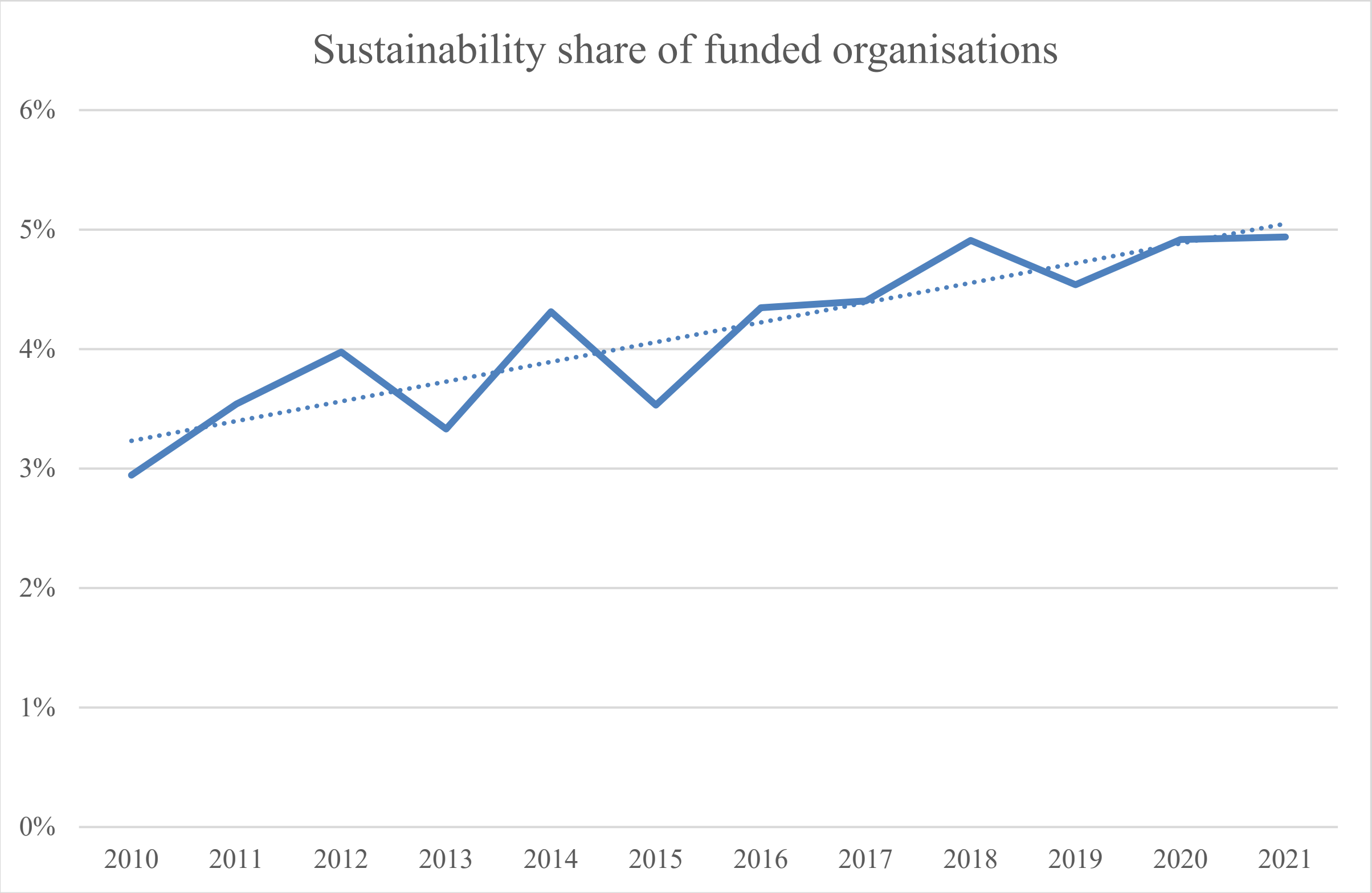

## 4.6 Funding activity

Of all the startup funding rounds in the past calendar year 2021 (*n*=34,884) four percent (*f*=1,397) were from the sustainability industry group.

Of those, slightly more than half (51.6%) were pre-seed (*f*=218) or seed (*f*=218) rounds, with a small number of angel rounds (1.2%, *f*=17), and slightly less than half (47.1%) being Series A (*f*=224), Series B (*f*=93), Series C (*f*=44) or other unspecified series (*f*=298).

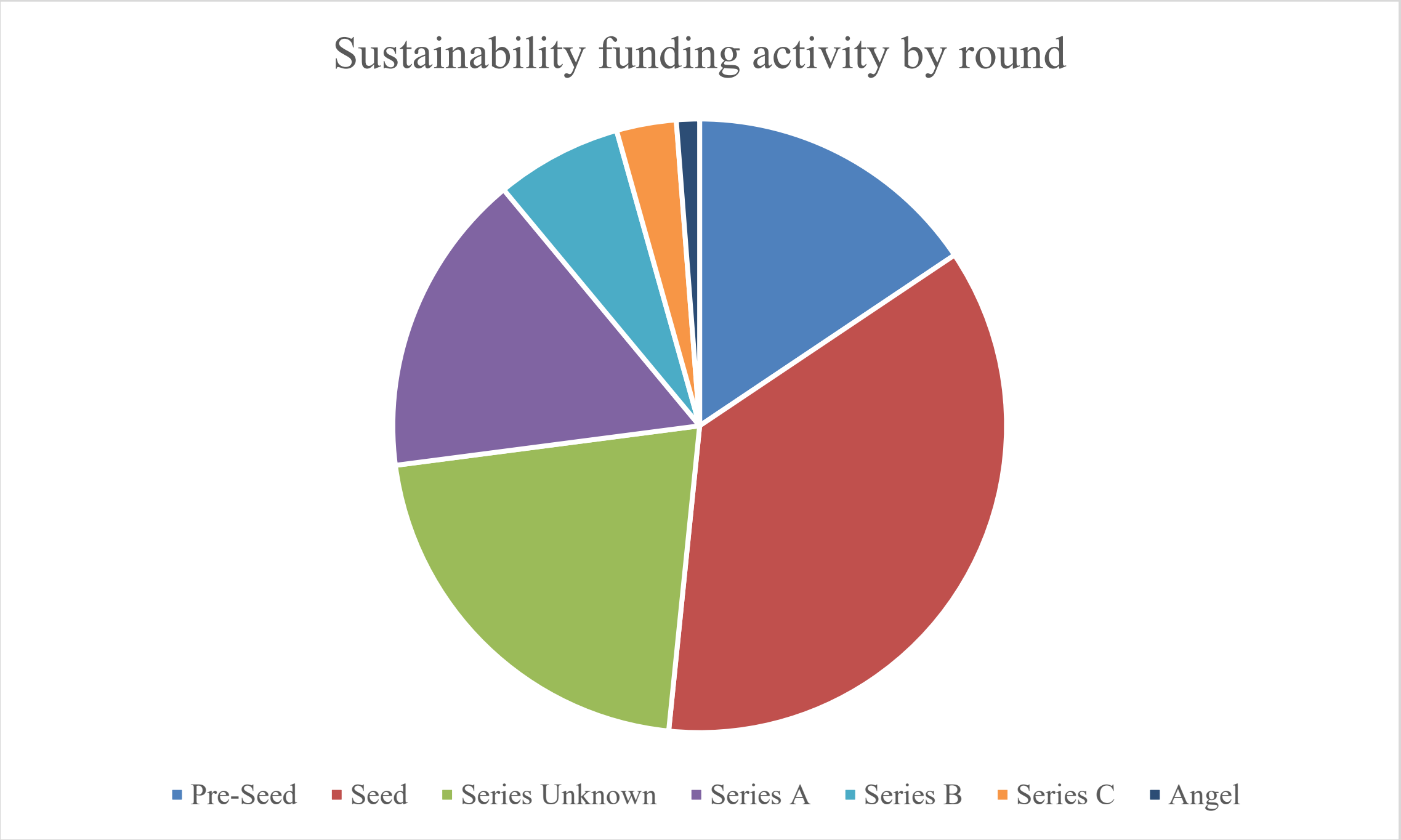

Figure 13 – Sustainability funding activity by round

## 4.7 Exits

Of the 1,000,000+ companies in the database, 167,131 companies had an exit date (i.e., IPO, M&A), and some 4,747 (2.84%) of those were in the sustainability industry group.

### 4.7.1 Unicorns

The database was found to contain a non-negligible number of unrealised sustainability "unicorns" (i.e., companies valued at over one billion US dollars), as valued by PrivCo, a provider of financial and business information on private companies.

These companies are in addition to those valued by investors at funding rounds, or at exit via initial public offering (IPO) or merger and acquisition (M&A), and may indicate the potential for future events not yet captured in the databases.

| Name | Founded (year) | Total Funding ($m) | Most Recent Value ($m) | MOIC (x) | CAGR (%) |
|---|---|---|---|---|---|
| Indigo | 2014 | 1,169 | 4,500 | 3.8 | 21.2% |
| Helion Energy | 2013 | 578 | 3,000 | 5.2 | 22.9% |
| BrightSource Energy | 2004 | 840 | 2,895 | 3.4 | 17.5% |
| Farmers Business Network | 2014 | 870 | 2,500 | 2.9 | 16.3% |
| Plenty | 2014 | 941 | 2,300 | 2.4 | 13.6% |
| Pivot Bio | 2011 | 617 | 2,000 | 3.2 | 12.5% |
| Goodnight Midstream | 2011 | 185 | 2,000 | 10.8 | 26.9% |
| SolarReserve | 2007 | 203 | 2,000 | 9.9 | 17.8% |
| Pegasus Resources | 2017 | 600 | 2,000 | 3.3 | 35.1% |
| Solugen | 2016 | 435 | 1,850 | 4.3 | 33.6% |
| Tall City Exploration | 2012 | 800 | 1,500 | 1.9 | 7.2% |
| Nanosolar | 2002 | 490 | 1,450 | 3.0 | 5.9% |
| Grove Collaborative | 2012 | 475 | 1,320 | 2.8 | 12.0% |
| LanzaTech | 2005 | 310 | 1,150 | 3.7 | 8.5% |
| Genomatica | 2000 | 387 | 1,150 | 3.0 | 5.3% |
| Rubicon | 2008 | 223 | 1,100 | 4.9 | 13.1% |
| Palmetto Clean Technology | 2009 | 478 | 1,000 | 2.1 | 6.3% |
| EcoFlow Tech | 2016 | 114 | 1,000 | 8.8 | 54.4% |
| Lotus Midstream | 2018 | 400 | 1,000 | 2.5 | 35.7% |

Figure 14 – Sustainability unicorns

### 4.7.2 Initial Public Offerings (IPOs)

#### 4.7.2.1 Sustainability IPOs

In the decade from 2012–2021 inclusive, 173 sustainability companies were identified, of which 27 reported both total funding and valuation at IPO, allowing the calculation of a Multiple of Invested Capital (MOIC) for sustainability IPOs.

Being a power law distribution, the usual measures of central tendency (*M*=13.95, *SD*=6.00) are of limited utility, however the median of approximately 5.2X may be indicative of relative performance (*Mdn*=5.19, *n*=27), per Figure 15.

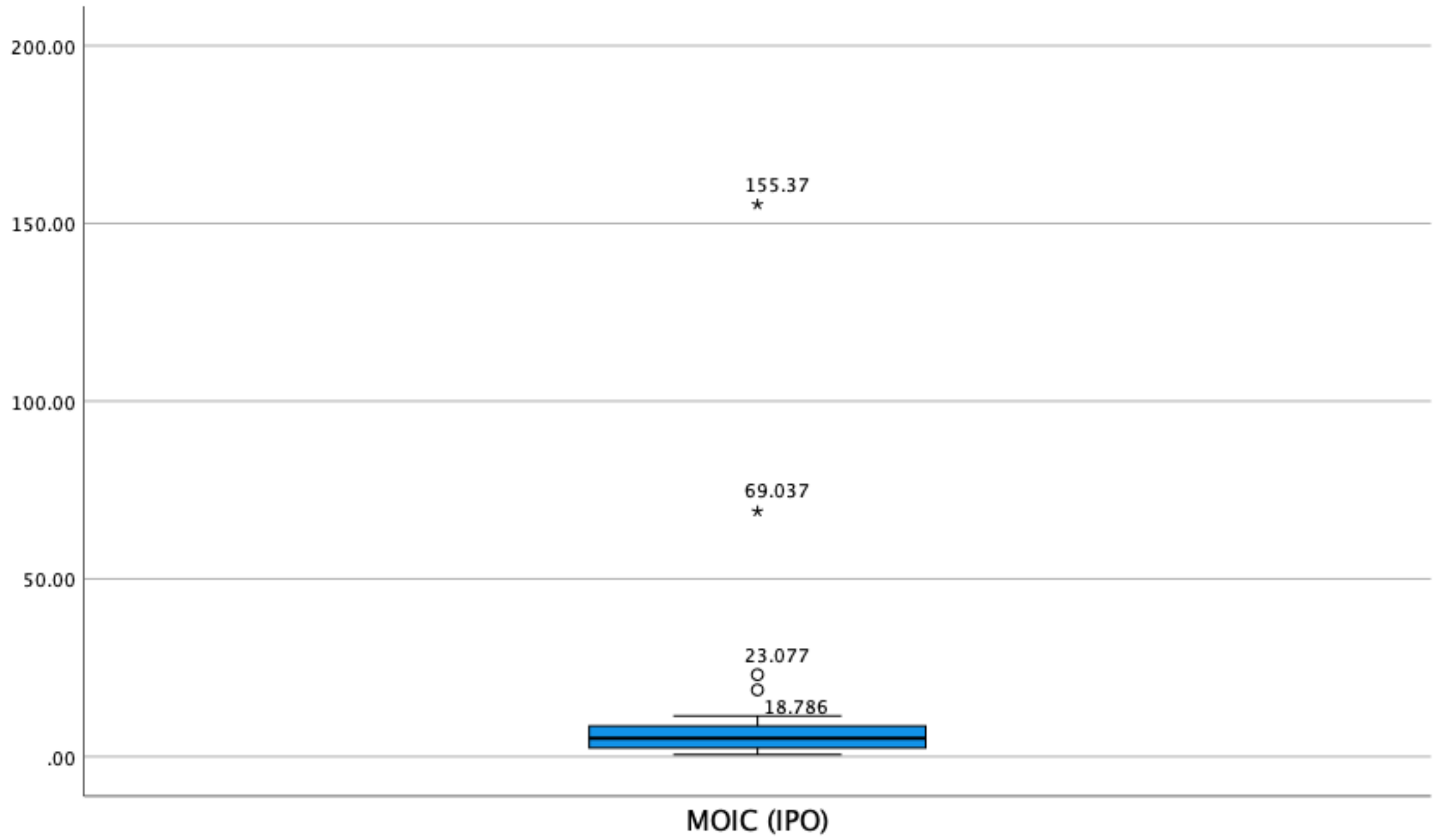

Figure 15 – Sustainability IPO Multiple of Invested Capital (MOIC)

#### 4.7.2.2 Baseline IPOs

In the same period 67 companies were sampled from the wider market that reported both total funding and valuation at IPO, allowing the calculation of a Multiple of Invested Capital (MOIC) for IPOs in general.

Being a power law distribution, the usual measures of central tendency (*M*=8.85, *SD*=1.29) are of limited utility, however the median of approximately 5.7X may be indicative of relative performance (*Mdn*=5.70, *n*=67), per Figure 16.

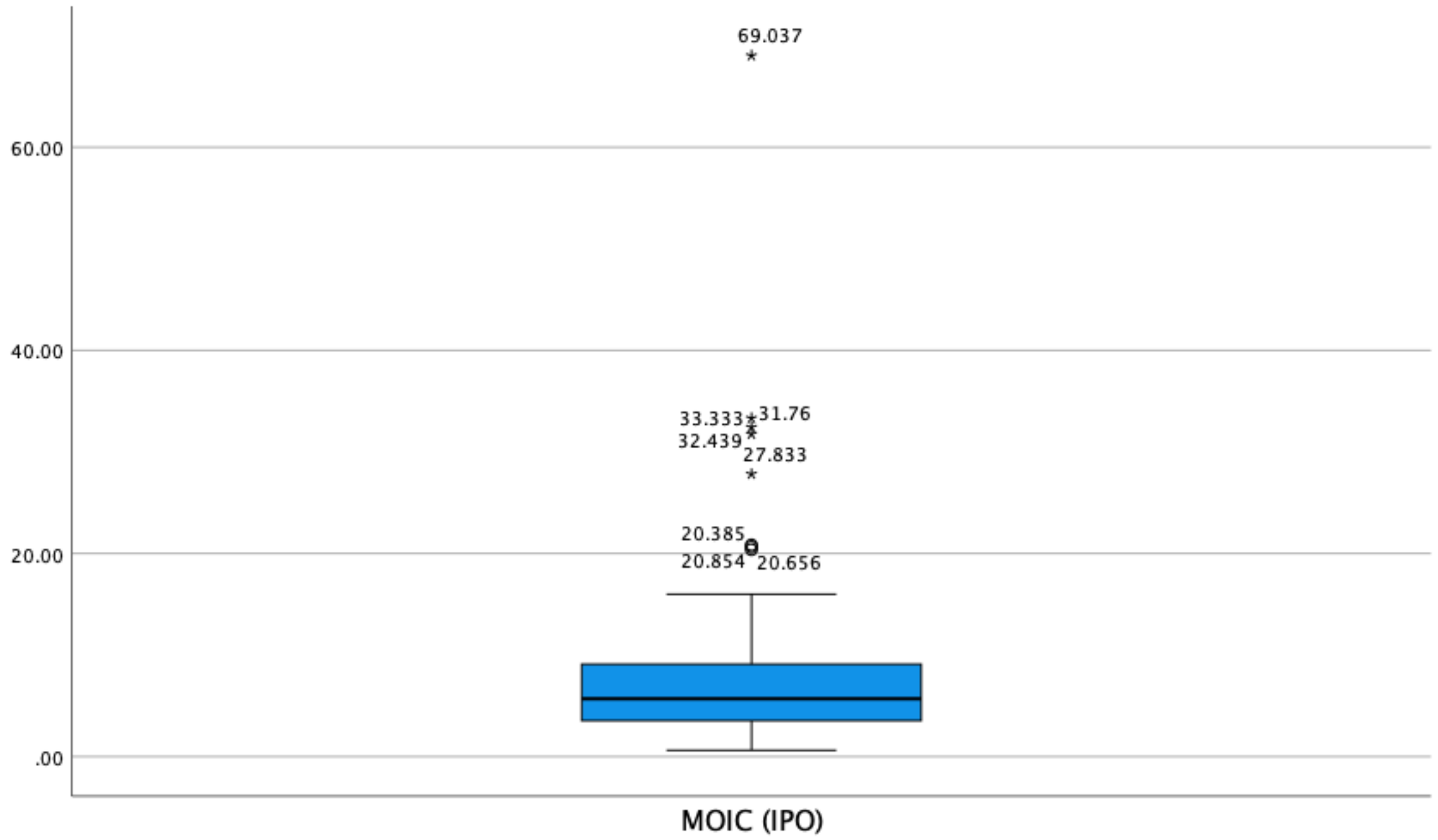

Figure 16 – Baseline IPO Multiple of Invested Capital (MOIC)

### 4.7.3 Mergers & Acquisitions (M&A)

#### 4.7.3.1 Sustainability M&A

In the decade from 2012–2021 inclusive, 173 sustainability companies were identified, of which 16 reported both total funding and acquisition price, allowing the calculation of a Multiple of Invested Capital (MOIC) for M&As.

Being a power law distribution, the usual measures of central tendency (*M*=32.64, *SD*=15.61) are of limited utility, however the median of approximately 3.5X may be indicative of relative performance (*Mdn*=3.54, *n*=16), per Figure 17.

A chi-square test of independence revealed a significant association between sustainability startups and being acquired, $\chi^2$ (1) = 7.10, $p$ = .008. No such association was observed for acquiring others.

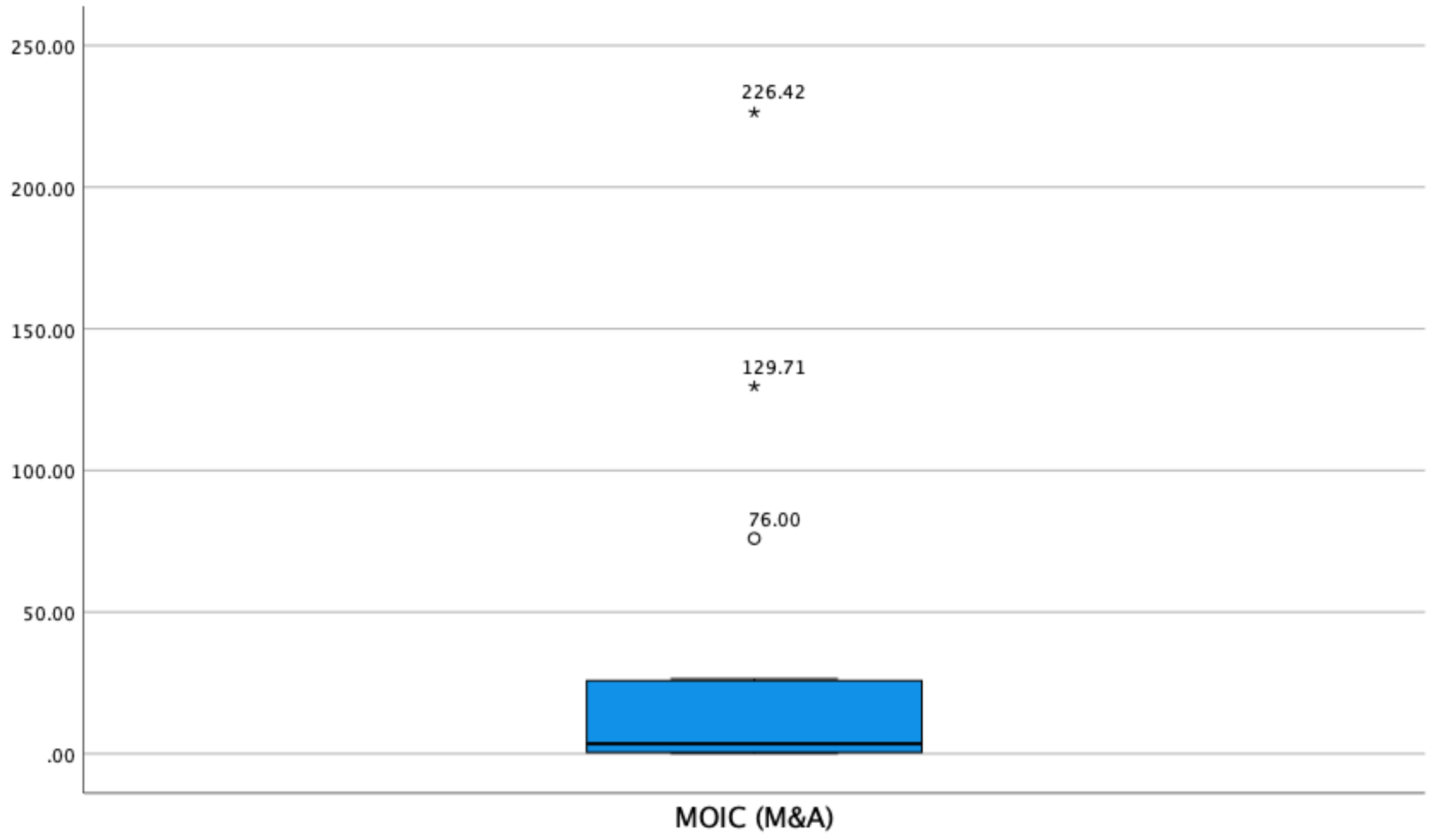

Figure 17 – Sustainability M&A Multiple of Invested Capital (MOIC)

#### 4.7.3.2 Baseline M&A

In the same period 81 companies were sampled from the wider market that reported both total funding and acquisition price, allowing the calculation of a Multiple of Invested Capital (MOIC) for M&As in general.

Being a power law distribution, the usual measures of central tendency (*M*=18.48, *SD*=3.45) are of limited utility, however the median of approximately 7.5X may be indicative of relative performance (*Mdn*=7.54, *n*=81), per Figure 18.

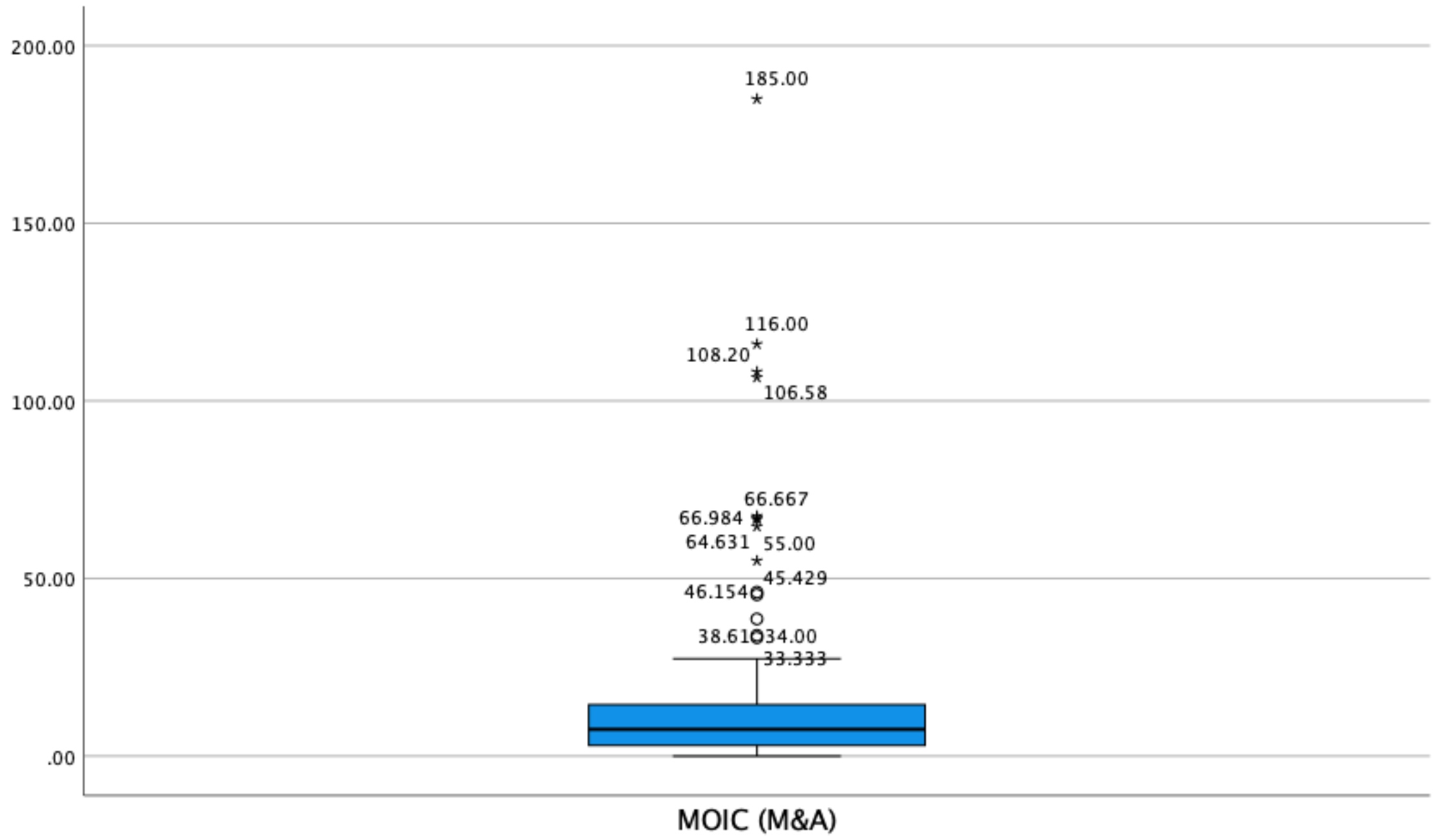

Figure 18 – Baseline M&A Multiple of Invested Capital (MOIC)

### 4.7.4 Sustainable venture capital firms (B Corp)

A number of active venture capital firms have already achieved B Corp certification from B Lab, per Table 16. A points-based assessment process, this assesses the sustainability profile of the firm itself rather than its portfolio companies, though the two are often linked in that B Corp VCs tend to invest in sustainable companies:

> "B Corp Certification is a designation that a business is meeting high standards of verified performance, accountability, and transparency on factors from employee benefits and charitable giving to supply chain practices and input materials."

While Bethnal Green Ventures positions itself as "*Europe's leading early-stage tech for good VC*" and Blisce is "*helping entrepreneurs build mission-driven global consumer technology companies*", claiming to have "*fine-tuned its investment strategy to fully embed ESG (environmental, social and corporate governance) practises across every level of its fund*",

Fifth Wall bills itself as "*Technology for the Built World*", and Foundry Group obtained certification "*to do more to align the Foundry brand with our commitment to give back to the communities in which we live and work*".

Examination of these firms demonstrates that B Corp certification is suitable and obtainable for active venture capital firms, however there is little to learn from their portfolio performance as they span the spectrum from full-blown sustainability investment theses to rubber stamping of operational activities, which is itself commendable.

Table 16 – B Corp venture capital firm examples

| Firm | Founded | Funds | Investments | Exits | Unicorns[a] |
|---|---|---|---|---|---|
| Bethnal Green Ventures | 2012 | 5 | 193 | 1 | 0 |
| Blisce | 2012 | 1 | 51 | 13 | 13 |
| Fifth Wall | 2016 | 15 | 110 | 7 | 20 |
| Foundry Group | 2007 | 7 | 337 | 74 | 7 |

[a] tagged against unicorn-related hubs in Crunchbase

### 4.7.5 Other observations

The following variables were not observed to be dependent on sustainability focus:

- Diversity
- Revenue
- Total funding
- Funding rounds
- Last funding amount
- Number of investors and lead investors

- Total employees
- IPO activity
- Number of active technologies (BuiltWith)
- Global traffic rank (SEMrush)

# Chapter 5 Findings and Discussion

The research led to several important findings that will impact the direction of the development of the company/s for which it was intended, in addition to confirming a number of assumptions.

## 5.1 Research findings

Analysing the data from surveying a sample of investors (*N*=114) resulted in several statistical findings:

### 5.1.1 Gender

1. Three quarters of the self-selecting participants were male despite an approximately even pool of candidates, suggesting participants in the industry are more likely to be male.
2. Nonetheless, females were significantly more likely to recommend sustainable investments to others, and to hold an investment for longer.

### 5.1.2 Generation

1. Older generations were significantly more likely to recommend sustainable investments to others than younger generations, and to prefer sustainable investments over others.

### 5.1.3 Country

1. Despite most participants coming from a small number of countries (e.g., Singapore, Australia, India, and the United States), no relationship was found between the country and any of the dependent variables analysed.

### 5.1.4 Investors

1. Professional investors were significantly more likely to have a higher risk appetite than amateurs, and vice versa.
2. Investors with a high risk appetite were less likely to hold investments for a long time, and to expect a higher return on their investment.

### 5.1.5 Sustainability

1. Two thirds of participants were likely or very likely to prefer sustainable investments.
2. A slight majority expected equivalent returns on sustainability investments compared to the wider market.
3. Very few investors expected much higher or much lower returns from a sustainability investment compared to the wider market.

### 5.1.6 Net Promoter Score (NPS)

1. In this sample, sustainability investments have a Net Promoter Score (NPS) of negative seven (-7.00).

a. This was a mandatory question and there was a peak at the mid-point (5.0) which suggests many who might otherwise not have answered opted for a sensible default which actually registered them as a detractor.
b. This contributed approximately -14 to the calculation, and were the question optional the result may have been as high as positive seven (+7.00).

### 5.1.7 Motivations

1. The most popular motivation for making sustainability investments was ***environment***, followed by ***performance***, then ***resilience*** and ***legacy*** coming in equal and very few selecting ***empowerment***.

### 5.1.8 Sustainable Development Goals (SDGs)

1. The top SDGs were ***SDG13 (Climate Action)*** and ***SDG4 (Education)***, followed by ***SDG7 (Affordable and Clean Energy)***, ***SDG6 (Clean Water and Sanitation)***, and ***SDG2 (Zero Hunger)***.
2. Females preferred ***SDG5 (Gender Equality)*** and ***SDG16 (Peace, Justice, and Strong Institutions)*** while males preferred ***SDG2 (Zero Hunger)***.
3. Older generations preferred ***SDG13 (Climate Action)*** while younger generations preferred ***SDG8 (Decent Work & Economic Growth)***.

## 5.2 Database findings

Analysing the data from a company database resulted in several statistical findings:

### 5.2.1 Rankings

1. Sustainability companies were significantly more likely to have higher rankings (CB Rank).

### 5.2.2 Taxonomy

1. Most sustainability companies were involved in renewable energy (24%), solar (13%), waste management (9%), or environmental engineering (8%).

### 5.2.3 Regions

1. North America (40%) and Europe (37%) were approximately even in sustainability startup activity, followed by Asia (14%).

### 5.2.4 Funding

1. The share of sustainability startups among all startups has been relatively stable over the past decade at around 1–in–40.
2. The share of sustainability startups among all *funded* startups is outsized and growing, reaching almost 5% in 2020.
3. The year predicted the number of funded sustainability startups, increasing from 72 in 2020 to hundreds in the 2030s.
4. More than half were pre-seed or seed rounds, with most of the remainder being venture capital.

### 5.2.5 Exits

1. There is a non-negligible number of unrealised "unicorns" (i.e., companies valued at over one billion dollars) that may proceed to exit in the future.
2. Sustainability IPOs returned a median multiple of invested capital (MOIC) of 5.2x compared to a baseline of 5.7x over the past decade.
    a. Measures of central tendency are unreliable for power law distributions (Crawford et al., 2015).
    b. MOIC does not factor in holding period, so IRR may be more appropriate.
3. Sustainability M&A returned a median multiple of invested capital (MOIC) of 3.5x compared to a baseline of 7.5x over the past decade.
    a. Measures of central tendency are unreliable for power law distributions (Crawford et al., 2015).
    b. MOIC does not factor in holding period, so IRR may be more appropriate.
4. A significant association between sustainability startups and being acquired was observed.

### 5.2.6 Others

Several other variables were examined and not found to be dependent on a sustainability focus (e.g. diversity, revenue, funding rounds & totals).

# Chapter 6 Conclusions

Billionaire investor “Chris Sacca believes the next trillion-dollar companies will be focused on climate change” (Harris, 2022). BlackRock CEO Larry Fink figures “it’s time to invest in the sustainability transition” (NASDAQ, 2022) and that “the next 1,000 billion-dollar start-ups will be in climate tech” (Clifford, 2021), further predicting a “renewable ‘investment boom’ as Ukraine war accelerates energy transition” (Agnew, 2022).

There are signs of increasing interest in sustainable investments across the industry, from individuals to pensions and sovereign funds. This is expected to continue as the effects of climate change become more apparent in the form of floods, fires, and rising sea levels, with necessity driving regulation and stakeholder engagement. Two thirds of survey participants were likely or very likely to prefer sustainable investments over traditional investments.

Sustainability is already an outsized and growing component of the global startup ecosystem, split almost evenly between North America and Europe followed by Asia according to company count. This survey was not able to reveal a relationship between country and any of the dependent variables analysed, but this is proposed as an area for future research with a larger audience, and the tools have been provided to facilitate it.

While IPO returns are comparable, M&A returns appear to have been comparatively modest as a lagging indicator, but neither were statistically tested in part because measures of central tendency are unreliable for power law distributions (Crawford et al., 2015). Leading indicators are more encouraging, including industry share, growth, and activity.

In terms of existing sustainable startup investing, current activity may be male dominated based on research participation, but women are more likely to recommend sustainable investments, and to hold them for longer periods to realise returns. Men are more motivated to work towards Zero Hunger (SDG2), while women prefer to focus on Gender Equality (SDG5) and Peace, Justice, and Strong Institutions (SDG16).

Older generations are also more likely to both recommend and prefer sustainable investments, and pursue Climate Action (SDG13), even if younger generations are more likely to be affected by it; they have more urgent priorities like Decent Work & Economic Growth (SDG8). No correlation was found between diversity (measured in the US only) and sustainability.

Professionals in general, who unsurprisingly tend to be older and have a higher risk appetite, expect higher returns and are less willing to hold investments for long periods to realise them. They were no more or less likely to recommend or prefer sustainable investments than others. This was a surprising finding, as it shows sustainable investments are already on the radar and in many cases on the books for professional investors. This is despite the under-developed ecosystem supporting them and suggests there is an opportunity to create one.

Overall, most investors expect similar returns to other investments, even if additional returns flow to other stakeholders as well as shareholders; sustainable investments with comparable financial outputs for shareholders are better in that they also produce often intangible outputs for other stakeholders. While environment was the primary motivator for sustainable investments and performance a distant second, most are unwilling to accept lower than average returns for doing the right thing. Resilience and legacy came in distant third.

Any startup or fund seeking to take advantage of the sustainability opportunity to deliver return to both shareholders and other stakeholders would do well to consider the priorities of same. Climate Action (SDG13) and Quality Education (SDG4) were the top priorities of this sample, followed by Affordable and Clean Energy (SDG7), Clean Water and Sanitation (SDG6), and Zero Hunger (SDG2).

It is important to take a nuanced view however, as topics like Decent Work and Economic Growth (SDG8) were almost dead last despite having statistically significant importance to a critical group: today's youth who are the future of tomorrow. Similarly, very few participants selected Life on Land (SDG15), which seemingly excludes existential threats to humanity. The tools have been provided to assess local and/or wider audiences with a view to delivering solutions to problems for which there is demand.

# Chapter 7 Recommendations

This applied research has resulted in several recommendations which will be adopted in the development of the studio and formation of any associated fund/s.

## 7.1 Investors

The primary findings of this research relate to investors in sustainability. It provides both an insight into the views of the sample and population it represents, and makes inferences from it using statistical testing. While the audience was necessarily tailored to the context of the application, being the professional network of the author and snowball samples from it, the tools have also been provided and appropriately licensed to allow other researchers to elaborate on the research or apply it in a different context.

Without restating the findings from previous sections, the relevant recommendation is to understand the needs and desires of your investors and to ensure there is a fit in terms of size, stage/s, and SDG/s addressed. It is also important to understand the personas who are more likely to be receptive to your investment thesis, such that you can help them deploy their funds in a way that is aligned with their own purpose.

For example, females were significantly more likely to recommend sustainable investments to others, and to hold and investment for longer. This may make women better advocates who are more amenable to longer horizon "deep tech" initiatives like direct air carbon capture. While Climate Action (SDG13) was widely popular, especially amongst older participants, women are also more likely to be amenable to those that address Gender Equality (SDG12) and Peace, Justice & Strong Institutions (SDG16), both of which are longer-term objectives. Younger and/or male audiences could be more interested in shorter-term goals such as Decent Work & Economic Growth (SDG8) and Zero Hunger (SDG2) respectively.

For a larger or anchor investor it may even be feasible to go in the direction of a donor advised fund, where a significant contribution grants a stronger voice in decision making on where to invest (in the case of venture capital) or what to build (in the case of a venture studio). This is similar to the approach taken by many non-profit organisations, but comes with the upside potential of a successful startup pursuing profit with purpose; this has the potential to be a very popular approach to sustainable financing.

## 7.2 Structured decision making

Humans tend to use shortcuts such as heuristics in decision making processes that can impair the resulting decisions. This is rarely life-or-death, but as portfolio company selection is the single most important function of venture capital, specific consideration should be given to the decision making processes deployed.

Several candidate processes have been considered and a consensus process proposed, including steps for the identification, screening, evaluation, structuring, managing, and exit of portfolio company deals (Maxwell et al., 2011). This is a good starting point for a fund manager, though it is one of many such sequences and one tailored to the individual firm is likely to have the best results. It needs further adaptation for novel business models including the venture studio.

It is also worth acknowledging that potential investors are likely to implicitly or explicitly follow a similar process, especially in the case of larger, more sophisticated, and/or institutional investors. A good fund manager will be cognisant of the process they are being subjected to, and be able to provide the right information at the right time for them to make the right decision. This applies even for negative outcomes, as having no investor on the cap table may be better than a bad investor.

## 7.3 Optimal diversification

Common sense would suggest that increasing specialisation would result in higher returns, but in venture capital the opposite is often, but not always, true (Buchner et al., 2017). Modern Portfolio Theory (MPT) would have an investor select investments to maximise returns within an acceptable level of risk. Indeed, venture capital firms deliver strong returns despite extremely high levels of risk at an individual startup level by carefully selecting 20–30 companies or more per fund.

Typically in the order of 1–in–20 startups ends up being a "home run" with many of the remainder going to zero, provided the portfolio is carefully selected and supported through exit and increasingly beyond. If all those startups are in one industry, say nuclear power, and there is an event that impacts that entire industry, say advances in fusion or a nuclear accident, then the performance of the entire fund suffers. Meanwhile a diversified fund may take a hit from that event, but other companies will continue to perform and may even move in the opposite direction, for example a solar company.

The recommendation for sustainable venture capital therefore is to diversify within sustainability, which is an industry group consisting of many industries. Renewable energy, solar, waste management, and environmental engineering form the majority of the group according to the company database studied, per Figure 10. A portfolio might choose to go to where the action is, or to blaze a new trail into relatively uncharted territories such as tackling the demand side of the energy equation by focusing on energy efficiency, or something altogether like water purification or organics. The same applies to development stage in that it may not make sense to limit oneself to, for example, seed-stage companies.

The exception to the rule is that one should remain within their domains of expertise, or risk a negative impact on fund performance. That is to say, diversify to the extent feasible rather than possible. It is also less applicable to emerging fund managers, perhaps because they have fewer opportunities to diversify in their deal flow.

## 7.4 Performance metrics

Venture capital firms are well aware their business model is built on power law distribution, and most will understand that the usual measures of central tendency such as min, max, and mean are meaningless in this context. It is worth restating however, as decisions made based on flawed information are arguably worse than decisions made on no information; at least in the latter case the decision-maker is aware they are venturing into the unknown.

It is less likely to be apparent however that other input (e.g., social capital, IP protection actions) and outcome (e.g. employee numbers and growth) variables typically used in entrepreneurial research also follow power law distributions (Crawford et al., 2015). With this information appearing in company databases it is tempting to use it as is, and the availability of "Statistics" functionality in the "Pro" version risks exacerbating the problem. However the logarithmic bands offered for e.g. employee counts may alleviate the issue somewhat, with the top bands accounting for orders of magnitude more employees.

In addition to specifically targeting companies exhibiting exponentially larger values for these variables, it may also be beneficial to actively increase the relevant figures to the extent possible, including the categories in Table 3. For example, a venture capital firm may, in addition to dramatically increasing financial capital, also increase social and human capital, we well as cognitions (e.g., expectations) and actions (e.g., patents).

## 7.5 Corporate innovation lessons

Corporate ventures are similar in many ways to venture capital, and particularly to venture studios, with a parent organisation providing human and financial capital with a view to building successful businesses. Extant research in this area is likely to be applicable to some extent to venture capital, and indeed the paper reviewed was instructive in this context.

While the goal of corporate ventures is often to escape the tyranny of the organisation itself with its rules and performance metrics, the dominant coalition (i.e., owners and managers) can play important roles from the identification of ideas via industry experience through execution through resourcing and strategy (Waldkirch et al., 2021).

The same applies to some extent in venture capital, and "hands off" firms are less likely to be successful than those who take a more active stake in the success of their portfolio companies. Several causal conditions at the firm, venture, and intersection levels were have been identified in earlier research, summarised in Table 4 – Corporate venture innovation theoretical framework.

"Open Business Models" are a related topic, where companies move away from the traditional "inside-out" model of innovation typified by companies like IBM, instead adopting a collaborative "outside-in" approach (Weiblen, 2016). While startups necessarily create intellectual property in the process of establishing themselves, this approach acknowledges that the overwhelming majority of innovation takes place outside the four walls of the firm and can be incorporated into it. This is particularly true of undifferentiating activities that can be identified through value chain or "Wardley" mapping (Wardley, 2013).

## 7.6 Use and abuse of UN SDGs

While society benefits from appropriate and liberal application of resources to addressing the UN Sustainable Development Goals (SDGs), social entrepreneurial ventures can also derive benefit from them provided they are used in constructive ways.

Using a similar framework to Net Promoter Scores (NPS) with its promoters, passives, and deniers, existing research proposes that social entrepreneurs fall into three categories: SDG Evangelism, SDG Opportunism, and SDG Denial (Günzel-Jensen et al., 2020). While denying the benefits of the SDG framework is unlikely to bring a portfolio company any value, using them opportunistically (e.g., to apply for government grants) is unlikely to realise their full potential.

In the absence of other suitable and well-known frameworks, it is recommended that sustainable venture capital firms embrace the framework and act as evangelists for it. The goals can be integrated into work processes at fund and/or company level, and used as common language to create a mutual understanding. For example, the goals could be used to diversify the portfolio and increase returns.

It is further recommended that fund managers and entrepreneurs take into consideration the differing demand for the different goals, in terms of what is required to meet society's needs, what is demanded by the markets, and what investors are willing and wanting to invest in.

This survey found that Climate Action (SDG13) and Quality Education (SDG4) are top priorities for investors, followed by Affordable and Clean Energy (SDG7), Clean Water and Sanitation (SDG6), and Zero Hunger (SDG2), but the focus may need to be adapted for different contexts and markets. For example, EdTech to address Quality Education (SDG4) is a notoriously difficult arena. The contrarian might also assume that a lack of focus in a given area translates to an opportunity for an early adopter.

## 7.7 B Corp certification by B Lab

B Corp certification is a certification program assessing for-profit companies for their "social and environmental performance". It is similar to but separate from the legal designation of a "benefit corporation" available in certain jurisdictions, bringing many of the same connotations by way of private certification marks.

Unlike non-profit status in which profit is strictly forbidden by rules that are actively enforced by authorities, benefit corporations are for-profit but factor in other stakeholders in addition to shareholders. They are typically established with a purpose that includes having a positive impact on their communities, environment, workers, and society in general, and are often branding with monikers like "profit with purpose".

Existing research examining the process of "imprinting" considers several pathways and windows, with differing outcomes depending which route an entrepreneur takes (Muñoz et al., 2018). The authors caution against imprinting purpose and receiving B Corp certification too early, as well as considering feedback from non-market actors (e.g., journalists, awards) to be business model validation.

They also describe a cluster of firms who consider their project a "vehicle for change" rather than a business, with neither market nor non-market actors engaging due to a lack of understanding of their motives and mechanisms. This is despite, or perhaps because of, their grandiose purpose. While they may transition to an SME phase, their disconnect from the outside world limits their value creation and delivery.

The recommendation therefore is to ensure the venturing is triggered by a business idea that responds to social or environmental problems, and to prioritise feedback from market (i.e., customers) over non-market (e.g., journalists, awards) actors. While the exact timing of the imprinting of purpose is up for debate, doing it too early or too late in the cycle may impair impact. To be more precise, while the findings of the research call into question B Lab's new B Corp Pending status which is available only to startups within their first year, it is this author's view that this is a valuable program that should be availed of towards the end of that short window unless the purpose is very clear at an earlier stage.

## 7.8 Non-profit status

One option available for social entrepreneurs is the adoption of non-profit status, foregoing the upside potential at exit typically associated with venture capital. This has some important advantages, such as the tax deductibility of donations available in many jurisdictions, but also some key disadvantages including profit motivation for social entrepreneurs an investors.

Research reviewed on the subject of Philanthropic Venture Capital (PhVC) goes into some detail on structuring with a view to maximising social impact rather than financial return, applying moral hazard and agency theory to describe investment behaviour. It finds that while the non-distribution constraint applicable in the case of non-profit funds and/or portfolio companies does act to align incentives and minimise moral hazard, with the investor themselves acting as steward rather than principal, this is not necessarily the case with for-profit counterparts which may require more active monitoring.

While there are some attractive aspects to more novel structures involving non-profit funds and/or portfolio companies, it is recommended that sustainable venture capital funds maximise returns for all stakeholders including shareholders. By retaining the profit incentive for social entrepreneurs and themselves, investors are better able to attract and retain the resources required to achieve their full potential. Non-profit structures still have their place in the sustainability ecosystem, but the venture capital model and its ability to attain extremely rapid growth rates not typical of non-profits is advantageous in social enterprises as it is in traditional startups.

# Chapter 8 Future Directions

This preliminary research identifies a number of avenues for future exploration, both to refine and/or solidify existing findings, as well as to follow new directions.

## 8.1 Sampling

### 8.1.1 Increased sample size

While the initial minimum target of a 10% margin of error at a 95% confidence interval with 114 respondents was adequate given the context, achieving the stretch goal of a 5% margin of error would have been preferable. This is particularly true for operations that further filter the result set, which in some instances risks the reliability of the results.

Convincing people to take even a few minutes out of their day to answer a survey is surprisingly difficult, and it's unlikely the minimum target would even have been achieved without promising to share results the following month. This is particularly true for the ideal survey respondent, professional investors who lack the time and inclination to engage in such activities.

### 8.1.2 Increased sample diversity

Convenience sampling was used, leveraging the professional network of the author to reach a subset of 35,000 social media followers. Algorithms on social media platforms are crafted to maximise engagement, and it is very likely that they penalise surveys specifically, meaning that it is necessary but not sufficient to publicly post requests.

A side effect of this decision was that the sample and results will be skewed towards the countries, industries, alumni networks, and other subgroups. Future research should endeavour to tap a wider network of participants, improving both quality and quantity of responses.

Another factor that increases the response rate at the risk of skewing results is the use of snowball sampling, where select participants are asked to further propagate the invitation within their networks. In this instance, the survey was directed to several directly accessible groups (i.e., classmates, incubator cohort-mates, ex-Googlers) as well as a number of indirectly accessible groups (e.g. sustainability professionals).

## 8.2 Statistics

"Lies, damned lies, and statistics" — Benjamin Disraeli

One of the risks of exploratory research and the liberal application of statistical tests is that the more tests one conducts, the more likely they are to stumble on Type I and Type II errors, where the null hypothesis is erroneously rejected or accepted respectively. Done deliberately, this process is referred to colloquially as data dredging or "p-hacking", and it can be problematic if not considered.

Indeed, if we accept the standard and select a maximum p-value of .05, we are specifying the chance that a test would find statistical significance in chance alone, and effectively saying that 1–in–20 tests being incorrect is acceptable. This may well be the case if we are conducting a specific test, but what if we conduct 20?

These odds happen to be the same base benchmark typically used in venture capital, only the 1–in–20 in this context is not an error, but a “home run” company that returns the entire fund despite constituting only a small part of it. By carefully choosing 20–30 companies a fund manager hopes one or more of them will be an outsized success even if the majority of the remainder go to zero. Similarly, by selecting 20–30 tests it is likely at least one will find statistical significance where there is none.

In order to avoid this phenomenon all tests have been reported, whether supportive of the hypotheses proposed or not, and whether found to be statistically significant or not. Future research may want to target specific tests, in which case the results will be more reliable.

## 8.3 Net Promoter Score (NPS)

The Net Promoter Score (NPS) is calculated by subtracting the detractors (who rated the subject 0-6) from the promoters (rating 9-10), with the middle ground being considered passive (7-8).

While a score of negative seven (-7.00) is not necessarily catastrophic, it does indicate that there are more participants acting as detractors than promoters, whether based on fact (e.g., past experience or performance) or perception (e.g., an assumption that returns will be lower).

This demands further enquiry, particularly given the impact of making the question required rather than optional. Almost 1–in–5 participants opted for a mid-point score of 5.00 which registered them as a detractor, while linear regression would have predicted a rate closer to 1–in–20. That being the case, not making this question optional may have contributed as much as -14 to the calculation, which otherwise would have been as high as positive seven (+7.00).

## 8.4 Company databases

Access to company databases for research purposes was expected to be straightforward given academia's important contribution to the startup community, but after approaching several of the leading providers it became apparent that expensive commercial licenses would be required should the short trials offered be inadequate. This may not be an impediment for funded research, but it does limit research done by others as well as the utility of the statistics tools provided with this paper (e.g. IBM SPSS Syntax).

Crunchbase was found to be readily accessible, with a 7–day Crunchbase Pro trial being offered. This was initially promising as the built-in "Statistics" function provided several measures of central tendency for any given search in Query Builder. Exports were disabled however, and an annual plan was required to export up to 1,000 records at a time and 5,000 records per month. As a result compromises had to be made, the most important of which was finding queries which would return as close as possible to 1,000 rows.

The solution used in the absence of systematic sampling functionality was to search for company and/or founder names starting with a given letter, which risks introducing unknown biases (for example, due to differences in culture or language resulting in uneven frequency distributions by alphabet).

While it is widely known that financial outcomes of high-growth startups follow a power law distribution rather than a normal distribution, the same applies to almost all other measures of resource, cognition, action, and environment-based variables pertinent to entrepreneurial research (Crawford et al., 2015). This renders the built-in "Statistics" function less useful at best and dangerous at worst, with most measures of central tendency (min, max, mean, etc.) being meaningless, with the possible limited exception of median.

## 8.5 Accelerators

Startup accelerators have long played an important role in the ecosystem, both to help the startups themselves build businesses and drive growth, and as deal flow for host and partner venture capital firms. These organisations typically provide physical and/or virtual spaces to operate in with access to an associated community, advice and/or advisors, training, introductions, and often funding in exchange of equity or the promise of same in a future round.

Founder Institute is one such accelerator, billing itself as "the world's most proven network to turn ideas into fundable startups, and startups into global businesses". Applying the same approach to venture capital firms as they do to startups, Founder Institute are now "working to launch 1,000 new venture capital firms over the next five years to support innovative companies worldwide that are working to make an impact". This is an exciting development in the industry and one that warrants further research in due course.

# Appendix A Reproducibility

This supporting information has been included for future reproducibility using different data sets and/or analytics.

## A.1 Survey (Typeform)

The survey was conducted using an online software as a service (SaaS) tool, Typeform.

### A.1.1 Questions

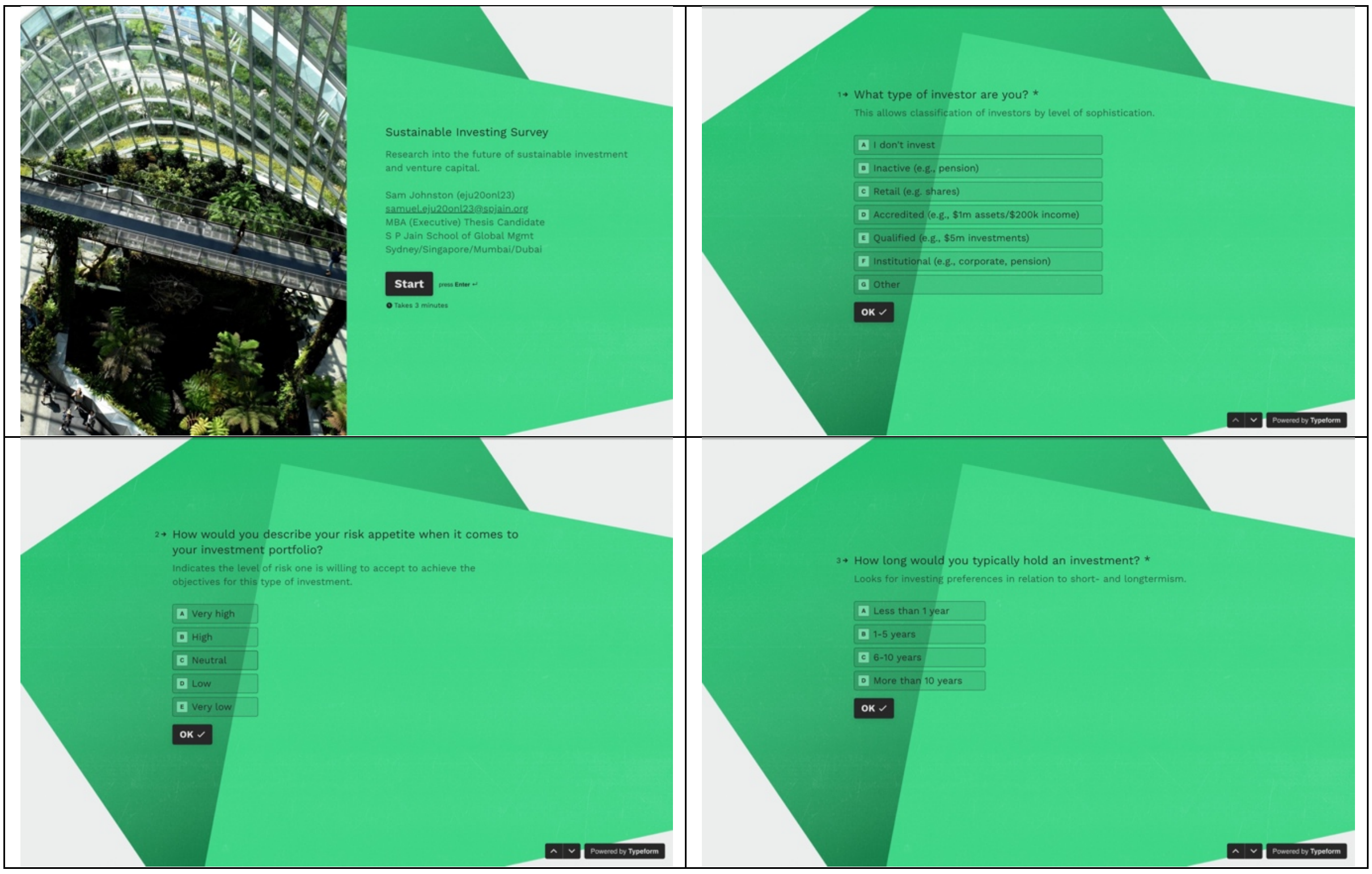

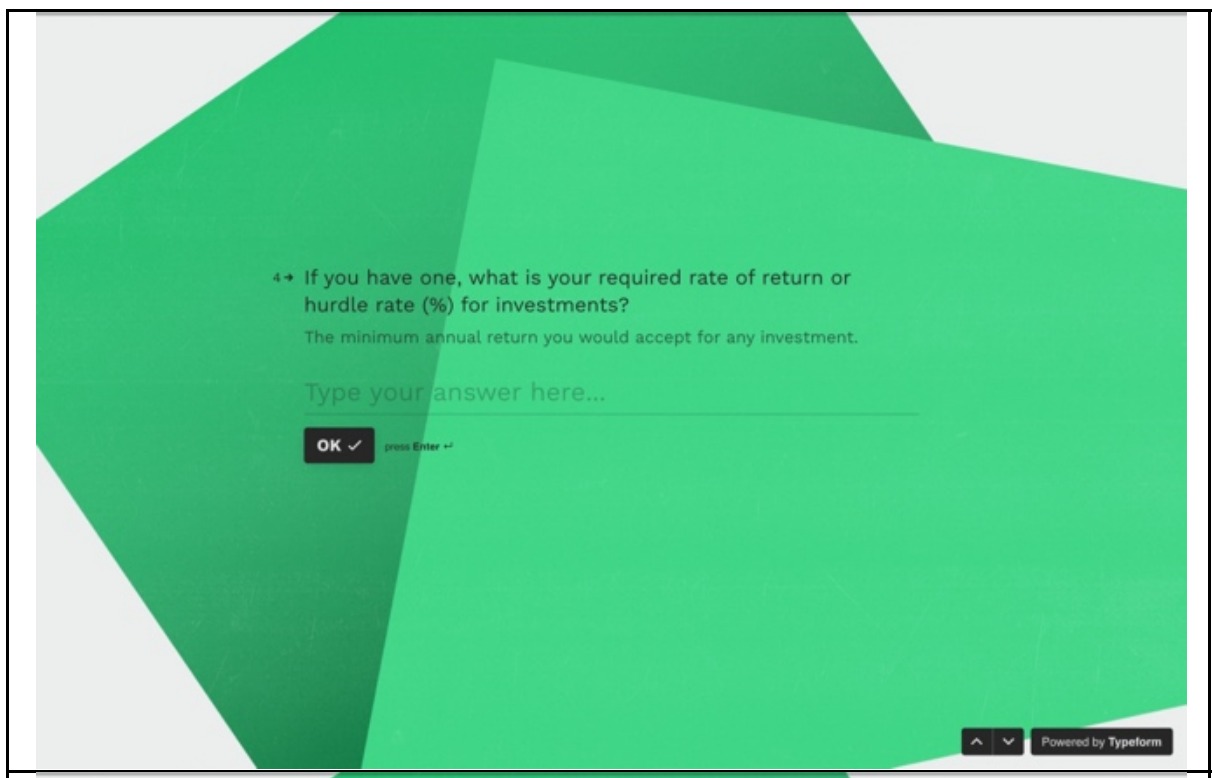
4→ If you have one, what is your required rate of return or hurdle rate (%) for investments?
The minimum annual return you would accept for any investment.
Type your answer here...
OK ✓
press Enter ↵
Powered by Typeform

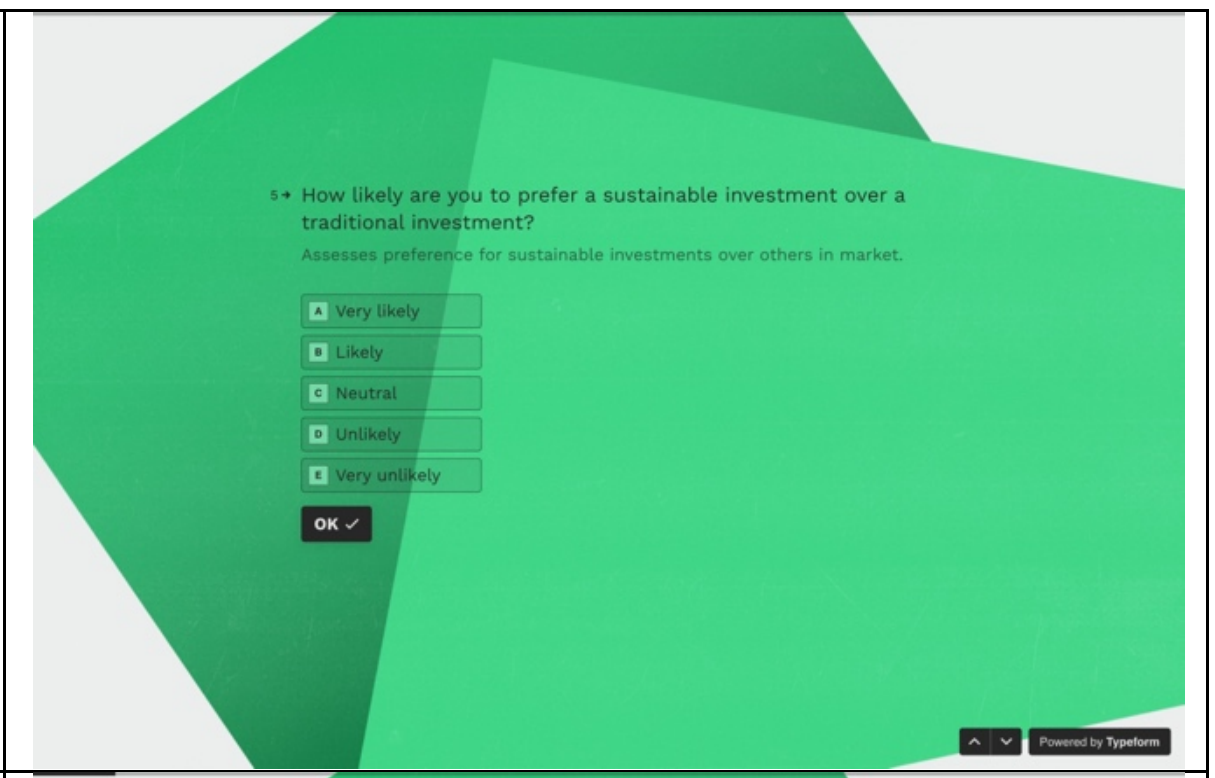
5→ How likely are you to prefer a sustainable investment over a traditional investment?
Assesses preference for sustainable investments over others in market.
A Very likely
B Likely
C Neutral
D Unlikely
E Very unlikely
OK ✓
Powered by Typeform

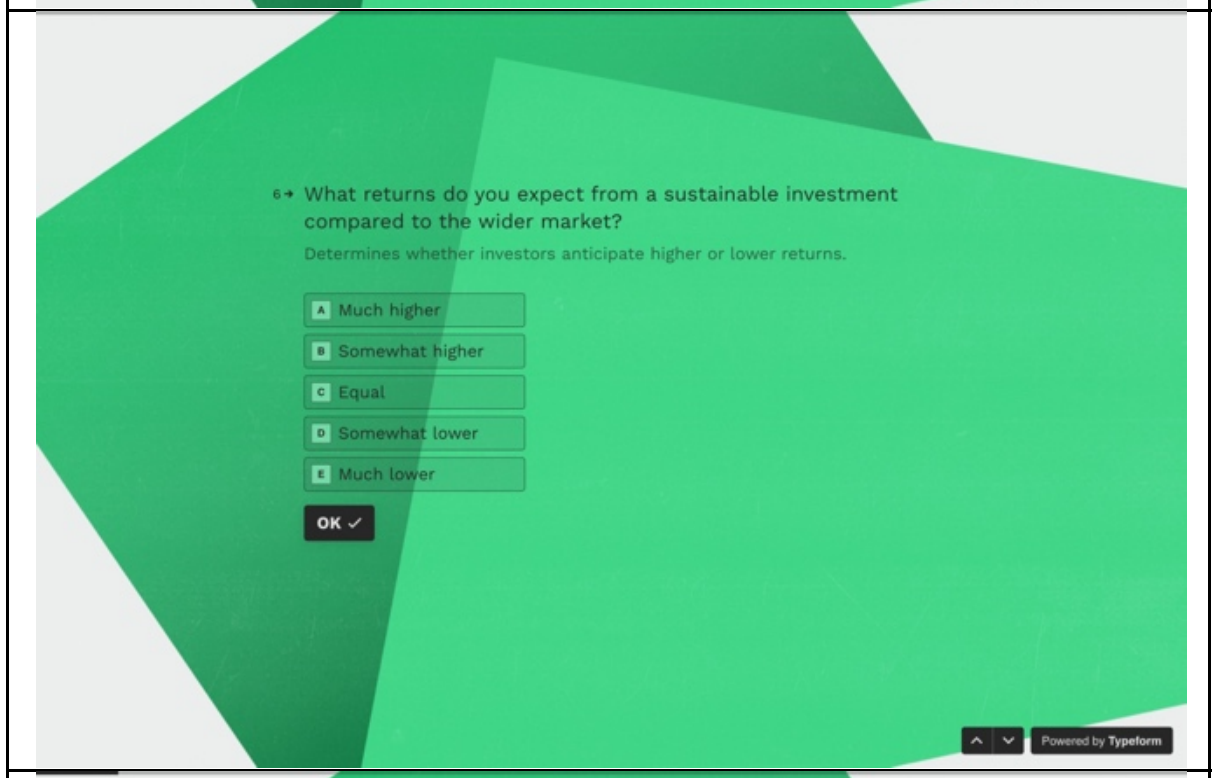
6→ What returns do you expect from a sustainable investment compared to the wider market?
Determines whether investors anticipate higher or lower returns.
A Much higher
B Somewhat higher
C Equal
D Somewhat lower
E Much lower
OK ✓
Powered by Typeform

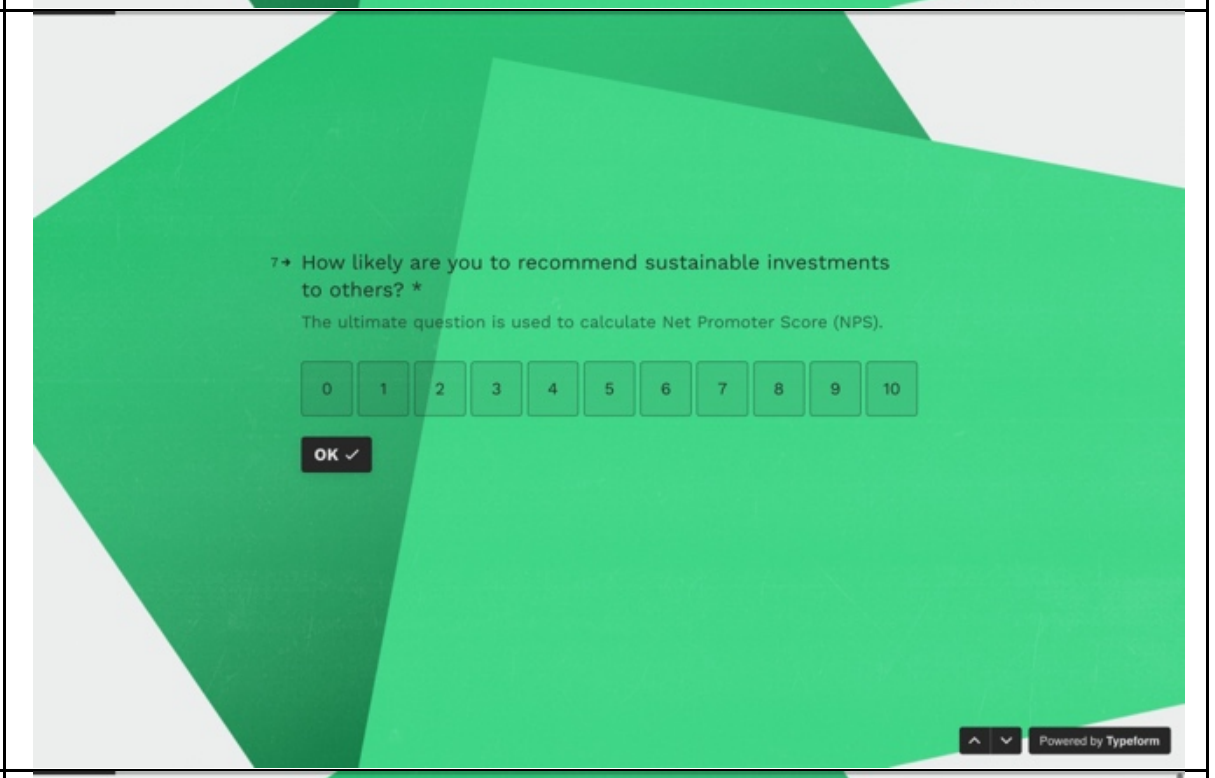
7→ How likely are you to recommend sustainable investments to others? *
The ultimate question is used to calculate Net Promoter Score (NPS).
0 1 2 3 4 5 6 7 8 9 10
OK ✓
Powered by Typeform

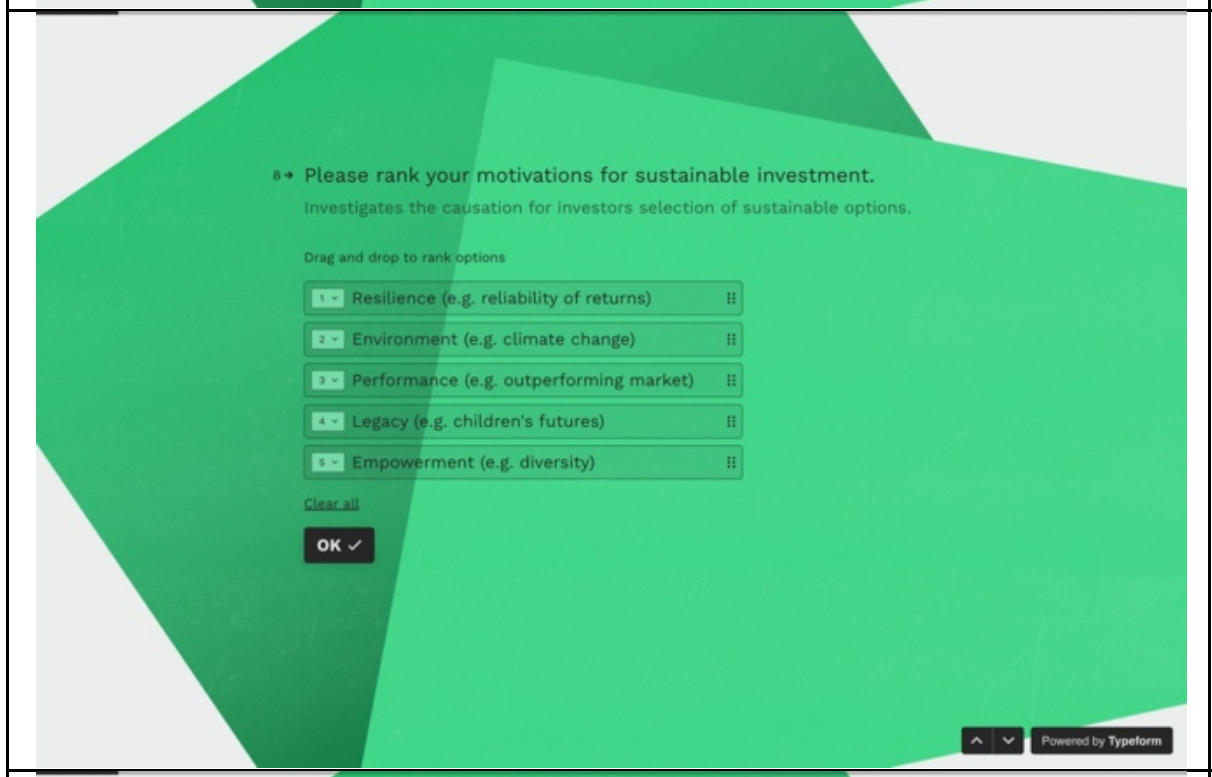
8→ Please rank your motivations for sustainable investment.
Investigates the causation for investors selection of sustainable options.
Drag and drop to rank options
1 Resilience (e.g. reliability of returns)
2 Environment (e.g. climate change)
3 Performance (e.g. outperforming market)
4 Legacy (e.g. children's futures)
5 Empowerment (e.g. diversity)
Clear all
OK ✓
Powered by Typeform

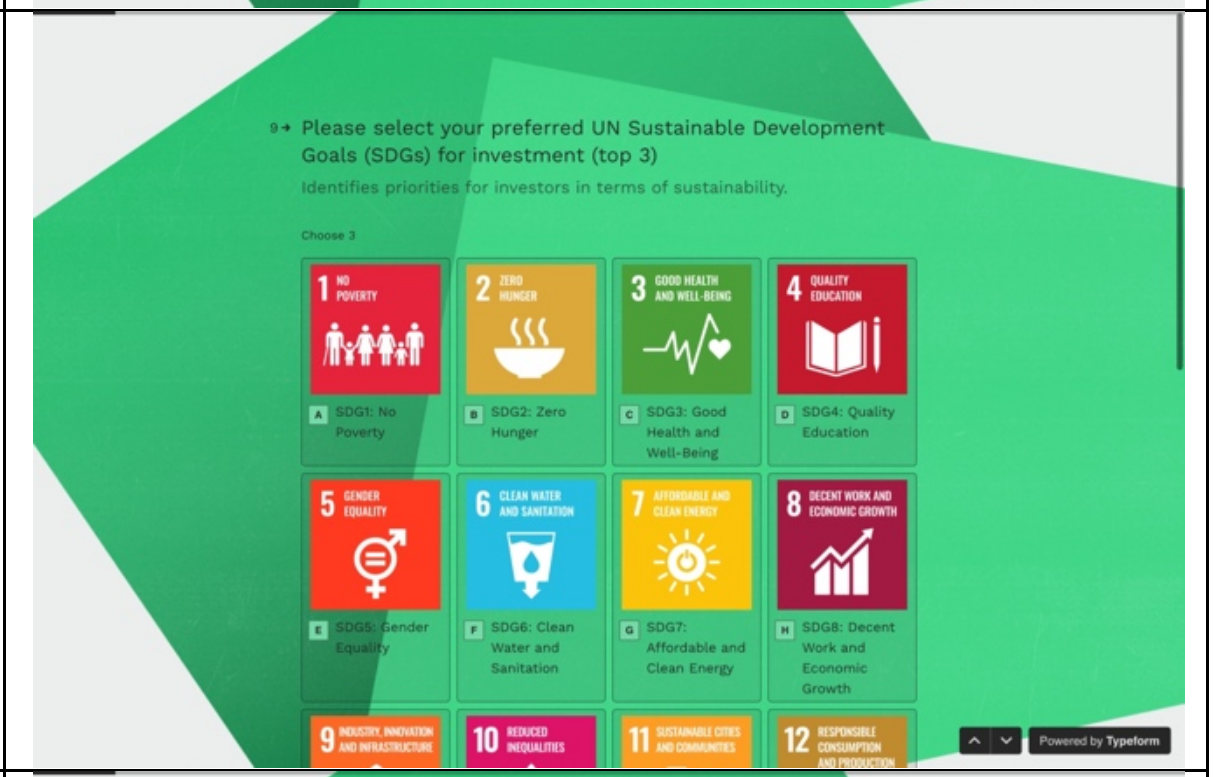
9→ Please select your preferred UN Sustainable Development Goals (SDGs) for investment (top 3)
Identifies priorities for investors in terms of sustainability.
Choose 3
A SDG1: No Poverty
B SDG2: Zero Hunger
C SDG3: Good Health and Well-Being
D SDG4: Quality Education
E SDG5: Gender Equality
F SDG6: Clean Water and Sanitation
G SDG7: Affordable and Clean Energy
H SDG8: Decent Work and Economic Growth
Powered by Typeform

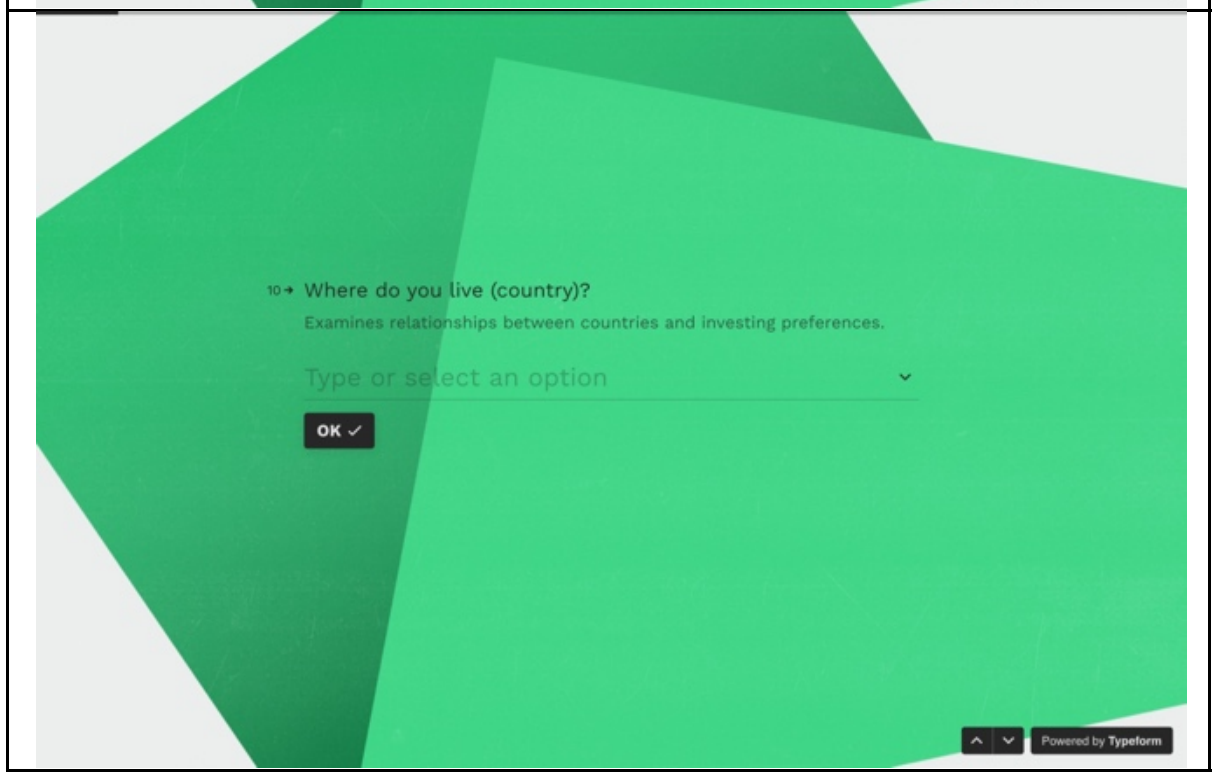
10→ Where do you live (country)?
Examines relationships between countries and investing preferences.
Type or select an option
OK ✓
Powered by Typeform

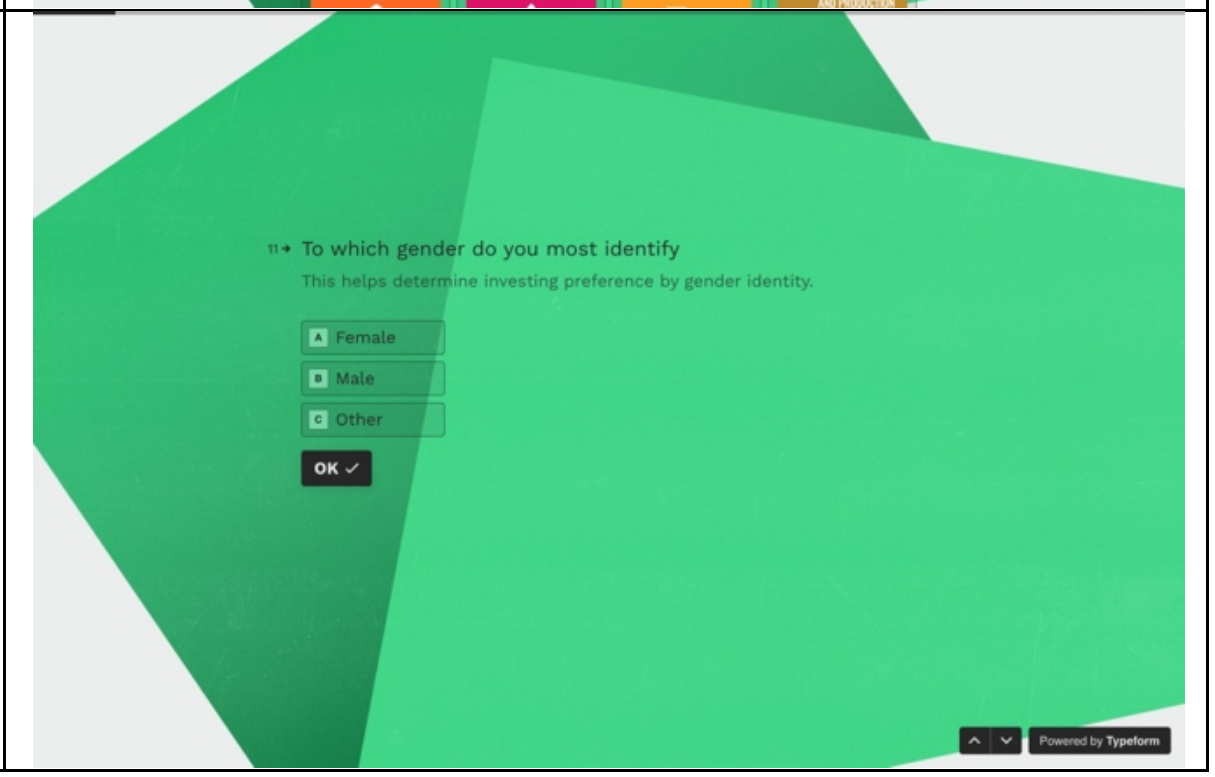
11→ To which gender do you most identify
This helps determine investing preference by gender identity.
A Female
B Male
C Other
OK ✓
Powered by Typeform

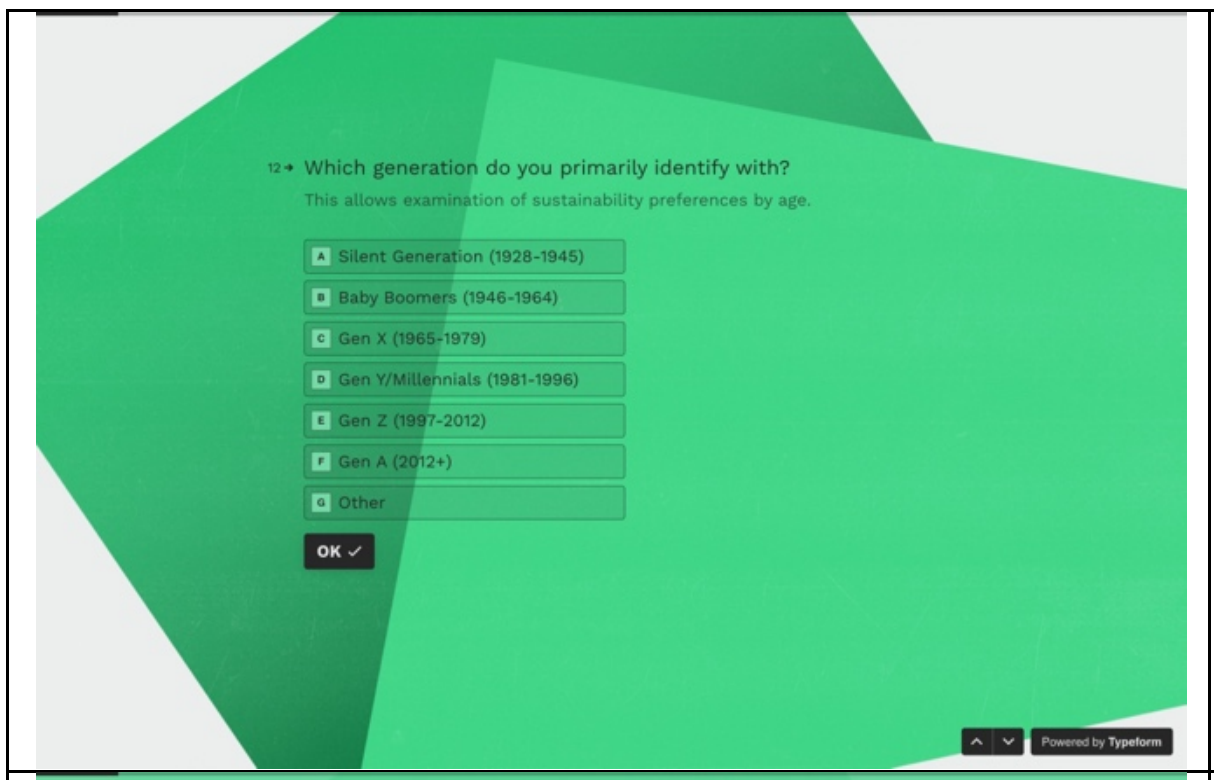
12→ Which generation do you primarily identify with?
This allows examination of sustainability preferences by age.
A Silent Generation (1928-1945)
B Baby Boomers (1946-1964)
C Gen X (1965-1979)
D Gen Y/Millennials (1981-1996)
E Gen Z (1997-2012)
F Gen A (2012+)
G Other
OK ✓
Powered by Typeform

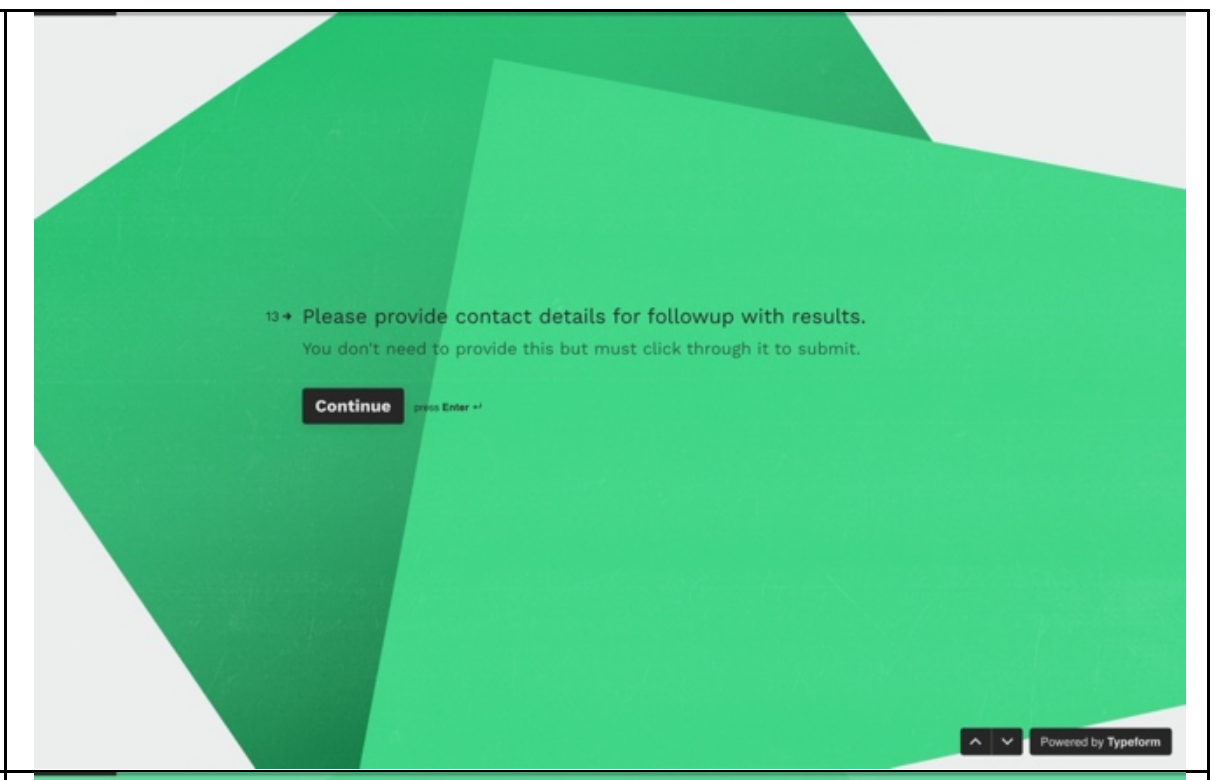
13→ Please provide contact details for followup with results.
You don't need to provide this but must click through it to submit.
Continue
press Enter ↵
Powered by Typeform

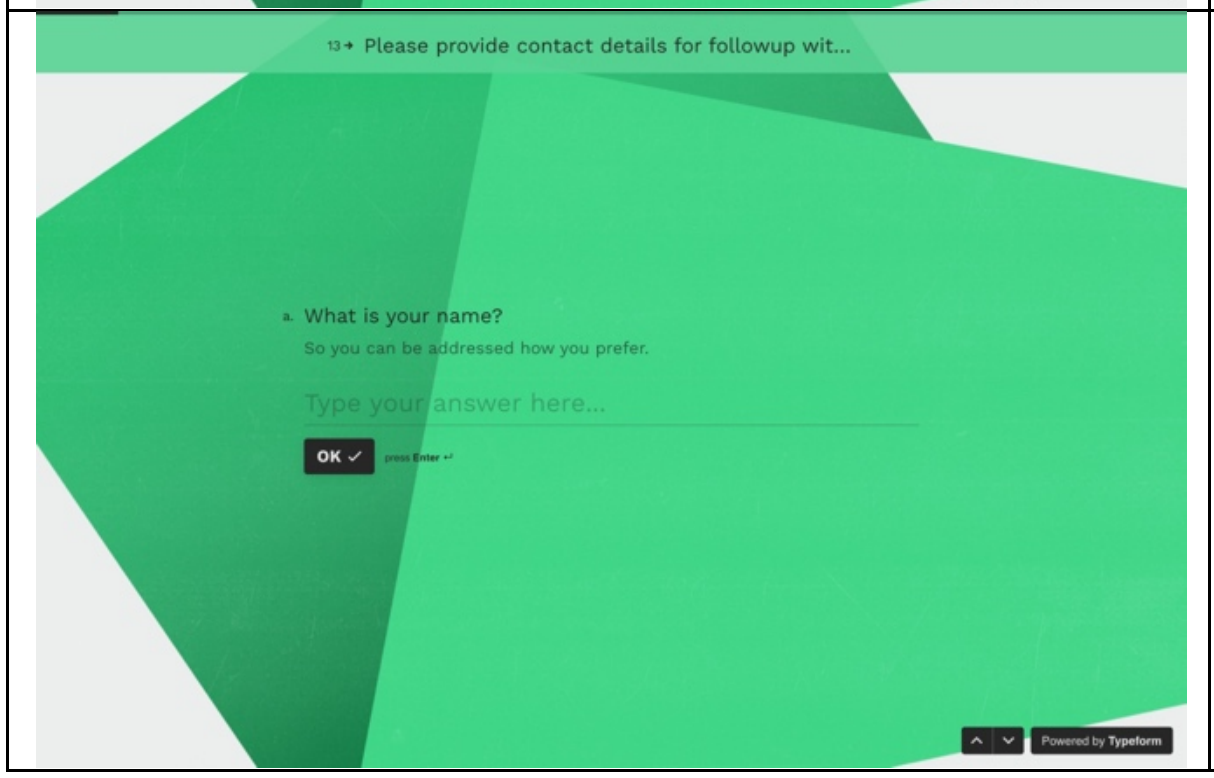
13→ Please provide contact details for followup wit...
a. What is your name?
So you can be addressed how you prefer.
Type your answer here...
OK ✓
press Enter ↵
Powered by Typeform

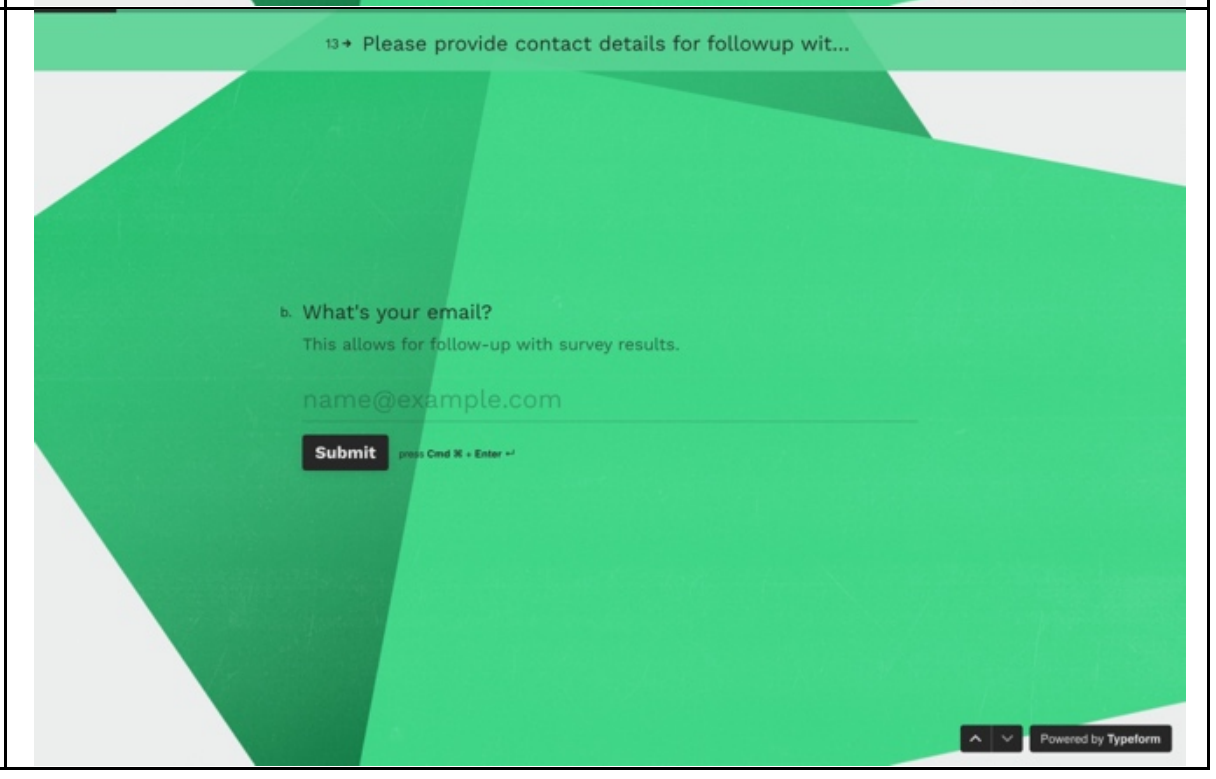
13→ Please provide contact details for followup wit...
b. What's your email?
This allows for follow-up with survey results.
name@example.com
Submit
Powered by Typeform

## A.2 Statistics (IBM SPSS)

IBM's SPSS 28.0.1.1 (14) was used for the analytics of survey data collected using Typeform.

### A.2.1 Survey

#### A.2.1.1 Recoding SPSS Syntax

The following SPSS Syntax was used to recode the data received from the survey tool (Typeform), exported as Microsoft Excel and/or transmitted to Google Workspace. The exported Excel file may need to be opened and saved in Excel to avoid an SPSS error ["(2003) All sheets in Excel file appear to be empty."].

```
* Encoding: UTF-8.

* Typeform metadata

rename variables (SubmittedAt=submitted_at) (Token=token).
variable labels
   submitted_at 'Typeform metadata: SubmittedAt'
   token 'Typeform metadata: Token'.

* Q1 type_of_investor: What type of investor are you?
  This allows classification of investors by level of sophistication

recode Whattypeofinvestorareyou
   ("I don't invest" = 0)
   ('Inactive (e.g., pension)' = 1)
   ('Retail (e.g. shares)'=2)
   ('Accredited (e.g., $1m assets/$200k income)'=3)
   ('Qualified (e.g., $5m investments)'=4)
   ('Institutional (e.g., corporate, pension)'=5)
   (else=6)
   into investor_type.
 execute.

value labels investor_type
   0 "I don't invest"
   1 'Inactive (e.g., pension)'
   2 'Retail (e.g. shares)'
   3 'Accredited (e.g., $1m assets/$200k income)'
   4 'Qualified (e.g., $5m investments)'
   5 'Institutional (e.g., corporate, pension)'
   6 'Others'.

delete variables Whattypeofinvestorareyou.

variable labels investor_type 'What type of investor are you?'.
formats investor_type(f2.0).
variable level investor_type (ordinal).
execute.

* Advanced investing scenarios => Accredited (e.g., $1m assets/$200k income)

if (Other = 'Active Startup investor owner') investor_type = 3.
if (Other = 'managed fund') investor_type = 3.
if (Other = 'Crypto') investor_type = 3.
execute.
delete variables Other.

* Create synthetic variables

recode investor_type (3 thru 4=1) (sysmis=sysmis) (else=0) into d_investor_type_pro.
formats d_investor_type_pro(f2.0).
execute.

* Q2 risk_appetite: How would you describe your risk appetite when it comes to your investment portfolio?
  Indicates the level of risk one is willing to accept to achieve the objectives for this type of investment.

recode Howwouldyoudescribeyourriskappetitewhenitcomestoyourinvestmentpo
   ('Very high' = 5)
   ('High' = 4)
   ('Neutral' = 3)
   ('Low' = 2)
   ('Very low' = 1)
   into risk_appetite.
 execute.

value labels risk_appetite
    5 'Very high'
    4 'High'
    3 'Neutral'
    2 'Low'
```

```
    1 'Very low'.

delete variables Howwouldyoudescribeyourriskappetitewhenitcomestoyourinvestmentpo.

variable labels risk_appetite 'How would you describe your risk appetite when it comes to your investment portfolio?'.
formats risk_appetite(f2.0).
variable level risk_appetite (ordinal).
execute.

recode risk_appetite (4 thru 5=1) (sysmis=sysmis) (else=0) into d_risk_appetite_high.
formats d_risk_appetite_high(f2.0).
execute.

* Q3 holding_period: How long would you typically hold an investment?
  Looks for investing preferences in relation to short- and longtermism.

recode Howlongwouldyoutypicallyholdaninvestment
   ('Less than 1 year' = 1)
   ('1-5 years' = 2)
   ('6-10 years' = 3)
   ('More than 10 years' = 4)
   into holding_period.
 execute.

value labels holding_period
    1 'Less than 1 year'
    2 '1-5 years'
    3 '6-10 years'
    4 'More than 10 years'.
 execute.

delete variables Howlongwouldyoutypicallyholdaninvestment.

variable labels holding_period 'How long would you typically hold an investment?'.
formats holding_period(f2.0).
variable level holding_period (ordinal).
execute.

recode holding_period (3 thru 4=1) (sysmis=sysmis) (else=0) into d_holding_period_long.
formats d_holding_period_long(f2.0).
execute.

* Q4 If you have one, what is your required rate of return or hurdle rate (%) for investments?
   The minimum annual return you would accept for any investment.

rename variables (Ifyouhaveonewhatisyourrequiredrateofreturnorhurdlerateforinvestm=hurdle_rate).
variable level hurdle_rate (scale).

* Q5 sustainability_preference: How likely are you to prefer a sustainable investment over a traditional investment?
  Assesses preference for sustainable investments over others in market.

* Excel export from Google Docs and/or import sets width to 11 but 'Very unlikely' is 13 - expand field to avoid warning:
   Warning # 4624 in column X: The preceding RECODE specifies a value to be recoded that is longer than some variable(s) in the recode
   The shorter values will be padded with blanks for the comparison.

recode Howlikelyareyoutopreferasustainableinvestmentoveratraditionalinv
   ('Very unlikely' = 1)
   ('Unlikely' = 2)
   ('Neutral' = 3)
   ('Likely' = 4)
   ('Very likely' = 5)
   into sustainability_preference.
 execute.

value labels sustainability_preference
    1 'Very unlikely'
    2 'Unlikely'
    3 'Neutral'
```

```
    4 'Likely'
    5 'Very likely'.

delete variables Howlikelyareyoutopreferasustainableinvestmentoveratraditionalinv.

variable labels sustainability_preference 'How likely are you to prefer a sustainable investment over a traditional investment?'.
formats sustainability_preference(f2.0).
variable level sustainability_preference (ordinal).
execute.

recode sustainability_preference (4 thru 5=1) (sysmis=sysmis) (else=0) into d_sustainability_preferred.
formats d_sustainability_preferred(f2.0).
execute.

* Q6 sustainability_returns: What returns do you expect from a sustainable investment compared to the wider market?
  Determines whether investors anticipate higher or lower returns.

recode Whatreturnsdoyouexpectfromasustainableinvestmentcomparedtothewid
   ('Much lower' = 1)
   ('Somewhat lower' = 2)
   ('Equal' = 3)
   ('Somewhat higher' = 4)
   ('Much higher' = 5)
   into sustainability_returns.
 execute.

value labels sustainability_returns
    1 'Much lower'
    2 'Lower'
    3 'Equal'
    4 'Somewhat higher'
    5 'Much higher'.

delete variables Whatreturnsdoyouexpectfromasustainableinvestmentcomparedtothewid.

variable labels sustainability_returns 'What returns do you expect from a sustainable investment compared to the wider market?'.
formats sustainability_returns(f2.0).
variable level sustainability_returns (ordinal).
execute.

* Q7 How likely are you to recommend sustainable investments to others?
   The ultimate question is used to calculate Net Promoter Score (NPS).

rename variables (Howlikelyareyoutorecommendsustainableinvestmentstoothers=nps).
variable level nps (scale).
variable labels nps 'How likely are you to recommend sustainable investments to others?'.

recode nps (0 thru 6=1) (sysmis=sysmis) (else=0) into nps_detractor.
recode nps (7 thru 8=1) (sysmis=sysmis) (else=0) into nps_passive.
recode nps (9 thru 10=1) (sysmis=sysmis) (else=0) into nps_promoter.
recode nps (0 thru 6=-1) (7 thru 8=0) (9 thru 10=1) (sysmis=sysmis) into nps_type.
formats nps_detractor(f2.0).
formats nps_passive(f2.0).
formats nps_promoter(f2.0).
formats nps_type(f2.0).
variable level nps_type (ordinal).
execute.

* Q8 motivation_ranking_1-5: Please rank your motivations for sustainable investment
  Investigates the causation for investors selection of sustainable options
  Split, recode, and label
  Each "," indicates a new answer
  rtrim eats spaces so we add them back in with value labels

string mr1 to mr5 (a50).

string #char(a1).
compute #index = 1.
vector mr = mr1 to mr5.
```

```
loop #pos = 1 to char.length(Pleaserankyourmotivationsforsustainableinvestment).
compute #char = char.substr(Pleaserankyourmotivationsforsustainableinvestment,#pos,1).
if(#char <> ",") mr(#index) = concat(rtrim(mr(#index)),#char).
if(#char = ",") #index = #index + 1.
end loop.
execute.

recode mr1 to mr5 (convert)
   ('Environment(e.g.climatechange)' = 1)
   ('Resilience(e.g.reliabilityofreturns)' = 2)
   ('Performance(e.g.outperformingmarket)' = 3)
   ('Legacy(e.g.children''sfutures)' = 4)
   ('Empowerment(e.g.diversity)' = 5)
   into motivation_ranking_1 to motivation_ranking_5.
execute.

 value labels motivation_ranking_1 to motivation_ranking_5
    1 'Environment'
    2 'Resilience'
    3 'Performance'
    4 'Legacy'
    5 'Empowerment'.

delete variables Pleaserankyourmotivationsforsustainableinvestment mr1 to mr5.

variable labels
   motivation_ranking_1 'Please rank your motivations for sustainable investment (1/5)'
   motivation_ranking_2 'Please rank your motivations for sustainable investment (2/5)'
   motivation_ranking_3 'Please rank your motivations for sustainable investment (3/5)'
   motivation_ranking_4 'Please rank your motivations for sustainable investment (4/5)'
   motivation_ranking_5 'Please rank your motivations for sustainable investment (5/5)'.
formats motivation_ranking_1 to motivation_ranking_5(f2.0).
execute.

* Q9 un_sdg_1-3: Please select your preferred UN Sustainable Development Goals (SDGs) for investment (top 3)
  Identifies priorities for investors in terms of sustainability
  sdg_x: Split, recode, and label
  Each "," indicates a new answer
  rtrim eats spaces so we add them back in with value labels

numeric sdg1 to sdg17 (f2.0).

recode SDG1NoPoverty (convert) (""=0) (else=1) into sdg1.
recode SDG2ZeroHunger (convert) (""=0) (else=1) into sdg2.
recode SDG3GoodHealthandWellBeing (convert) (""=0) (else=1) into sdg3.
recode SDG4QualityEducation (convert) (""=0) (else=1) into sdg4.
recode SDG5GenderEquality (convert) (""=0) (else=1) into sdg5.
recode SDG6CleanWaterandSanitation (convert) (""=0) (else=1) into sdg6.
recode SDG7AffordableandCleanEnergy (convert) (""=0) (else=1) into sdg7.
recode SDG8DecentWorkandEconomicGrowth (convert) (""=0) (else=1) into sdg8.
recode SDG9IndustryInnovationandInfrastructure (convert) (""=0) (else=1) into sdg9.
recode SDG10ReducedInequalities (convert) (""=0) (else=1) into sdg10.
recode SDG11SustainableCitiesandCommunities (convert) (""=0) (else=1) into sdg11.
recode SDG12ResponsibleConsumptionandProduction (convert) (""=0) (else=1) into sdg12.
recode SDG13ClimateAction (convert) (""=0) (else=1) into sdg13.
recode SDG14LifeBelowWater (convert) (""=0) (else=1) into sdg14.
recode SDG15LifeonLand (convert) (""=0) (else=1) into sdg15.
recode SDG16PeaceJusticeandStrongInstitutions (convert) (""=0) (else=1) into sdg16.
recode SDG17PartnershipsfortheGoals (convert) (""=0) (else=1) into sdg17.
execute.

delete variables
   SDG1NoPoverty
   SDG2ZeroHunger
   SDG3GoodHealthandWellBeing
   SDG4QualityEducation
   SDG5GenderEquality
   SDG6CleanWaterandSanitation
```

```
   SDG7AffordableandCleanEnergy
   SDG8DecentWorkandEconomicGrowth
   SDG9IndustryInnovationandInfrastructure
   SDG10ReducedInequalities
   SDG11SustainableCitiesandCommunities
   SDG12ResponsibleConsumptionandProduction
   SDG13ClimateAction
   SDG14LifeBelowWater
   SDG15LifeonLand
   SDG16PeaceJusticeandStrongInstitutions
   SDG17PartnershipsfortheGoals.

recode sdg1 to sdg17 (sysmis=0).
execute.

 variable labels
    sdg1 'SDG1: No Poverty'
    sdg2 'SDG2: Zero Hunger'
    sdg3 'SDG3: Good Health and Well-Being'
    sdg4 'SDG4: Quality Education'
    sdg5 'SDG5: Gender Equality'
    sdg6 'SDG6: Clean Water and Sanitation'
    sdg7 'SDG7: Affordable and Clean Energy'
    sdg8 'SDG8: Decent Work and Economic Growth'
    sdg9 'SDG9: Industry, Innovation, and Infrastructure'
    sdg10 'SDG10: Reduced Inequalities'
    sdg11 'SDG11: Sustainable Cities and Communities'
    sdg12 'SDG12: Responsible Consumption and Production'
    sdg13 'SDG13: Climate Action'
    sdg14 'SDG14: Life Below Water'
    sdg15 'SDG15: Life on Land'
    sdg16 'SDG16: Peace, Justice, and Strong Institutions'
    sdg17 'SDG17: Partnerships for the Goals'.
 execute.

* Q10 Where do you live (country)?
   Examines relationships between countries and investing preferences.

* Fix cases first

if(Wheredoyoulivecountry="Botswana") Wheredoyoulivecountry="Germany".

autorecode variables=Wheredoyoulivecountry
   /into country
   /blank=missing
   /print.
execute.
delete variables Wheredoyoulivecountry.
variable labels country 'Where do you live (country)?'.

* Q11 gender: To which gender do you most identify?
  This helps determine investing preference by gender identity.

recode Towhichgenderdoyoumostidentify
   ('Female' = 1)
   ('Male' = 2)
   ('' = sysmis)
   (else = 3)
   into gender.
 execute.

value labels gender
    1 'Female'
    2 'Male'
    3 'Other'.

delete variables Towhichgenderdoyoumostidentify.

* No gender and sexual diversity cases :(
```

```
delete variables Other_A.

variable labels gender 'To which gender do you most identify?'.
formats gender(f2.0).
variable level gender (nominal).
execute.

* Q12 generation: Which generation do you primarily identify with?
  This allows examination of sustainability preferences by age.

recode Whichgenerationdoyouprimarilyidentifywith
    ('Silent Generation (1928-1945)' = 1)
    ('Baby Boomers (1946-1964)' = 2)
    ('Gen X (1965-1979)' = 3)
    ('Gen Y/Millennials (1981-1996)' = 4)
    ('Gen Z (1997-2012)' = 5)
    ('Gen A (2012+)' = 6)
    (else = sysmis)
    into generation.
 execute.

value labels generation
     1 'Silent Generation (1928-1945)'
     2 'Baby Boomers (1946-1964)'
     3 'Gen X (1965-1979)'
     4 'Gen Y/Millennials (1981-1996)'
     5 'Gen Z (1997-2012)'
     6 'Gen A (2012+)'.

delete variables Whichgenerationdoyouprimarilyidentifywith.

variable labels generation 'Which generation do you primarily identify with?'.
formats generation(f2.0).
variable level generation (ordinal).
execute.

* Fix cases: pedants are millennials

if (Other_B = 'Between Gen X and Gen Y (fix your survey)') generation = 4.
execute.
delete variables Other_B.

* Create synthetic variables

recode generation (1 thru 3=1) (4 thru 6=0) (sysmis=sysmis) into d_generation_older.
formats d_generation_older(f2.0).
execute.

* Q13 Please provide contact details for followup with results
    You dont need to provide this but must click through it to submit

* Q13a What is your name?
    So you can be addressed how you prefer

rename variables (Whatisyourname=name).
variable labels name 'Please provide contact details for followup with results'.

* Q13b What's your email?
    This allows for follow-up with survey results

rename variables (Whatsyouremail=email).
variable labels email 'What''s your email?'.
execute.

* Purge personally identifialbe information (PII)

delete variables @# name email StartDateUTC SubmitDateUTC NetworkID.
```

sort variables by name.

### A.2.1.2 Analysis SPSS Syntax

The following SPSS Syntax was used to automatically execute the analysis itself, such that the survey data could be periodically updated as new responses were submitted.

```
* Encoding: UTF-8.

* Establish working directory

cd '/Users/samj/Library/CloudStorage/OneDrive-Personal/Education/MBA/ABR/Survey/'.

* Specify file path (make sure extension is included)

get file='dataset.sav'.

* Demographics

frequencies variables=gender generation investor_type
  /piechart percent
  /format=limit(10)
  /order=analysis.

* Countries

FREQUENCIES VARIABLES=country
  /PIECHART PERCENT
  /FORMAT=DFREQ LIMIT(10)
  /ORDER=ANALYSIS.

* Hurdle rate

EXAMINE VARIABLES=hurdle_rate
  /PLOT BOXPLOT STEMLEAF
  /COMPARE GROUPS
  /STATISTICS DESCRIPTIVES
  /CINTERVAL 95
  /MISSING LISTWISE
  /NOTOTAL
  /ID=hurdle_rate.

* Net Promoter Score (SNPS)

FREQUENCIES VARIABLES=nps
  /BARCHART PERCENT
  /ORDER=ANALYSIS.

FREQUENCIES VARIABLES=nps_type
  /BARCHART PERCENT
  /ORDER=ANALYSIS.

* Sustainable Development Goals (SDGs)

GRAPH
  /BAR(SIMPLE)=SUM(sdg17) SUM(sdg16) SUM(sdg15) SUM(sdg14) SUM(sdg13) SUM(sdg12) SUM(sdg11)
    SUM(sdg10) SUM(sdg9) SUM(sdg8) SUM(sdg7) SUM(sdg6) SUM(sdg5) SUM(sdg4) SUM(sdg3) SUM(sdg2)
    SUM(sdg1)
  /MISSING=LISTWISE.

CROSSTABS
  /TABLES=gender BY sdg1 sdg2 sdg3 sdg4 sdg5 sdg6 sdg7 sdg8 sdg9 sdg10 sdg11 sdg12 sdg13 sdg14
    sdg15 sdg16 sdg17
  /FORMAT=AVALUE TABLES
  /STATISTICS=CHISQ
  /CELLS=COUNT EXPECTED ROW COLUMN
  /COUNT ROUND CELL.
```

```
CROSSTABS
  /TABLES=d_generation_older BY sdg1 sdg2 sdg3 sdg4 sdg5 sdg6 sdg7 sdg8 sdg9 sdg10 sdg11 sdg12 sdg13 sdg14
    sdg15 sdg16 sdg17
  /FORMAT=AVALUE TABLES
  /STATISTICS=CHISQ
  /CELLS=COUNT EXPECTED ROW COLUMN
  /COUNT ROUND CELL.

* Motivation for sustainable investments is independent of age (χ2 (1) = 2.030; P = 0.154)

CROSSTABS
  /TABLES=d_generation_older BY motivation_ranking_1
  /FORMAT=AVALUE TABLES
  /STATISTICS=CHISQ
  /CELLS=COUNT ROW COLUMN TOTAL
  /COUNT ROUND CELL.

* Preference for sustainable investments is independent of gender (χ2 (1) = 1.432; P = 0.231)

CROSSTABS
  /TABLES=gender BY d_sustainability_preferred
  /FORMAT=AVALUE TABLES
  /STATISTICS=CHISQ
  /CELLS=COUNT ROW COLUMN TOTAL
  /COUNT ROUND CELL.

* Preference for sustainable investments is independent of age (χ2 (1) = 2.030; P = 0.154)

CROSSTABS
  /TABLES=d_generation_older BY d_sustainability_preferred
  /FORMAT=AVALUE TABLES
  /STATISTICS=CHISQ
  /CELLS=COUNT ROW COLUMN TOTAL
  /COUNT ROUND CELL.

* Preference for sustainable investments is independent of holding period (χ2 (1) = 0.835; P = 0.361)

CROSSTABS
  /TABLES=d_holding_period_long BY d_sustainability_preferred
  /FORMAT=AVALUE TABLES
  /STATISTICS=CHISQ
  /CELLS=COUNT ROW COLUMN TOTAL
  /COUNT ROUND CELL.

* Preference for sustainable investments is dependent on professionalism of investor (χ2 (1) = 4.966; P = 0.040)

CROSSTABS
  /TABLES=d_investor_type_pro BY d_sustainability_preferred
  /FORMAT=AVALUE TABLES
  /STATISTICS=CHISQ
  /CELLS=COUNT ROW COLUMN TOTAL
  /COUNT ROUND CELL.

* Preference for sustainable investments is independent of risk appetite (χ2 (1) = 1.419; P = 0.234)

CROSSTABS
  /TABLES=d_risk_appetite_high BY d_sustainability_preferred
  /FORMAT=AVALUE TABLES
  /STATISTICS=CHISQ
  /CELLS=COUNT ROW COLUMN TOTAL
  /COUNT ROUND CELL.

* Preference for sustainable investments is dependent on NPS answer (χ2 (2) = 33.670; P < 0.001)

CROSSTABS
  /TABLES=nps_type BY d_sustainability_preferred
  /FORMAT=AVALUE TABLES
  /STATISTICS=CHISQ
  /CELLS=COUNT ROW COLUMN TOTAL
  /COUNT ROUND CELL.
```

```
* One sample t-test (Gender)
   Even popluation of females (1) and males (2) with no others (3) averages 1.5

T-TEST
  /TESTVAL=1.5
  /MISSING=ANALYSIS
  /VARIABLES=gender
  /ES DISPLAY(TRUE)
  /CRITERIA=CI(.95).

* Independent samples t-test (Gender)

T-TEST GROUPS=gender(1 2)
  /MISSING=ANALYSIS
  /VARIABLES=nps risk_appetite investor_type holding_period sustainability_preference
    sustainability_returns
  /ES DISPLAY(TRUE)
  /CRITERIA=CI(.95).

* Independent samples t-test (Generation)

T-TEST GROUPS=d_generation_older(0 1)
  /MISSING=ANALYSIS
  /VARIABLES=nps risk_appetite investor_type holding_period sustainability_preference
    sustainability_returns
  /ES DISPLAY(TRUE)
  /CRITERIA=CI(.95).

* Independent samples t-test (Professional)

T-TEST GROUPS=d_investor_type_pro(0 1)
  /MISSING=ANALYSIS
  /VARIABLES=nps risk_appetite investor_type holding_period sustainability_preference
    sustainability_returns
  /ES DISPLAY(TRUE)
  /CRITERIA=CI(.95).

* Independent samples t-test (Sustainability Preferred)
   Those who prefer sustainable investments also expect higher returns from them

T-TEST GROUPS=d_sustainability_preferred(0 1)
  /MISSING=ANALYSIS
  /VARIABLES=nps risk_appetite investor_type holding_period sustainability_preference
    sustainability_returns
  /ES DISPLAY(TRUE)
  /CRITERIA=CI(.95).

* APA formatting

OUTPUT MODIFY
  /SELECT TABLES
  /IF COMMANDS=["Frequencies(LAST)"] SUBTYPES="Frequencies"
  /TABLECELLS SELECT=[VALIDPERCENT CUMULATIVEPERCENT] APPLYTO=COLUMN HIDE=YES
  /TABLECELLS SELECT=[TOTAL] SELECTCONDITION=PARENT(VALID MISSING) APPLYTO=ROW HIDE=YES
  /TABLECELLS SELECT=[VALID] APPLYTO=ROWHEADER UNGROUP=YES
  /TABLECELLS SELECT=[PERCENT] SELECTDIMENSION=COLUMNS FORMAT="PCT" APPLYTO=COLUMN
  /TABLECELLS SELECT=[COUNT] APPLYTO=COLUMNHEADER REPLACE="N"
  /TABLECELLS SELECT=[PERCENT] APPLYTO=COLUMNHEADER REPLACE="%".

* Sustainable Development Goals (SDG) histogram

MRSETS
  /MDGROUP NAME=$sdgs LABEL='UN Sustainable Development Goals (SDGs)' CATEGORYLABELS=VARLABELS
    VARIABLES=sdg1 sdg2 sdg3 sdg4 sdg5 sdg6 sdg7 sdg8 sdg9 sdg10 sdg11 sdg12 sdg13 sdg14 sdg15 sdg16
    sdg17 VALUE=1
  /DISPLAY NAME=[$sdgs].
```

```
CTABLES
  /VLABELS VARIABLES=$sdgs DISPLAY=LABEL
  /TABLE $sdgs [COUNT F40.0]
  /CATEGORIES VARIABLES=$sdgs ORDER=D KEY=COUNT EMPTY=INCLUDE
  /CRITERIA CILEVEL=95.
```

### A.2.2 Database

A company database (Crunchbase) was used to examine the state of the industry, however the data itself is subject to proprietary licenses and cannot be shared.

#### A.2.2.1 Recoding SPSS Syntax

The following SPSS Syntax was used to recode the data from the company database (Crunchbase), exported as CSV.

```
* Encoding: UTF-8.

* Extract year from dates (XDATE)

compute LastFundingYear = xdate.year(LastFundingDate).
compute FoundedYear = xdate.year(FoundedDate).
compute ExitYear = xdate.year(ExitDate).
compute ClosedYear = xdate.year(ClosedDate).
formats LastFundingYear FoundedYear ExitYear ClosedYear(f4.0).

compute d_sustainability = (char.index(upcase(IndustryGroups),'SUSTAINABILITY') > 0).
compute d_region_us = (char.index(upcase(HeadQuartersLocation),'UNITED STATES') > 0).
compute d_region_latam = (char.index(upcase(HeadQuartersRegions),'LATIN AMERICA') > 0).
compute d_region_eu = (char.index(upcase(HeadQuartersRegions),'EUROPEAN UNION') > 0).
compute d_region_apac = (char.index(upcase(HeadQuartersRegions),'APAC') > 0).
execute.

formats d_sustainability d_region_us d_region_latam d_region_eu d_region_apac (f4.0).

* Acquisition status

compute acquirer = (char.index(upcase(AcquisitionStatus),'MADE ACQUISITIONS') > 0).
compute acquiree = (char.index(upcase(AcquisitionStatus),'WAS ACQUIRED') > 0).
execute.
formats acquirer acquiree(f2.0).

* Diversity

compute diversity_women = (char.index(upcase(DiversitySpotlightUSOnly),'WOMEN') > 0).
compute diversity_black = (char.index(upcase(DiversitySpotlightUSOnly),'BLACK') > 0).
compute diversity_hispanic = (char.index(upcase(DiversitySpotlightUSOnly),'HISPANIC') > 0).
compute diversity_asian = (char.index(upcase(DiversitySpotlightUSOnly),'ASIAN') > 0).
formats diversity_women diversity_black diversity_hispanic diversity_asian(f2.0).
execute.

compute diversity=0.
if diversity_women=1 diversity=1.
if diversity_black=1 diversity=1.
if diversity_hispanic=1 diversity=1.
if diversity_asian=1 diversity=1.
formats diversity(f2.0).
execute.

* Number of employees

recode NumberofEmployees ('1-10'=1) ('11-50'=2) ('51-100'=3) ('101-250'=4) ('251-500'=5)
```

```
   ('501-1000'=6) ('1001-5000'=7) ('5001-10000'=8) ('10001+'=9) INTO employees.
execute.
formats employees(f2.0).

value labels employees
   1 '1-10'
   2 '11-50'
   3 '51-100'
   4 '101-250'
   5 '251-500'
   6 '501-1000'
   7 '1001-5000'
   8 '5001-10000'
   9 '10001+'.

* Estimated revenue range

recode EstimatedRevenueRange ('Less than $1M'=1) ('$1M to $10M'=2) ('$10M to $50M'=3) ('$50M to $100M'=4) ('$100M to $500M'=5)
   ('$500M to $1B'=6) ('$1B to $10B'=7) INTO revenue.
execute.
formats revenue(f2.0).

value labels revenue
   1 'Less than $1M'
   2 '$1M to $10M'
   3 '$10M to $50M'
   4 '$50M to $100M'
   5 '$100M to $500M'
   6 '$500M to $1B'
   7 '$1B to $10B'.

* Initial Public Offerings (IPOs)

recode IPOStatus ('Private'=0) ('Public'=1) into ipo.
variable labels  ipo 'IPO'.
formats ipo(f2.0).

compute moic_ipo = ValuationatIPOCurrencyinUSD / TotalEquityFundingAmountCurrencyinUSD.
compute moic_acq = PriceCurrencyinUSD / TotalEquityFundingAmountCurrencyinUSD.
execute.
```

### A.2.2.2 Analysis SPSS Syntax

The following SPSS Syntax was used to automatically execute the analysis itself, such that the survey data could be periodically updated as new responses were submitted.

```
* Encoding: UTF-8.

* Crunchbase analysis
   Load the appropriate dataset before running each command

* Total funding independent of sustainability

T-TEST GROUPS=d_sustainability(0 1)
  /MISSING=ANALYSIS
  /VARIABLES=TotalEquityFundingAmountCurrencyinUSD
  /ES DISPLAY(TRUE)
  /CRITERIA=CI(.95).

* Total employees independent of sustainability

T-TEST GROUPS=d_sustainability(0 1)
```

```
  /MISSING=ANALYSIS
  /VARIABLES=employees
  /ES DISPLAY(TRUE)
  /CRITERIA=CI(.95).

* Sustainability companies rank higher in CB Rank

T-TEST GROUPS=d_sustainability(0 1)
  /MISSING=ANALYSIS
  /VARIABLES=CBRankCompany
  /ES DISPLAY(TRUE)
  /CRITERIA=CI(.95).

* CB Rank 90-day trend independent of sustainability.

T-TEST GROUPS=d_sustainability(0 1)
  /MISSING=ANALYSIS
  /VARIABLES=TrendScore90Days
  /ES DISPLAY(TRUE)
  /CRITERIA=CI(.95).

* IPO independent of sustainability? Only 6 total/2 sustainability IPOs though

CROSSTABS
  /TABLES=d_sustainability BY ipo
  /FORMAT=AVALUE TABLES
  /STATISTICS=CHISQ
  /CELLS=COUNT EXPECTED ROW COLUMN TOTAL
  /COUNT ROUND CELL.

* Sustainability companies were not more likely to acquire others, but they were significantly less likely to be acquired.

CROSSTABS
  /TABLES=d_sustainability BY acquirer acquiree
  /FORMAT=AVALUE TABLES
  /STATISTICS=CHISQ
  /CELLS=COUNT EXPECTED ROW COLUMN TOTAL
  /COUNT ROUND CELL.

* Sustainability companies were not more likely to be active in IP.

T-TEST GROUPS=d_sustainability(0 1)
  /MISSING=ANALYSIS
  /VARIABLES=IPqweryIPActivityScore
  /ES DISPLAY(TRUE)
  /CRITERIA=CI(.95).

* Sustainability companies were not (quite) built with less tech.

T-TEST GROUPS=d_sustainability(0 1)
  /MISSING=ANALYSIS
  /VARIABLES=BuiltWithActiveTechCount
  /ES DISPLAY(TRUE)
  /CRITERIA=CI(.95).

* SEMrush Global Traffic Rank was (barely) independent of sustainability status.

T-TEST GROUPS=d_sustainability(0 1)
  /MISSING=ANALYSIS
  /VARIABLES=SEMrushGlobalTrafficRank
  /ES DISPLAY(TRUE)
  /CRITERIA=CI(.95).

* Number of investors & lead investors and funding rounds & last funding amount were independent of sustainability status.

T-TEST GROUPS=d_sustainability(0 1)
  /MISSING=ANALYSIS
  /VARIABLES=NumberofInvestors NumberOfLeadInvestors NumberOfFundingRounds LastEquityFundingAmount
```

```
  /ES DISPLAY(TRUE)
  /CRITERIA=CI(.95).

* Number of articles was independent of sustainability status.

T-TEST GROUPS=d_sustainability(0 1)
  /MISSING=ANALYSIS
  /VARIABLES=NumberofArticles
  /ES DISPLAY(TRUE)
  /CRITERIA=CI(.95).

* Sustainability companies had significantly fewer founders.

T-TEST GROUPS=d_sustainability(0 1)
  /MISSING=ANALYSIS
  /VARIABLES=NumberofFounders
  /ES DISPLAY(TRUE)
  /CRITERIA=CI(.95).

* Revenue independent of sustainability status.

CROSSTABS
  /TABLES=d_sustainability BY revenue
  /FORMAT=AVALUE TABLES
  /STATISTICS=CHISQ
  /CELLS=COUNT EXPECTED ROW COLUMN TOTAL
  /COUNT ROUND CELL.

* Diversity independent of sustainability status.

CROSSTABS
  /TABLES=d_sustainability BY diversity BY d_region_us
  /FORMAT=AVALUE TABLES
  /STATISTICS=CHISQ
  /CELLS=COUNT EXPECTED ROW COLUMN TOTAL
  /COUNT ROUND CELL.

* 5.19X (vs 5.70X0) median moic at sustainability IPO (40/10,704 vs 812/)

EXAMINE VARIABLES=moic_ipo
  /PLOT BOXPLOT STEMLEAF
  /COMPARE GROUPS
  /STATISTICS DESCRIPTIVES
  /CINTERVAL 95
  /MISSING LISTWISE
  /NOTOTAL
  /ID moic_ipo.

* 3.54x (vs 7.54x) median moic at sustainability acquisition

EXAMINE VARIABLES=moic_acq
  /PLOT BOXPLOT STEMLEAF
  /COMPARE GROUPS
  /STATISTICS DESCRIPTIVES
  /CINTERVAL 95
  /MISSING LISTWISE
  /NOTOTAL
  /ID moic_acq.
```

# Appendix B Data

The survey data analysed is provided for validation and further research.

country,d_generation_older,d_holding_period_long,d_investor_type_pro,d_risk_appetite_high,d_sustainability_preferred,gender,generat
ion,holding_period,hurdle_rate,investor_type,motivation_ranking_1,motivation_ranking_2,motivation_ranking_3,motivation_ranking_4,
motivation_ranking_5,nps,nps_detractor,nps_passive,nps_promoter,nps_type,risk_appetite,sdg1,sdg2,sdg3,sdg4,sdg5,sdg6,sdg7,sdg8,sdg
9,sdg10,sdg11,sdg12,sdg13,sdg14,sdg15,sdg16,sdg17,sustainability_preference,sustainability_returns,rank_environment,rank_resilience,r
ank_performance,rank_legacy,rank_empowerment
6,1,0,1,1,1,2,3,1, ,5,1,3,4,2,5,7,0,1,0,0,4,0,0,0,1,0,0,0,0,1,0,0,0,0,1,0,0,0,4,3,5,2,4,3,1
1,1,0,1,1,1,2,3,2, ,4,4,1,2,5,3,10,0,0,1,1,4,0,1,0,0,0,1,0,0,0,0,1,0,0,0,0,0,0,5,3,4,3,1,5,2
8,1,0,0,1,1,2,3,2,10,2,1,3,2,5,4,9,0,0,1,1,4,0,1,0,0,0,1,1,0,0,0,0,0,0,0,0,0,0,5,3,5,3,4,1,2
16,0,1,1,1,0,2,4,4, ,3,2,5,1,3,4,1,1,0,0,-1,5,0,1,0,1,0,0,0,0,0,0,0,1,0,0,0,0,0,2,2,3,5,2,1,4
13,0,0,0,0,1,1,4,2,15,2,4,1,2,3,5,6,1,0,0,-1,3,0,1,0,1,0,1,0,0,0,0,0,0,0,0,0,0,0,4,4,4,3,2,5,1
13,0,0,1,1,1,1,4,2, ,3,5,1,2,3,4,10,0,0,1,1,4,0,0,0,0,0,0,0,0,1,0,1,1,0,0,0,0,5,3,4,3,2,1,5
13,1,1,1,1,0,1,3,3,6,4,1,2,5,3,4,7,0,1,0,0,4,0,1,0,0,0,1,0,0,0,1,0,0,0,0,0,0,0,3,3,5,4,2,1,3
13,1,0,1,1,1,2,3,2,10,3,1,3,4,2,5,9,0,0,1,1,5,0,0,0,0,0,0,0,0,0,1,1,0,1,0,0,0,0,5,3,5,2,4,3,1
1, ,1,1,0,1,1, ,3,40,3,1,2,4,5,3,10,0,0,1,1,2,0,0,0,0,1,0,0,0,0,0,1,0,1,0,0,0,0,5,3,5,4,1,3,2
1,1,1,1,1,1,1,2,4,10,3,1,4,2,3,5,10,0,0,1,1,4,0,0,0,0,0,0,0,0,0,0,0,1,1,1,0,0,0,5,3,5,3,2,4,1
16,0,1,0,1,1,2,4,3,7,1,4,1,5,2,3,8,0,1,0,0,4,1,0,0,1,0,0,0,0,0,0,1,0,0,0,0,0,0,4,3,4,2,1,5,3
13,0,0,0,0,0,1,4,2,10,2,1,4,5,2,3,7,0,1,0,0,3,0,0,1,1,1,0,0,0,0,0,0,0,0,0,0,0,0,3,3,5,2,1,4,3
1,0,0,0,1,0,1,4,1, ,2,3,2,5,1,4,4,1,0,0,-1,4,1,0,0,0,0,0,1,0,0,1,0,0,0,0,0,0,0,2,3,2,4,5,1,3
20,0,0,0,1,1,1,4,2,5,2,1,4,2,3,5,7,0,1,0,0,4,0,1,1,0,0,1,0,0,0,0,0,0,0,0,0,0,0,4,3,5,3,2,4,1
2,1,0,0,0,1,2,3,2,8,2,3,2,5,1,4,8,0,1,0,0,3,0,0,1,0,0,0,0,1,0,1,0,0,0,0,0,0,0,4,2,2,4,5,1,3
16,0,1,0,0,1,2,4,3, ,2,1,4,2,3,5,9,0,0,1,1,3,0,0,0,0,0,0,1,1,0,0,0,0,1,0,0,0,0,5,4,5,3,2,4,1
20,0,1,0,1,1,2,4,3,15,2,3,2,1,4,5,5,1,0,0,-1,4,0,1,0,0,0,0,0,0,0,0,0,1,0,0,1,0,4,5,3,4,5,2,1
16,0,1,0,1,0,2,4,4,8,2,1,4,5,2,3,0,1,0,0,-1,5,0,0,0,0,0,0,0,0,0,0,1,1,1,0,0,0,0,2,1,5,2,1,4,3
20,1,0,0,0,1,2,2,2,0,2,4,1,5,2,3,9,0,0,1,1,3,1,0,0,0,0,1,0,0,0,0,0,0,1,0,0,0,0,4,2,4,2,1,5,3
16,0,0,0,1,1,2,4,2,12,2,4,1,2,3,5,5,1,0,0,-1,4,0,0,1,1,0,0,0,0,0,0,0,0,0,1,0,0,0,5,4,4,3,2,5,1
1,1,0,1,1,1,2,3,2,12,5,3,4,1,2,5,5,1,0,0,-1,4,1,0,0,0,0,0,1,0,0,0,0,0,1,0,0,0,0,4,3,3,2,5,4,1
1,1,1,0,1,1,2,3,3, ,1,1,4,3,2,5,10,0,0,1,1,5,0,0,0,0,0,0,1,0,0,1,0,0,1,0,0,0,0,4,3,5,2,3,4,1
15,1,1,0,0,1,2,3,3,4,1,1,3,2,4,5,7,0,1,0,0,2,0,0,0,1,0,0,1,0,0,0,0,0,1,0,0,0,0,4,4,5,3,4,2,1
5,1,1,0,0,1,2,3,3, ,2,4,1,5,3,2,9,0,0,1,1,3,0,1,0,1,0,0,0,0,0,0,0,0,1,0,0,0,0,4,3,4,1,2,5,3
6,0,1,0,0,1,2,4,3,4,2,1,4,2,3,5,9,0,0,1,1,3,0,0,0,1,0,0,0,0,0,1,0,1,0,0,0,0,0,5,4,5,3,2,4,1
8,0,0,0,0,1,2,4,2,8,2,4,3,2,5,1,7,0,1,0,0,3,0,1,0,1,0,1,0,0,0,0,0,0,0,0,0,0,0,4,4,1,3,4,5,2
13,1,0,0,1,1,2,3,2,25,2,4,1,2,3,5,8,0,1,0,0,4,0,0,0,0,0,1,0,0,0,0,1,0,1,0,0,0,0,4,3,4,3,2,5,1
13,0,0,0,1,0,2,4,2,15,2,1,3,4,2,5,5,1,0,0,-1,4,0,0,1,1,0,0,0,1,0,0,0,0,0,0,0,0,0,3,4,5,2,4,3,1
12,1,0,0,1,0,2,3,1,10,2,4,2,3,1,5,5,1,0,0,-1,4,1,1,0,1,0,0,0,0,0,0,0,0,0,0,0,0,0,3,2,2,4,3,5,1
9,1,0,0,1,0,2,3,2,50,2, , , , , ,2,1,0,0,-1,4,0,0,1,0,0,0,0,0,0,1,0,0,1,0,0,0,3,2, , , , ,
8,0,1,0,1,0,1,5,4,12,2,2,3,1,4,5,8,0,1,0,0,4,0,0,0,0,1,0,0,0,0,0,1,0,0,0,0,1,0,3,4,3,5,4,2,1
13,0,0,0,1,1,2,4,1,10,2,4,1,3,2,5,5,1,0,0,-1,5,0,0,0,0,0,1,0,0,0,0,0,0,1,0,1,0,0,4,4,4,2,3,5,1
13,0,0,0,0,0,1,4,1,10,2,5,4,1,3,2,5,1,0,0,-1,2,1,0,0,1,0,1,0,0,0,0,0,0,0,0,0,0,0,3,3,3,1,2,4,5
8,0,0,0,0,0,2,4,2,8,2,2,1,4,3,5,7,0,1,0,0,3,0,0,0,1,0,1,1,0,0,0,0,0,0,0,0,0,0,3,2,4,5,2,3,1
20,1,0,0,1,1,2,3,2,7,2,1,3,2,4,5,9,0,0,1,1,4,0,0,0,0,0,1,0,1,0,0,0,1,0,0,0,0,0,4,3,5,3,4,2,1
9,0,0,0,0,1,2,4,2,5,2,1,5,4,2,3,7,0,1,0,0,3,0,0,0,1,0,0,1,1,0,0,0,0,0,0,0,0,0,4,3,5,2,1,3,4
20,1,1,0,1,1,2,2,4,7,1,4,1,2,3,5,8,0,1,0,0,4,1,1,1,0,0,0,0,0,0,0,0,0,0,0,0,0,0,4,3,4,3,2,5,1
9,1,1,1,0,0, ,3,3,8,3,2,3,4,1,5,4,1,0,0,-1,3,0,1,0,0,0,0,0,0,1,1,0,0,0,0,0,0,0,3,3,2,5,4,3,1
13,1,1,1,0,0,1,3,3, ,4,3,2,1,5,4,6,1,0,0,-1,3,0,0,0,0,0,0,0,0,0,0,1,1,0,0,1,0,3,4,3,4,5,1,2
9,1,1,0,0,1,1,3,4,10,0,1,5,4,2,3,5,1,0,0,-1,2,0,0,0,1,1,0,0,0,0,0,0,1,0,0,0,0,4,2,5,2,1,3,4
18,0,0,0,1,0,2,4,2,20,2,3,2,1,4,5,7,0,1,0,0,4,0,0,0,0,0,1,1,0,1,0,0,0,0,0,0,0,0,3,3,3,4,5,2,1
1,0,0,0,0,1,1,4,2, ,0,1,2,4,3,5,10,0,0,1,1,2,0,0,0,0,0,0,1,0,0,0,1,0,1,0,0,0,0,5,2,5,4,2,3,1
11,0,1,0,0,1,1,4,3, ,2,4,5,2,1,3,9,0,0,1,1,3,0,0,0,0,1,0,0,1,0,0,0,1,0,0,0,0,0,4,3,2,3,1,5,4

13,1,1,0,0,1,2,3,3,7,2,1,4,2,3,5,8,0,1,0,0,3,1,0,0,0,0,0,0,0,0,0,0,1,1,0,0,0,0,5,3,5,3,2,4,1
5,0,0,0,1,1,2,4,1, ,0, , , , , ,8,0,1,0,0,5,0,0,1,0,0,0,0,0,0,0,0,1,0,0,0,0,1,5,3, , , , ,
1,1,0,1,1,0,2,3,2,9,3,3,2,5,1,4,5,1,0,0,-1,4,0,1,0,1,0,1,0,0,0,0,0,0,0,0,0,0,0,3,3,2,4,5,1,3
5,1,0,0,0,1,2,2,2,5,2,2,3,4,1,5,7,0,1,0,0,3,0,0,0,1,0,0,1,0,1,0,0,0,0,0,0,0,0,4,4,2,5,4,3,1
1,0,1,0,0,1,2,4,3, ,2,3,2,1,5,4,8,0,1,0,0,3,0,1,1,0,0,0,1,0,0,0,0,0,0,0,0,0,0,5,4,3,4,5,1,2
6,0,0,0,1,0,2,4,2,8,2,3,1,2,4,5,0,1,0,0,-1,4,0,0,0,0,0,0,0,0,0,0,1,0,0,1,1,0,0,3,2,4,3,5,2,1
1,1,1,1,1,0,2,3,3,15,3,1,4,2,3,5,5,1,0,0,-1,5,0,0,0,1,0,1,0,0,0,0,0,1,0,0,0,0,0,3,2,5,3,2,4,1
13,0,0,0,1,0,2,4,2,10,2,1,5,4,2,3,5,1,0,0,-1,4,1,1,0,0,0,1,0,0,0,0,0,0,0,0,0,0,0,2,2,5,2,1,3,4
17,0,0,0,1,0,2,4,2,10,2,3,1,2,4,5,5,1,0,0,-1,4,0,0,0,1,0,0,0,1,0,0,0,0,0,0,0,1,0,3,3,4,3,5,2,1
13,0,0,0,0,0,2,4,2,10,2,5,2,1,4,3,6,1,0,0,-1,3,0,0,0,0,0,0,0,0,1,1,0,1,0,0,0,0,0,3,3,3,4,1,2,5
1,1,1,1,1,1,2,3,3,8,5,1,2,4,3,5,8,0,1,0,0,4,1,1,0,1,0,0,0,0,0,0,0,0,0,0,0,0,0,5,3,5,4,2,3,1
1,1,0,0,1,1,2,3,2, ,2,1,5,4,3,2,9,0,0,1,1,4,0,0,0,0,0,1,1,0,0,0,0,1,0,0,0,0,0,5,3,5,1,2,3,4
10,1,1,0,0,0,2,3,4, ,2,2,3,1,4,5,5,1,0,0,-1,3,1,1,1,0,0,0,0,0,0,0,0,0,0,0,0,0,0,3,4,3,5,4,2,1
13,0,1,1,0,0,2,4,4,3,3,3,2,5,1,4,0,1,0,0,-1,3,1,0,0,0,1,0,0,0,0,0,0,0,0,0,1,0,3,3,2,4,5,1,3
13,0,1,1,0,0,1,4,3,15,3,3,1,2,4,5,8,0,1,0,0,3,0,0,0,0,0,0,0,0,1,0,0,0,0,1,1,0,0,3,3,4,3,5,2,1
1,1,1,0,0,1,1,3,4, ,2,1,4,3,5,2,9,0,0,1,1,2,0,0,0,1,0,0,1,0,0,0,0,0,1,0,0,0,0,5,3,5,1,3,4,2
3,0,0,0,1,0,2,4,1,25,2,2,3,4,1,5,3,1,0,0,-1,5,0,0,1,0,0,1,0,1,0,0,0,0,0,0,0,0,0,2,3,2,5,4,3,1
13,1,1,0,0,1,2,3,3,6,2,4,2,3,1,5,7,0,1,0,0,3,0,0,0,1,0,0,0,0,0,0,1,0,1,0,0,0,0,4,3,2,4,3,5,1
20,1,1,1,1,1,1,3,4,10,5,1,4,2,3,5,7,0,1,0,0,4,0,0,0,1,1,1,0,0,0,0,0,0,0,0,0,0,0,5,4,5,3,2,4,1
7,1,0,1,1,0,2,3,2,7,3,2,3,1,4,5,7,0,1,0,0,4,0,0,1,0,0,0,1,0,0,1,0,0,0,0,0,0,0,3,3,3,5,4,2,1
16,0,1,0,1,1,1,4,3,3,1,1,2,4,3,5,9,0,0,1,1,4,0,0,0,0,1,0,0,0,0,0,0,0,1,0,0,1,0,4,3,5,4,2,3,1
13,0,0,0,1,1,2,4,2, ,2,1,5,3,4,2,3,1,0,0,-1,4,0,0,0,1,0,0,0,0,0,0,0,0,1,0,0,1,0,4,3,5,1,3,2,4
4,0,0,0,1,1,2,4,2, ,2,1,4,5,2,3,4,1,0,0,-1,4,0,0,0,0,0,0,1,0,0,1,0,1,0,0,0,0,0,4,2,5,2,1,4,3
21,0,0,0,0,0,2,4,2, ,1,2,1,4,5,3,3,1,0,0,-1,2,1,0,0,1,0,0,0,0,0,0,0,0,1,0,0,0,0,3,2,4,5,1,3,2
13,0,0,1,1,1,2,4,2,7,3,3,2,1,4,5,9,0,0,1,1,4,0,0,0,0,0,0,1,0,0,0,0,1,1,0,0,0,4,3,3,4,5,2,1
21,0,1,0,1,1,2,4,3,5,2,3,1,2,4,5,6,1,0,0,-1,4,0,1,0,0,0,1,1,0,0,0,0,0,0,0,0,0,0,4,3,4,3,5,2,1
1,0,0,0,0,1,2,4,2,5,0,4,5,1,2,3,5,1,0,0,-1,3,0,0,1,1,0,1,0,0,0,0,0,0,0,0,0,0,0,4,4,3,2,1,5,4
13,0,0,0,0,0,2,4,2,7,2,1,4,3,2,5,8,0,1,0,0,3,1,1,0,0,0,0,0,0,0,1,0,0,0,0,0,0,0,3,3,5,2,3,4,1
6,1,1,0,0,1,1,2,4,3,2,1,4,5,2,3,8,0,1,0,0,2,0,0,0,0,1,0,0,0,0,0,0,0,1,1,0,0,0,5,2,5,2,1,4,3
1,0,0,1,1,1,2,4,2, ,3,1,3,2,4,5,9,0,0,1,1,4,0,0,0,1,0,0,1,0,0,0,1,0,0,0,0,0,0,4,3,5,3,4,2,1
13,1,0,0,1,1,2,3,2,20,2,1,2,3,4,5,9,0,0,1,1,4,0,0,0,0,0,1,1,0,0,0,0,0,1,0,0,0,0,4,3,5,4,3,2,1
13,0,1,0,1,1,2,4,3,8,2,4,1,5,2,3,10,0,0,1,1,4,0,0,0,0,0,0,1,0,0,0,1,1,0,0,0,0,0,5,3,4,2,1,5,3
8,1,0,0,1,0,2,3,2,35,2, , , , , ,5,1,0,0,-1,4,0,1,0,1,0,0,1,0,0,0,0,0,0,0,0,0,0,3,4, , , , ,
1,0,1,0,0,1,2,4,3,5,1,3,1,2,4,5,2,1,0,0,-1,1,0,1,0,0,0,0,0,1,0,0,0,0,0,1,0,0,0,4,3,4,3,5,2,1
1,0,1,0,0,1,2,4,3,10,2,1,4,2,3,5,5,1,0,0,-1,2,0,1,0,1,0,1,0,0,0,0,0,0,0,0,0,0,0,4,3,5,3,2,4,1
1,0,1,0,1,1,1,4,4,15,1,3,2,4,1,5,8,0,1,0,0,4,1,0,0,0,0,0,0,0,0,1,0,0,0,0,0,1,0,5,2,2,4,5,3,1
13,1,1,1,0,1,2,3,3,25,3,5,1,2,3,4,10,0,0,1,1,3,0,0,1,0,0,0,0,0,1,0,1,0,0,0,0,0,0,4,3,4,3,2,1,5
13,0,1,0,1,0,2,4,4,10,2,3,2,1,4,5,4,1,0,0,-1,4,0,1,0,1,0,0,0,0,0,0,1,0,0,0,0,0,0,3,3,3,4,5,2,1
1,1,1,1,1,1,2,3,3,8,5,1,4,3,2,5,9,0,0,1,1,4,0,0,0,0,0,0,1,0,0,0,0,1,1,0,0,0,0,5,3,5,2,3,4,1
13,1,0,0,0,1,2,3,2,15,2,3,2,5,1,4,10,0,0,1,1,2,1,0,0,1,0,0,1,0,0,0,0,0,0,0,0,0,0,5,4,2,4,5,1,3
1,0,0,0,0,1,2,4,2,8,2,2,3,1,4,5,5,1,0,0,-1,1,1,0,0,1,0,0,0,0,1,0,0,0,0,0,0,0,0,4,2,3,5,4,2,1
20,1,0,1,1,1,2,3,2,5,3,1,3,2,5,4,10,0,0,1,1,4,0,0,0,0,0,0,1,0,0,0,0,1,1,0,0,0,0,4,3,5,3,4,1,2
8,0,0,0,0,1,2,4,1,12,2,3,2,1,4,5,10,0,0,1,1,3,1,0,1,0,0,1,0,0,0,0,0,0,0,0,0,0,0,5,3,3,4,5,2,1
14,0,1,0,0,1,1,4,4, ,1,1,4,3,2,5,9,0,0,1,1,3,1,0,0,1,0,0,0,0,0,0,0,1,0,0,0,0,0,5,3,5,2,3,4,1
16,0,1,0,1,0,2,4,3,7,2,3,2,1,5,4,5,1,0,0,-1,4,1,1,0,0,0,0,0,0,0,0,1,0,0,0,0,0,0,3,3,3,4,5,1,2
8,0,1,0,0,0,2,4,4,10,1,3,4,2,1,5,7,0,1,0,0,2,1,1,0,0,0,0,0,0,1,0,0,0,0,0,0,0,0,3,3,2,3,5,4,1
16,0,0,0,1,1,1,4,2,8,2,3,1,2,4,5,9,0,0,1,1,5,0,0,0,0,0,0,1,0,0,1,0,1,0,0,0,0,0,4,4,4,3,5,2,1
8,0,0,0,1,1,2,4,2,10,2,3,2,4,1,5,10,0,0,1,1,4,0,0,1,0,0,1,0,0,1,0,0,0,0,0,0,0,0,5,4,2,4,5,3,1
8,0,0,0,1,1,1,4,1,35,2,1,5,2,3,4,7,0,1,0,0,4,0,0,0,0,1,0,0,0,0,0,0,0,1,0,0,1,0,4,4,5,3,2,1,4
19,1,0,0,1,1,2,3,2,25,1,2,3,1,5,4,10,0,0,1,1,4,0,0,1,0,0,1,1,0,0,0,0,0,0,0,0,0,0,5,3,3,5,4,1,2
20,0,0,0,1,0,2,4,2,5,2,2,1,3,4,5,6,1,0,0,-1,4,0,0,0,0,0,0,1,0,0,0,0,0,1,1,0,0,0,3,3,4,5,3,2,1
8,0,1,0,1,1,2,4,3,16,2, , , , , ,9,0,0,1,1,4,0,0,1,0,0,1,1,0,0,0,0,0,0,0,0,0,0,5,4, , , , ,
16,0,1,0,0,1,1,4,3, ,1,1,2,3,4,5,10,0,0,1,1,3,0,0,1,0,0,1,0,0,0,1,0,0,0,0,0,0,0,5,3,5,4,3,2,1
8,0,1,0,1,0,2,4,4,30,2,2,4,1,3,5,6,1,0,0,-1,4,0,1,0,0,0,0,0,0,0,0,0,1,1,0,0,0,0,3,3,3,5,2,4,1
20,0,0,0,0,1, ,4,2,5,1,1,3,2,5,4,0,1,0,0,-1,3,0,0,0,0,0,0,1,1,0,0,0,1,0,0,0,0,0,4,3,5,3,4,1,2
19,0,0,0,1,1,1,4,2,0,2,1,5,4,3,2,8,0,1,0,0,5,0,0,1,1,1,0,0,0,0,0,0,0,0,0,0,0,0,4,4,5,1,2,3,4
8,0,0,0,1,1,2,4,1,20,2,2,4,5,3,1,8,0,1,0,0,5,0,1,0,1,0,0,0,0,0,0,0,0,0,0,1,0,0,4,2,1,5,2,4,3
8,0,0,0,0,1,2,4,2,20,2,1,2,5,3,4,5,1,0,0,-1,2,0,1,1,0,0,1,0,0,0,0,0,0,0,0,0,0,0,4,4,5,4,2,1,3
20,1,0,1,0,1,2,2,2,8,5,1,5,4,2,3,7,0,1,0,0,3,0,1,0,0,0,1,0,0,0,1,0,0,0,0,0,0,0,4,2,5,2,1,3,4
13,0,0,0,0,1,2,4,2,6,2,2,1,4,3,5,7,0,1,0,0,3,0,1,1,0,0,1,0,0,0,0,0,0,0,0,0,0,0,4,2,4,5,2,3,1
8,0,0,0,0,1,2,4,2,11,0,1,4,5,2,3,10,0,0,1,1,3,0,1,0,0,0,0,1,0,0,0,0,0,1,0,0,0,0,5,4,5,2,1,4,3
19,1,1,0,0,1,1,3,3,0,2,1,5,4,3,2,7,0,1,0,0,2,0,0,0,0,1,0,0,0,0,0,0,0,1,0,0,1,0,4,3,5,1,2,3,4
13,0,0,0,1,1,2,4,2,20,2,1,3,2,5,4,10,0,0,1,1,4,0,0,0,1,0,1,0,0,0,0,0,0,1,0,0,0,0,4,4,5,3,4,1,2
13,0,0,0,0,0,2,4,2,10,0,1,4,2,3,5,8,0,1,0,0,2,0,0,1,1,0,1,0,0,0,0,0,0,0,0,0,0,0,3,3,5,3,2,4,1
13,0,0,0,0,0,2,4,2,1,2,1,4,2,3,5,5,1,0,0,-1,3,1,1,0,0,0,0,0,0,0,0,0,0,0,0,0,1,0,2,3,5,3,2,4,1
13,0,0,0,1,0,2,4,2,8,2,3,2,1,5,4,6,1,0,0,-1,4,0,1,1,0,0,0,0,0,0,0,0,1,0,0,0,0,0,3,3,3,4,5,1,2
13,0,0,0,1,0,2,4,2,30,2,2,3,1,4,5,5,1,0,0,-1,5,1,1,0,0,0,1,0,0,0,0,0,0,0,0,0,0,0,3,3,3,5,4,2,1
13,0,1,0,0,1,1,4,3,4,2,1,2,5,4,3,7,0,1,0,0,3,0,0,0,0,0,0,0,0,0,0,0,1,1,1,0,0,0,4,2,5,4,1,2,3

21,0,0,0,1,0,1,4,2, ,2,3,1,5,2,4,5,1,0,0,-1,5,0,0,0,0,0,0,0,1,0,1,0,1,0,0,0,0,0,3,3,4,2,5,1,3
13,0,1,0,0,1,1,4,4,4,1,1,2,4,5,3,10,0,0,1,1,3,0,0,0,0,0,0,1,0,1,0,0,0,1,0,0,0,0,5,4,5,4,1,3,2
13,1,0,1,1,1,2,3,2,10,3,4,1,2,3,5,9,0,0,1,1,5,0,0,0,0,0,0,1,0,0,1,0,0,1,0,0,0,0,4,3,4,3,2,5,1

# Appendix C License